\documentclass[letterpaper,12pt,onecolumn]{article}
\usepackage{multicol}
\usepackage{graphicx}
\usepackage{amssymb}
\usepackage{amsmath}
\usepackage{multirow}
\usepackage{booktabs}
\usepackage{hyperref}
\usepackage{comment}
\usepackage{systeme}
\usepackage{footmisc}
\usepackage{comment}
\usepackage{xcolor}
\usepackage{subfig}
\usepackage{caption}
\usepackage{rotating}
\usepackage{sidecap}
\usepackage{multibib}
\usepackage[utf8]{inputenc}
\usepackage[italian,english]{babel}
\usepackage[T1]{fontenc}
\usepackage{geometry}
\newcommand{\blue}{\textcolor{blue}}

\definecolor{OliveGreen}{rgb}{0,0.6,0}

\definecolor{orange}{rgb}{1,0.5,0}

\begin{document}

\pagenumbering{gobble}

\pagenumbering{arabic}

\begin{center}
{\Large{\bf  
\noindent{Fast magnetic field amplification in distant galaxy clusters}
}}
\end{center}

%\medskip
\noindent 
\textbf{Gabriella Di Gennaro$^1$, 
Reinout J. van Weeren$^1$, 
Gianfranco Brunetti$^2$, 
Rossella Cassano$^2$, 
Marcus Br\"uggen$^3$, 
Matthias Hoeft$^4$, 
Timothy W. Shimwell$^{1,5}$, 
Huub J.A. R\"ottgering$^1$, 
Annalisa Bonafede$^{2,3,6}$, 
Andrea Botteon$^{1,2}$,
Virginia Cuciti$^3$, 
Daniele Dallacasa$^{2,6}$, 
Francesco de Gasperin$^3$, 
Paola Dom\'{i}nguez-Fern\'{a}ndez$^3$, 
Torsten A. En{\ss}lin$^7$, 
Fabio Gastaldello$^8$, 
Soumyajit Mandal$^1$, 
Mariachiara Rossetti$^8$ \& 
Aurora Simionescu$^{1,9,10}$}

\medskip
\noindent$^1$\emph{Leiden Observatory, Leiden University, PO Box 9513, 2300 RA Leiden, The Netherlands}

\noindent$^2$\emph{Istituto Nazionale di Astrofisica-Istituto di Radioastronomia, Bologna Via Gobetti 101, I40129 Bologna, Italy}

\noindent$^3$\emph{Hamburger Sternwarte, Universit\"at Hamburg, Gojenbergsweg 112, 21029 Hamburg, Germany}

\noindent$^4$\emph{Th\"uringer Landessternwarte, Sternwarte 5, 07778 Tautenburg, Germany}

\noindent$^5$\emph{ASTRON, The Netherlands Institute for Radio Astronomy, Postbus 2, 7990 AA, Dwingeloo, The Netherlands}

\noindent$^6$\emph{Dipartimento di Fisica e Astronomia, Universit{\`a} degli Studi di Bologna, Via Gobetti 93/2, I40129 Bologna, Italy}

\noindent$^7$\emph{MPI for Astrophysics, Karl Schwarzschildstr. 1, 85741 Garching}

\noindent$^8$\emph{IASF-Milano/INAF, via Corti 12, 20133 Milano}

\noindent$^9$\emph{SRON Netherlands Institute for Space Research, Sorbonnelaan 2, 3584 CA Utrecht, The Netherlands}

\noindent$^{10}$\emph{Kavli Institute for the Physics and Mathematics of the Universe (WPI), The University of Tokyo, Kashiwa, Chiba 277-8583, Japan}

\medskip

\medskip

\renewcommand\thefigure{\arabic{figure}}
\setcounter{figure}{0}

{\bf In the present-day Universe, magnetic fields pervade galaxy clusters \cite{carilli+taylor02}, with strengths of a few microGauss obtained from Faraday Rotation \cite{bonafede+10}. 
Evidence for cluster magnetic fields is also provided by Megaparsec-scale radio emission, namely radio halos and relics \cite{vanweeren+19}. These are commonly found in merging systems \cite{cassano+10a} and are characterized by a steep radio spectrum ($\alpha<-1$, where $S_\nu\propto\nu^\alpha$).
It is widely believed that magneto-hydrodynamical turbulence and shock-waves (re-)accelerate cosmic rays \cite{brunetti+jones14}, producing halos and relics.
The origin and the amplification of magnetic fields in clusters is not well understood. It has been proposed that turbulence drives a small-scale dynamo \cite{dolag+05,subramanian+06,ryu+08,miniati+beresnyak15,vazza+18,dominguez-fernandez+19} that amplifies seed magnetic fields (primordial and/or injected by galactic outflows, as active galactic nuclei, starbursts, or winds \cite{donnert+18}).
At high redshift, radio halos are expected to be faint, due to the Inverse Compton losses and dimming effect with distance. Moreover, Faraday Rotation measurements are difficult to obtain. If detected, distant radio halos provide an alternative tool to investigate magnetic field amplification. 
Here, we report 
LOFAR observations which reveal diffuse radio emission in massive clusters when the Universe was only half of its present age, with a sample occurrence fraction of about 50\%. The high radio luminosities indicate that these clusters have similar magnetic field strengths to those in nearby clusters, and suggest that magnetic field amplification is fast during the first phases of cluster formation.
}

\bigskip
\bigskip

To investigate this unexplored territory, we present a systematic investigation of magnetic fields in distant galaxy clusters, using the low-frequency radio telescope LOFAR \cite{vanhaarlem+13}. Our observations were taken from the LOFAR Two-metre Sky Survey (LoTSS \cite{shimwell+19}). The 120--168~MHz LoTSS survey has a spatial resolution of approximately $6''$ and reaches a median sensitivity of about $70~\mu$Jy beam$^{-1}$ \cite{shimwell+19}. Currently, LoTSS covers about 40\% of the Northern sky. The clusters were selected from the latest {\it Planck} Sunyaev-Zel'dovich (SZ) PSZ2 catalog \cite{planckcoll16}, at redshift above 0.6 and declination above 20 degrees. The advantage of SZ-selection is that the SZ-signal ($y \propto \int n_e T_e dl$, i.e. the line of sight integral of the product of the electron number density, $n_e$, and the electron temperature, $T_e$) does not suffer from redshift dimming and the integrated cluster?s SZ-signal is closely related to the cluster mass %\footnote{$M_{\rm SZ,500}$ is } 
($M_{\rm SZ,500}$). The cluster masses are taken from the {\it Planck} catalog, and are obtained from the integrated cluster's SZ-signal within $R_{500}$, where $R_{500}$ is the radius with a density 500 times the critical density of the Universe at the given redshift.

\bigskip
\bigskip

Using this selection and taking the available LoTSS observations, we obtain a sample of 19~galaxy clusters in a redshift range of $\sim0.6-0.9$ and with masses $M_{\rm SZ, 500}\sim4-8\times10^{14}$ M$_\odot$. These objects are among the most massive structures at these redshifts. In Figure 1, %~\ref{fig:bestcases}, 
we show a number of full-resolution (i.e. about $6''$) LOFAR images of the detected diffuse radio sources from our sample, with radio contours tracing the low-resolution (approximately $15''$) emission, without compact sources (for the full sample, see Extended Data Figure 1). We find that at least nine out of the 19 clusters host diffuse radio emission (see Table~1). The diffuse emission is mostly centrally located in the clusters, with an overall roundish shape, and extending over scales spanning from few hundreds of kiloparsec to about one Megaparsec. In those systems that have targeted X-ray observations, the radio emission follows the thermal radiation,
see Figure 2 (for the full sample, see Extended Data Figure 2). %~\ref{fig:xray}. 
We therefore classify these diffuse sources as radio halos. From the X-ray images, most clusters look dynamically disturbed. Particularly noteworthy among the clusters in our sample are PSZ2\,G091.83+26.11 at $z=0.822$, where a very bright Megaparsec-sized radio halo and relic are present, and  PSZ2\,G160.83+81.66 at $z=0.888$, which hosts the most distant radio halo discovered to date, with a size of 0.7 Mpc.

\bigskip
\bigskip

Based on the radio flux density measurements and assuming $\alpha=-1.5\pm0.3$, %\cite{cassano+06} 
we compute radio luminosities in the range $P_{\rm 1.4GHz}\sim0.7-14\times10^{24}$ W Hz$^{-1}$ (Table 1). %\ref{tab:occurrence}).
Although most of the radiation in high-redshift halos is expected to be emitted via Inverse Compton (IC), the radio luminosities of these halos fall surprisingly within the scatter of those observed in nearby (median value $z\sim0.2$ \cite{cassano+13}) galaxy clusters of the same mass range (see Figure 3a). %\ref{fig:prad_m500_1.3}a). 

\bigskip
\bigskip

For synchrotron emission, the similar radio luminosities imply that the product of the number of the radio-emitting electrons and magnetic field ($N_e \times B^2$) in high-redshift clusters is similar to that in lower-redshift systems of comparable mass. Furthermore, this key observable may also suggests that both the magnetic field strength and the number of electrons in high-redshift clusters are comparable to those in low-$z$ systems, otherwise the energetics
of the halos would be very different while generating similar radio luminosities.

\bigskip
\bigskip

In re-acceleration models, the synchrotron luminosity is \cite{cassano+19,brunetti+vazza20}
 
\begin{equation}
P_{\rm rad}\propto \eta_{\rm rel} \frac{\rho v_t^3}{L_{\rm inj}} \frac{B^2}{B^2+B_{\rm CMB}^2}\, ,
\label{eq:power}
\end{equation}
where $\rho v_t^3/L_{\rm inj}$ is the turbulent energy flux (with $\rho$ the gas density, and $v_t$ and $L_{\rm inj}$ the turbulent velocity and injection scale, respectively), $\eta_{\rm rel}$ is the fraction of the turbulent energy flux that is dissipated in the re-acceleration of seed relativistic electrons, $B$ is the magnetic field averaged over the halo volume and $B_{\rm CMB}=3.25(1+z)^2~\mu$G is the CMB equivalent magnetic field strength. The ratio $\frac{B^2}{B^2+B_{\rm CMB}^2}$ sets the fraction of non-thermal luminosity that is emitted into synchrotron radiation. The stronger energy losses due to IC  ($dE/dt\propto(1+z)^4$) are expected to balance the effect of a potentially larger injection rate of electrons at higher redshift (due to enhanced activity of active galactic nuclei and/or star-forming galaxies). For this reason we can assume that the budget of seed particles that accumulate in the ICM at high redshift is similar to that at lower redshift (as detailed in the Methods). For a fixed budget of seed particles to re-accelerate, $\eta_{\rm rel}$ depends only on the interplay between turbulence and particles \cite{brunetti+jones14}, and %thus 
we assume it is independent of redshift. From the comparison of their radio luminosities, and
taking into account that mergers at {$z\sim0.7-0.8$} generate more turbulent energy flux compared to the $z=0.2$ sample (by a factor $\sim 3$, see Methods), we find that the magnetic field in high-$z$ halos has to be similar to that observed for local clusters, i.e. $B \geq 1 ~\mu$G. 
This suggests that the magnetic field amplification has been surprisingly efficient, producing a microGauss-level field in a $\rm Mpc^3$ volume already at $z\sim0.7$, i.e. within few Gyr from cluster formation (see Figure 3b). %\ref{fig:prad_m500_1.3}b). 
This result provides important insights on the origin and evolution of magnetic fields in galaxy clusters.

\bigskip
\bigskip

It is widely accepted that the small-scale turbulent dynamo plays a role in the amplification of the initial magnetic field from primordial seeds or galactic outflows \cite{donnert+18}. Initially, the amplification operates in a kinematic regime, where the magnetic field grows exponentially with time $(B^2(t) \sim B^2_0 \exp (t \, \Gamma))$. The competition of turbulent stretching and diffusion makes this phase initially very slow, with a growth time-scale $\Gamma^{-1}=30 L_{\rm inj} / ({\rm Re}^{1/2} v_t)$ \cite{cho14,beresnyak+miniati16}, where $\rm Re$ is the Reynolds number.
When the magnetic and the kinetic energy densities become comparable at the viscous dissipation scale, the turbulent dynamo becomes faster and transits to a phase where the magnetic field grows linearly with time. 
During this phase, the magnetic field reaches equipartition with the kinetic turbulent energy at increasingly larger scales, and saturates after several eddy turnover times (several Gyrs).
In this scenario, our observations then require a fast magnetic amplification during the initial exponential phase, as it needs to be much shorter than a few Gyrs. This constrains the initial field and the ICM Reynolds number, being ${\rm Re}> 4\times 10^4$ and $>5\times10^3$ for $B_0 \sim 1$ nG and $\sim 0.1~\mu$G, respectively, assuming a continuous injection of turbulence with $v_t=500$ km s$^{-1}$ at the injection scale $L_{\rm inj}=1$ Mpc \cite{dolag+05,donnert+18,hitomicoll18}, the ICM number density of $n=3\times10^{-3}$ cm$^{-3}$ \cite{markevitch+vikhlinin07}, and a time available for the magnetic field amplification of $\sim3.7$ Gyr (see Figure 3b and Methods). %\ref{fig:prad_m500_1.3}b and Methods). 
These values of the Reynolds number are much larger than the classical value obtained assuming Coulomb collisions ($\approx 100$ \cite{brunetti+lazarian07,cho14}), and would suggest that kinetic effects and instabilities play an important role in the weakly-collisional magnetized ICM \cite{schekochihin+cowley06}. This is also in line with recent X-ray observations in the ICM of local systems that have suggested a much smaller viscosity than the isotropic value obtained only considering Coulomb collisions \cite{zhuravleva+19}. In case of a smaller Reynolds number, our observations would require that the activity of galactic outflows and AGN in high redshift clusters is sufficient to generate microGauss fields spread on $>100$ kpc scales \cite{xu+11}. 
This would generate a clumpier distribution of the radio emission in the cluster volume. This can be tested with deeper observations, in combination with predictions of the spatial distribution of radio emission from cosmological magneto-hydrodynamic simulations that explore different magnetogenesis scenarios.

\newpage

\begin{figure}[t!]
\centering
{\includegraphics[width=\textwidth]{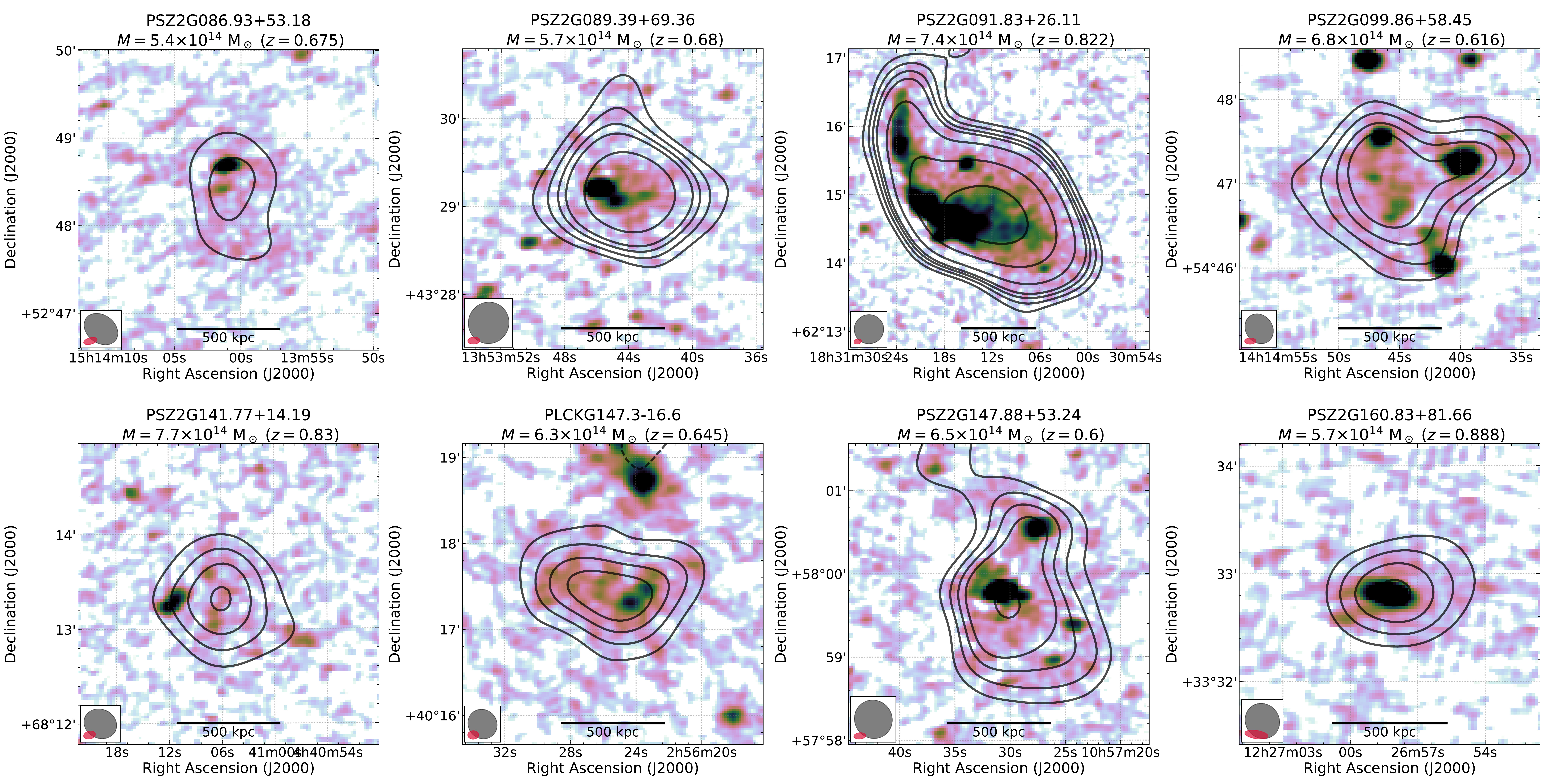}}
\vspace{-6mm}
\caption*{{\bf Figure 1:  Examples of observed radio emission in our high-$z$ galaxy cluster sample.} In colorscale we show the full-resolution LOFAR images, while radio contours show the low-resolution emission after the subtraction of compact sources, displayed at the $[-2,2,3,4,5,8,16]\times\sigma_{\rm rms}$ level (with $\sigma_{\rm rms}$ the individual map noise; the negative contour levels are indicated with a short-dashed line style). The full- and low-resolution LOFAR beams are displayed in the bottom left corner (in pink and grey colors, respectively). In the header of each image, the galaxy cluster name, mass and redshift are reported.}
\label{fig:bestcases}
\end{figure}

\begin{figure*}
\centering
\hspace{-4mm}
{\includegraphics[height=0.22\textwidth]{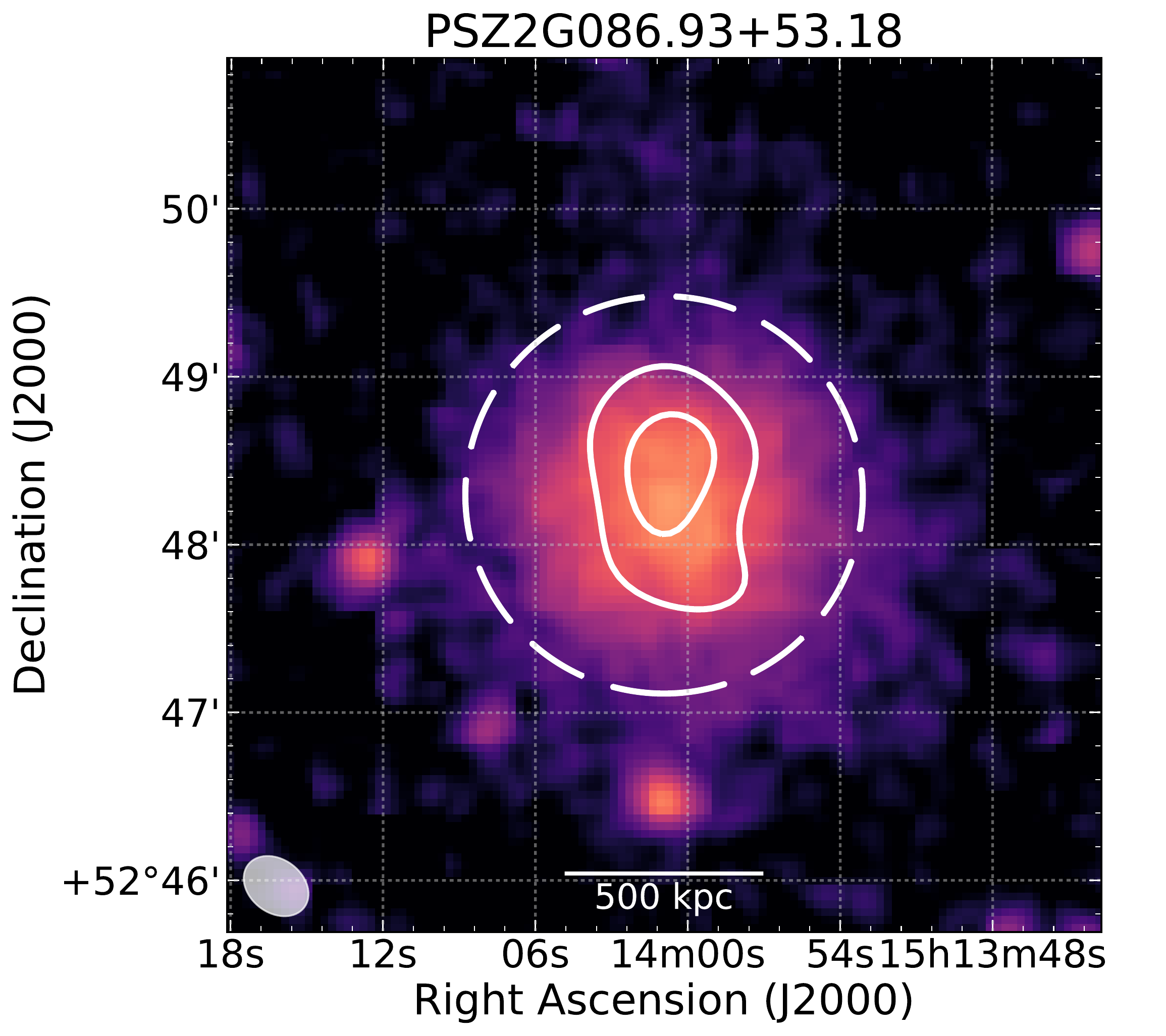}}
\hspace{-4mm}
{\includegraphics[height=0.22\textwidth]{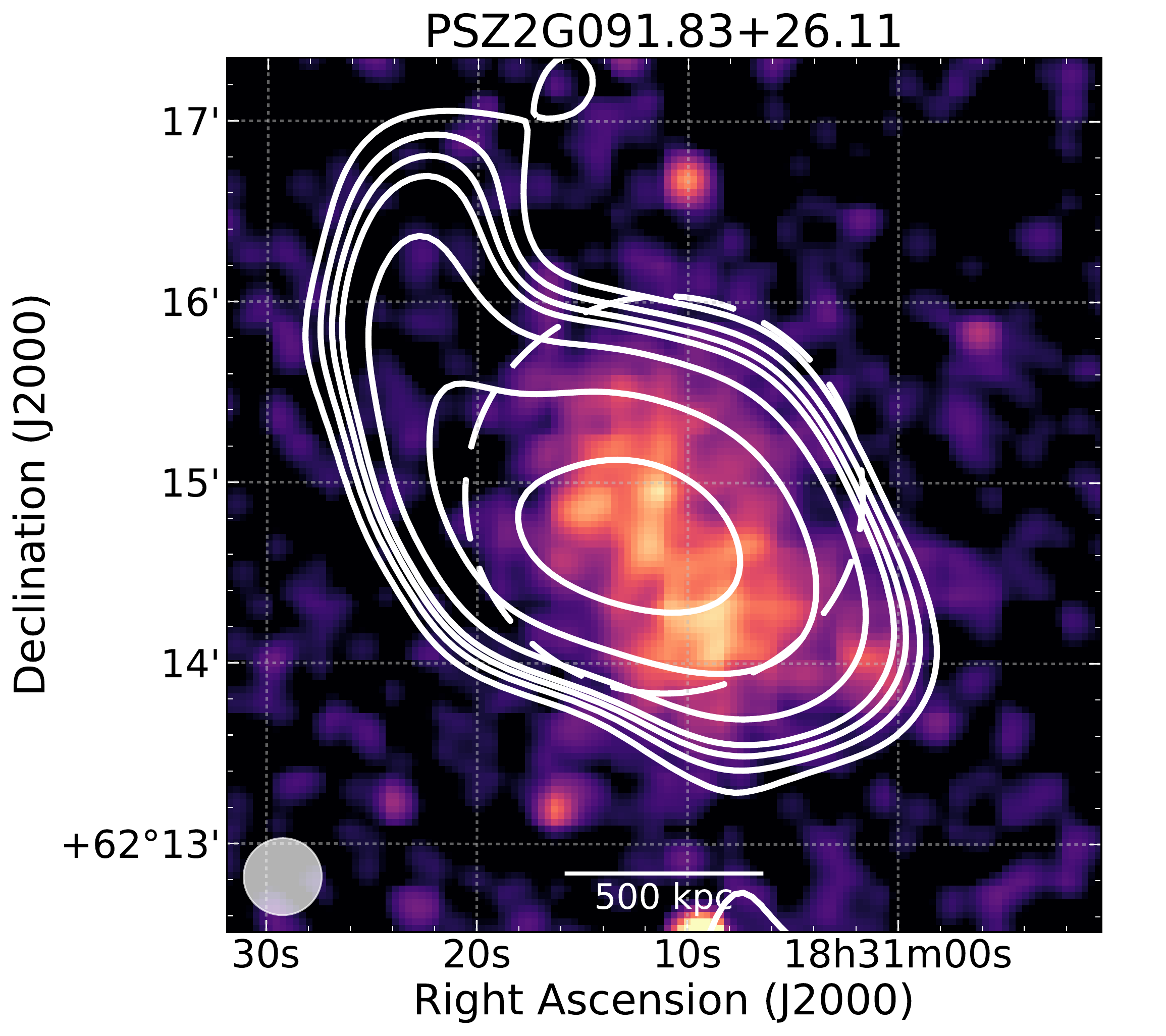}}
\hspace{-4mm}
{\includegraphics[height=0.22\textwidth]{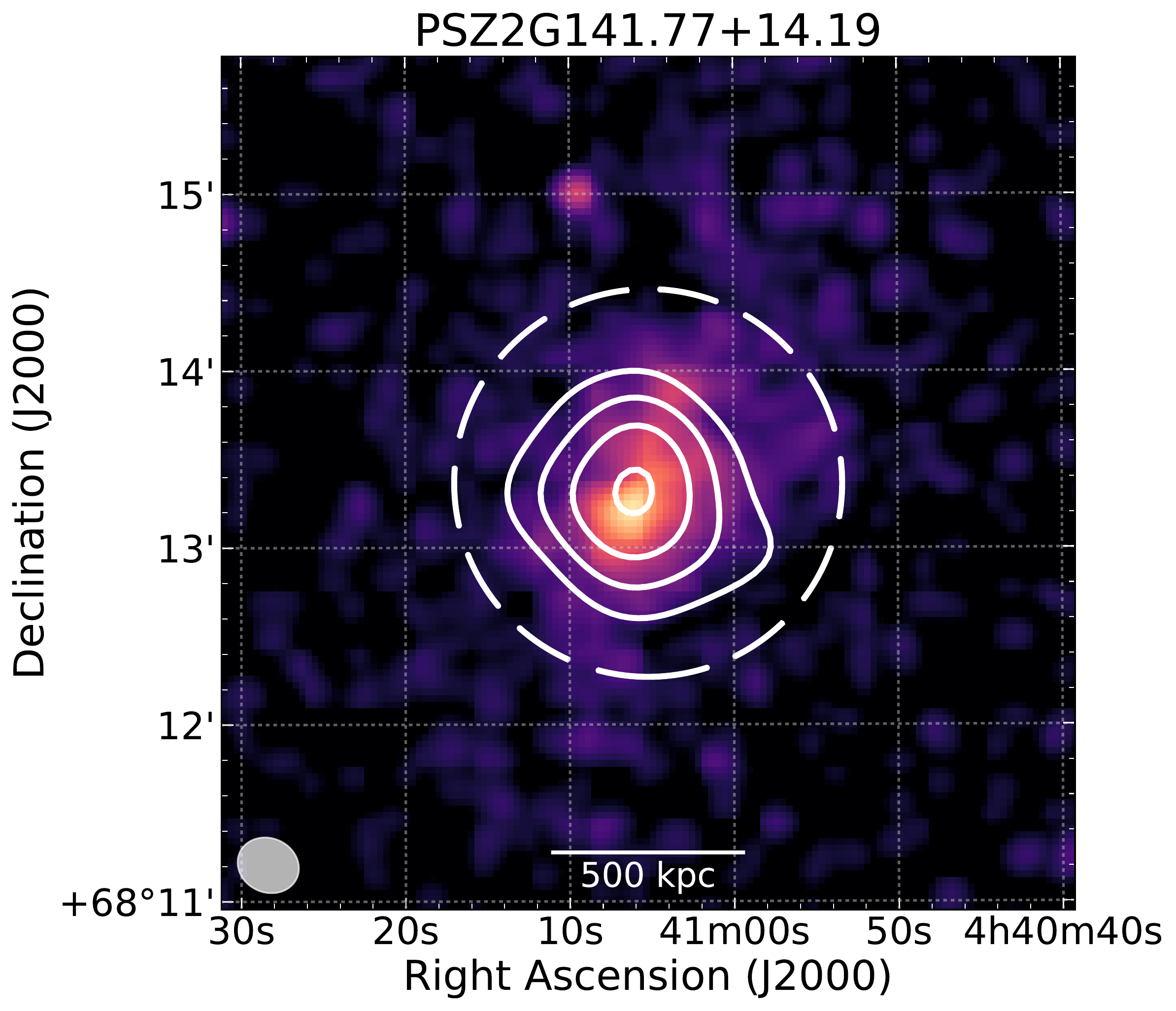}}
\hspace{-4mm}
{\includegraphics[height=0.22\textwidth]{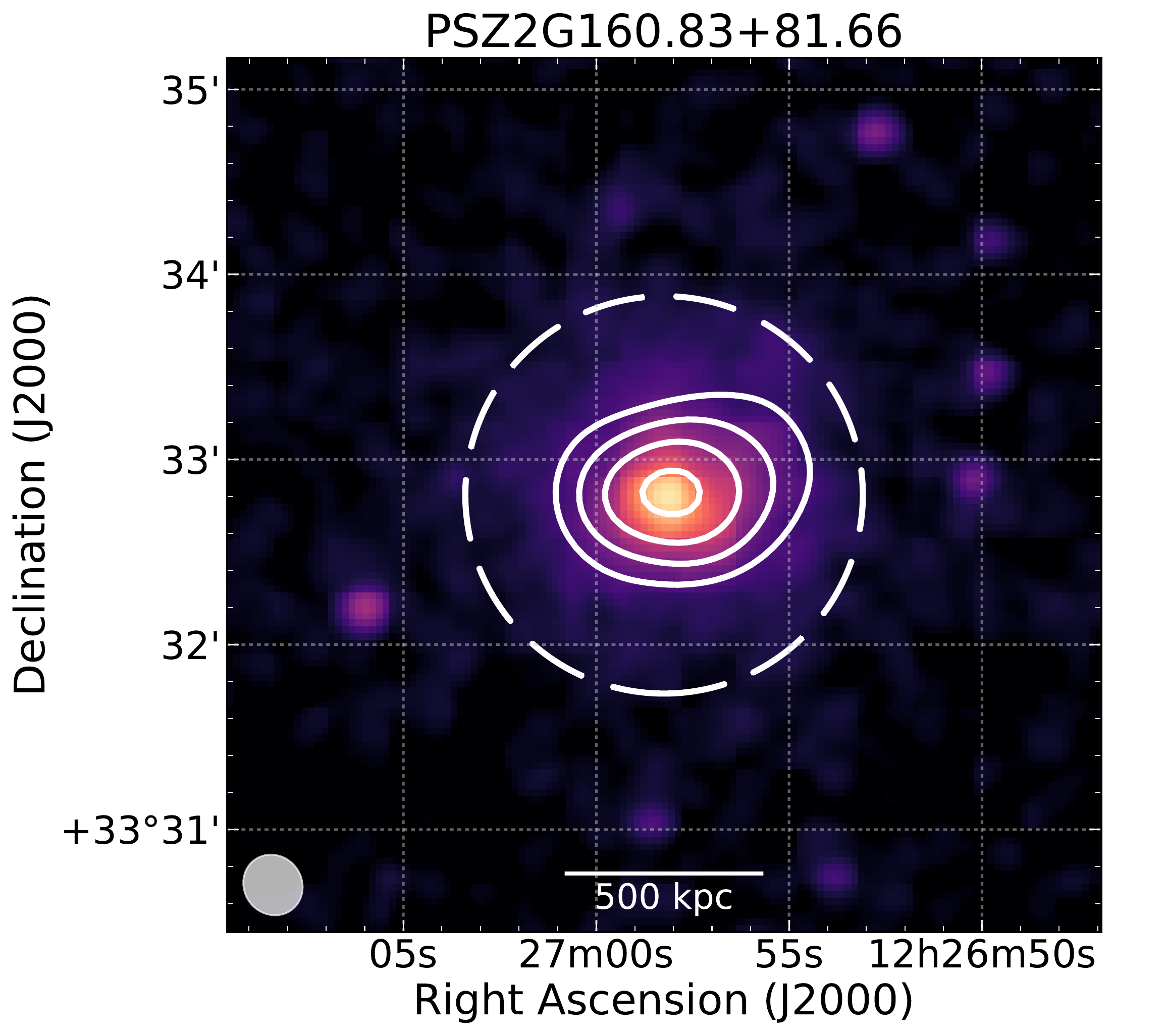}}
\caption*{{\bf Figure 2: X-ray images of a sub-sample of Figure 1.} LOFAR radio contours are drawn as Figure 1, with the LOFAR beam displayed in the bottom left corner. The dashed white circle in each map shows the $R=0.5R_{\rm SZ,500}$ region, obtained from $M_{\rm SZ,500}$. In the header of each image, the galaxy cluster name is reported.}
\label{fig:xray}
\end{figure*}

\begin{figure*}[t!]
\centering
{\includegraphics[width=0.52\textwidth]{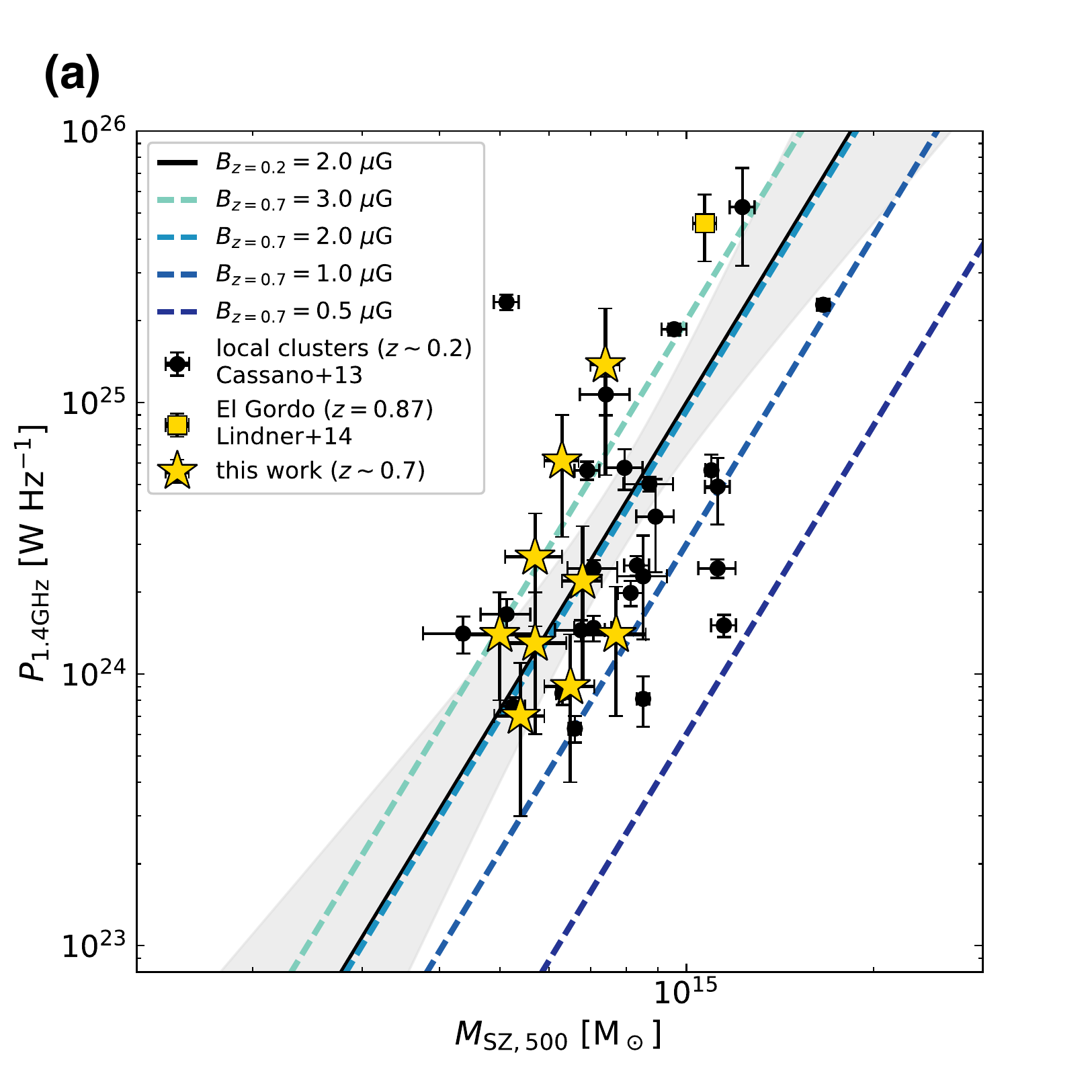}}
\hspace{-10mm}
{\includegraphics[width=0.52\textwidth]{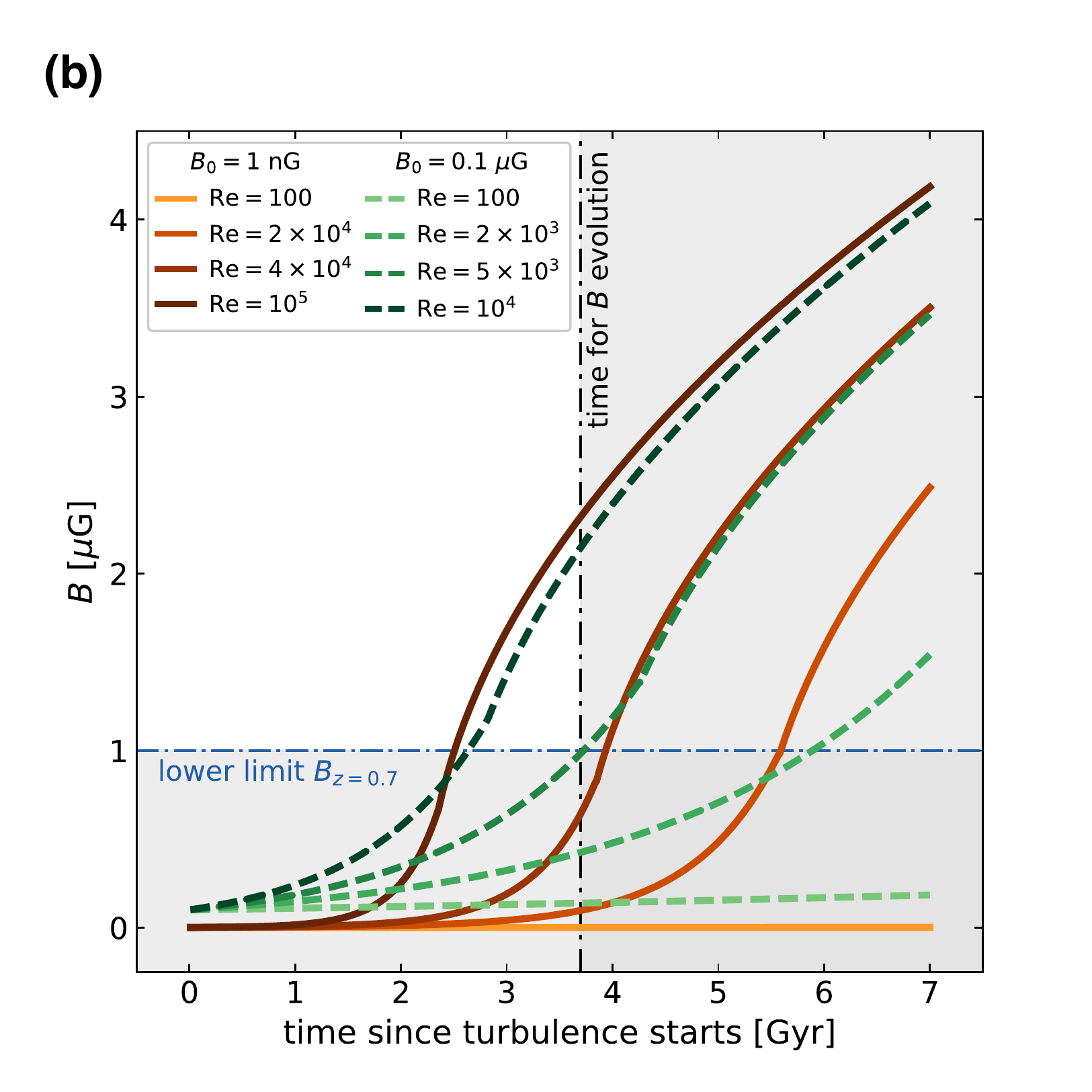}}
\vspace{-8mm}
\caption*{{\bf Figure 3: Cluster magnetic field estimation and theoretical magnetic field evolution.}
\textit{Panel (a):} $P_{\rm 1.4GHz}$--$M_{500}$ diagram for nearby ($median(z)\sim0.2$, black dots \cite{cassano+13}) and distant ($median(z)\sim0.7$, yellow starts) radio halos, including that in El Gordo ($z=0.87$ \cite{lindner+14}, yellow square). Error bars on the yellow stars are obtained with 100 Monte Carlo realization of the radio luminosity, including flux density and spectral index uncertainties, while error bars on the black dots and yellow square are given by the literature. The black solid line represents the best-fit relation from the low-redshift sample, with the 95\% level confidence (grey area). The blue dashed lines represent, from light to darker colors, the theoretical relations at high-$z$ taking magnetic field strengths of 3.0, 2.0, 1.0 and 0.5 $\mu$G (assuming a typical local cluster magnetic field of 2 $\mu$G). \textit{Panel (b):} Magnetic field growth since the start of the turbulence, based on the small-scale dynamo theory. The initial fields, $B_0$, are set at 1~nG (solid orange lines; from light to darker colors: $\rm Re=[100,~ 2\times10^4,~ 4\times10^4,~10^5]$) and 0.1 $\mu$G (dashed green lines; from light to darker colors: $\rm Re=[100,~2\times10^3,~ 5\times10^3,~10^4]$). The horizontal blue dot-dashed line sets the lower limit of the magnetic field strength in distant galaxy clusters (see panel \textit{(a)}). The vertical black dot-dashed line shows the upper limit on the approximate time available for the magnetic field growth (see Methods).}
\label{fig:prad_m500_1.3}
\end{figure*}

\begin{table*}[h!]
\begin{center}
\caption*{{\bf Table 1: Integrated flux density and radio luminosity of the galaxy clusters observed with LOFAR.} Column 1: {\it Planck} cluster name. Column 2: Cluster redshift. Column 3: Largest linear size (LLS) of the diffuse radio emission. Column 4: Classification of the diffuse radio emission. Column 5: Integrated flux density of the diffuse emission (compact sources removed). Column 6: 1.4 GHz ($k$-corrected) radio luminosity.}
\resizebox{0.8\textwidth}{!}{%0.48\textwidth
\begin{tabular}{lcclccc}
\hline
\hline
\noalign{\smallskip}

{\it Planck} (PSZ) name & $z$ & LLS & Classification	&  $S_{\rm 144MHz}$ & $P_{\rm 1.4GHz}$ \\
%\cline{2-4}
		 			&   & [Mpc] & & [mJy] 			& [$10^{24}$ W Hz$^{-1}$] \\
\hline
\noalign{\smallskip}
PSZ2G045.87+57.70	& 0.611 & -- & Uncertain	    & $-$ & $-$ \\
PSZ2G070.89+49.26   & 0.610 & -- & \blue{--}       	& $-$	& $-$ \\ 
PSZ2G084.10+58.72	& 0.731 & -- & Uncertain        & $-$	& $-$ \\ 
PSZ2G086.28+74.76	& 0.699 & -- & Uncertain        & $-$  & $-$ \\
PSZ2G086.93+53.18	& 0.675 & 0.5 & Halo            & $7.2\pm1.5$ & $0.7\pm0.4$ \\ 
PSZ2G087.39+50.92	& 0.748 & -- & --            	& $-$ 			    & $-$\\ 
PSZ2G089.39+69.36	& 0.680 & 1.0 & Halo          	& $12.5\pm1.9$ 	    & $1.3\pm0.7$ \\
\noalign{\smallskip}
\multirow{2}{*}
{PSZ2G091.83+26.11}	&
\multirow{2}{*}{0.822} & 1.2 & Halo              & $84.3\pm12.7$     & $13.8\pm8.4$\\ 
					& & 1.2 & Relic	            & $259.4\pm38.9$    & $-$ \\
\noalign{\smallskip}
PSZ2G092.69+59.92	& 0.848 & -- & --	            & $-$ 				& $-$\\
PSZ2G099.86+58.45	& 0.616 & 1.2 & Halo              & $27.8\pm4.3$	    & $2.2\pm1.3$ \\
PSZ2G104.74+40.42	& 0.690 & -- & Uncertain         & $-$		    	& $-$ \\ 
PSZ2G126.28+65.62	& 0.820 & 0.8 & Halo       	    & $8.8\pm1.7$		& $1.4\pm0.6$ \\
PSZ2G127.01+26.21	& 0.630 & -- & Uncertain         & $-$ & $-$ \\
PSZ2G139.00+50.92	& 0.600 & -- & --        	    & $-$				& $-$ \\ 
PSZ2G141.77+14.19	& 0.830 & 0.6 & Halo      	    & $8.8\pm1.4$ 	   	& $1.4\pm0.7$ \\ 
PSZ2G141.98+69.31	& 0.714 & -- & --        	    & $-$				& $-$ \\
PLCKG147.3--16.6	& 0.645 & 0.8 & Halo      	    & $22.5\pm3.7$		& $6.4\pm3.4^\star$\\
PSZ2G147.88+53.24	& 0.600 & 0.6 & Halo              & $14.4\pm2.3$	& $0.9\pm0.5$\\ 
PSZ2G160.83+81.66	& 0.888 & 0.7 & Halo      	    & $13.0\pm2.1$    	& $2.7\pm1.5$ \\
\noalign{\smallskip}
\hline
\end{tabular}\label{tab:occurrence}}
\end{center}
{Note: Uncertainties on the 1.4 GHz radio luminosity include the flux density and spectral index uncertainties, assuming a Gaussian distribution of spectral indices ($\alpha=-1.5\pm0.3$), see Method. $^\star$We used the spectral index obtained combining our LOFAR observation with literature GRMT flux measurement (\cite{vanweeren+14} see the Supplementary).}
\end{table*}

\begin{figure}[t!]
\centering
{\includegraphics[width=\textwidth]{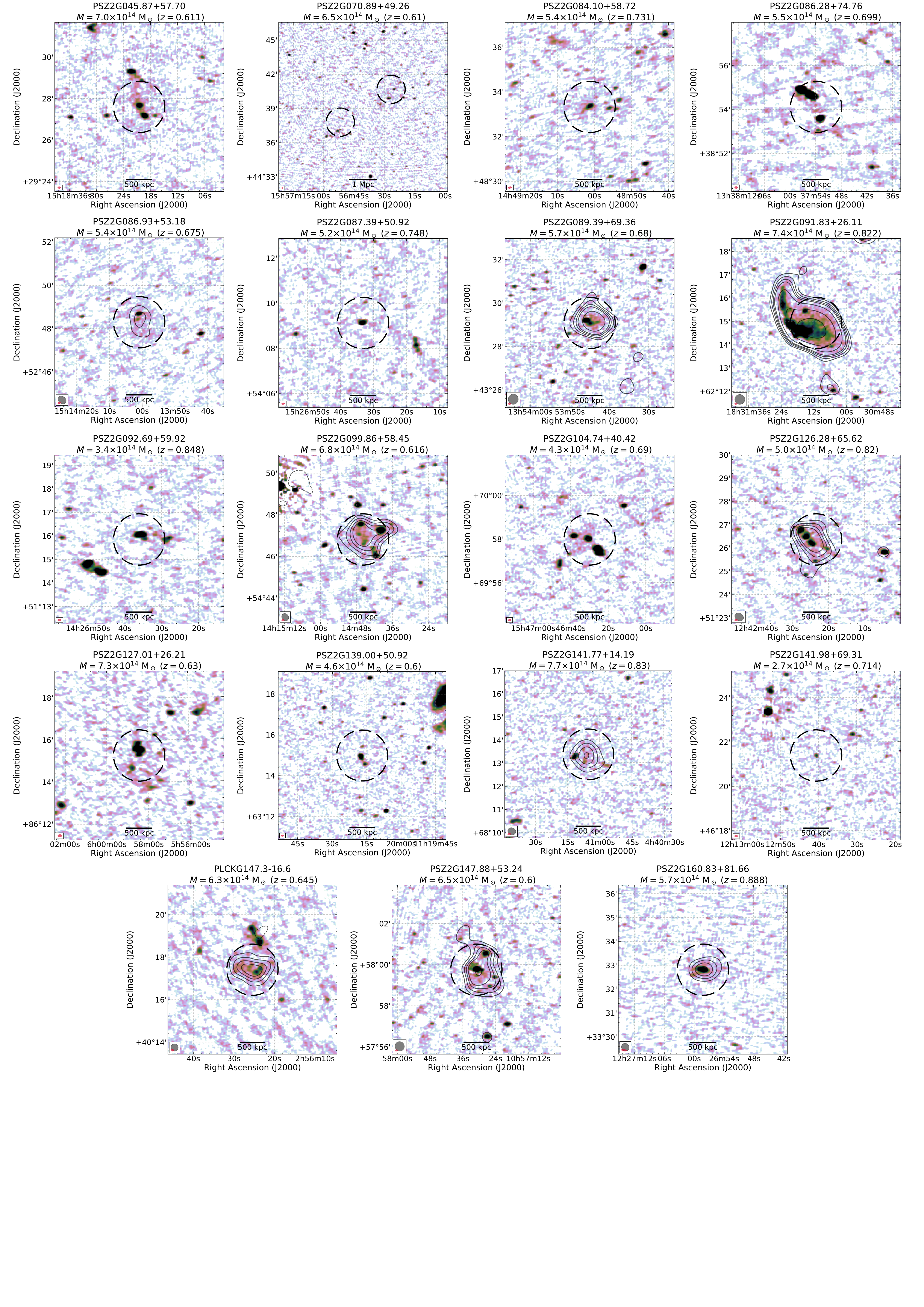}}
\vspace{-6mm}
\caption*{{\bf Extended Data Figure 1: Observed radio emission in our high-z galaxy cluster sample.} In colorscale we show the full-resolution LOFAR images. Low-resolution source-subtracted radio contours, displayed at the $[-2,2,3,4,5,8,16]\times \sigma_{\rm rsm}$ level, are shown only for clusters that host diffuse radio emission (with $\sigma_{\rm rsm}$ the individual map noise; the negative contour levels are indicated with a short-dashed line style). The full- and low-resolution LOFAR beams are displayed in the bottom left corner (in pink and grey colors, respectively). In the header of each image, the galaxy cluster name, mass and redshift are reported. The dashed black circle in each map shows the $R=0.5R_{\rm SZ,500}$ region, obtained from $M_{\rm SZ,500}$.}
\label{fig:bestcases}
\end{figure}

\begin{figure}[t!]
\centering
{\includegraphics[width=\textwidth]{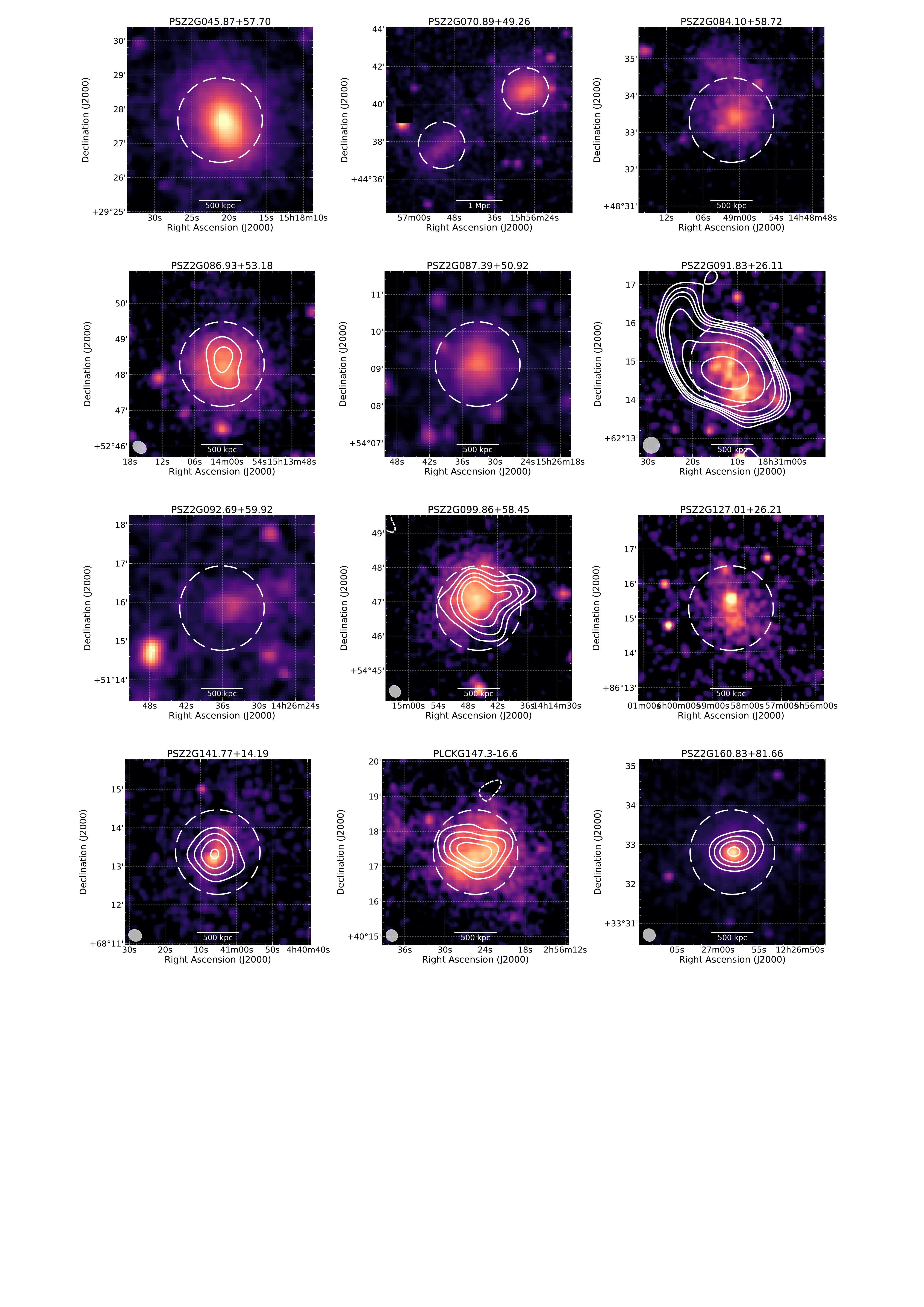}}
\vspace{-6mm}
\caption*{{\bf Extended Data Figure 2: X-ray images of all the galaxy clusters in our sample.} In colorscale we show the {\it Chandra}/{\it XMM-Newton} images. LOFAR radio contours are drawn as Figure 1, with the LOFAR beam displayed in the bottom left corner. The dashed white circle in each map shows the $R=0.5R_{\rm SZ,500}$ region, obtained from $M_{\rm SZ,500}$. In the header of each image, the galaxy cluster name is reported.
}
\label{fig:bestcases}
\end{figure}

\clearpage
\newpage

\section*{Method}
\noindent\textbf{Galaxy cluster sample selection.} 
We construct our sample of high-$z$ galaxy clusters using the latest {\it Planck} Sunyaev-Zel'dovich (SZ) catalog, i.e. PSZ2 \cite{planckcoll16}. This catalog provides a reliable cluster mass estimation for galaxy clusters up to $z\sim1$. Although the exact effect of selection biases is still being debated and may overestimate the observed fraction of clusters with radio halos, 
there is no clear consensus in the literature that suggests the SZ sample chosen here disproportionately favours merging over relaxed clusters \cite{eckert+11,rossetti+17,andrade-santos+2017}. We select all the objects at $\rm DEC\geq 20^\circ$, to match the part of the sky with the best LOFAR sensitivity, and $z\geq0.6$. No mass-threshold has been applied in our cluster selection. 
These selection criteria result in a sample of 30 objects (see Supplementary Table 1). The sample includes clusters where precise redshift measurements became available recently \cite{amodeo+18,barrena+18,burenin+18,sereno+18,streblyanska+18,vanderburg+16,zohren+19} and a previous-discovered {\it Planck} cluster, i.e. PLCK\,G147.3--16.6 \cite{vanweeren+14}. We used the optical images from the Panoramic Survey Telescope and Rapid Response System (Pan-STARRS \cite{panstarss16}) to refine the accuracy of the cluster center coordinates, using the brightest cluster galaxy (BCG), when detected, or approximately the center of the distribution of the galaxies. We also inspected the {\it Chandra} and/or {\it XMM-Newton} X-ray image, when available, to verify the correspondence of the cluster coordinates with the center of the thermal gas distribution. Among these 30 {\it Planck} clusters, 21 objects were already covered as part of the LOFAR Two-metre Sky Survey (LoTSS \cite{shimwell+19}). From these 21 clusters one was affected by bad ionospheric observing conditions (i.e. PSZ2\,G088.98+55.07) and one lays on the same line of sight of a cluster at $z=0.3$ (i.e. PSZ2\,G097.52+51.70). Those are therefore excluded from the final sample (see Extended Data Figure 1), consisting of 19 objects at median redshift of $z=0.7$.

\bigskip
\bigskip

\noindent\textbf{LOFAR observations, data reduction, images and flux measurements.} 
The LOFAR observations were carried out together with the LOFAR Two-metre Sky Survey (LoTSS \cite{shimwell+19}) in the 120--168 MHz frequency range. The survey consists in 8 hours of observation for each pointing, with a field of view of $\sim2.6^\circ$, full resolution of about $6''$, and median noise of about $70~\mu$Jy beam$^{-1}$. Given the large sky coverage already achieved by the survey, our targets were often observed by more than one pointing (see Supplementary Table 2), which further improved our signal to noise. For each LoTSS pointing, we performed standard data reduction \cite{shimwell+19}, which includes direction-independent and dependent calibration, and imaging of the full LOFAR field of view using \texttt{prefactor} \cite{vanweeren+16,williams+16,degasperin+19}, \texttt{killMS} \cite{tasse14,smirnov+tasse15} and \texttt{DDFacet} \cite{tasse+18}. We additionally improved the quality of the calibration, using the products of the pipeline, subtracting all the sources outside a region of $\sim15^\prime\times15^\prime$ surrounding the target, and performing extra phase and amplitude self-calibration loops in each target sub-field. To verify the quality of the calibration, each pointing is imaged independently using \texttt{WSClean} \cite{offringa+14}, with the wideband deconvolution mode (\texttt{channelsout=6}). For all our imaging, we employ automatic clean masks for the deconvolution. We utilize a $3\sigma_{\rm rms}$ masking threshold, with $\sigma_{\rm rms}$ the local map noise, and cleaning down to the $1\sigma_{\rm rms}$ level inside the mask. The images have a central frequency of 144 MHz, and are provided in the Extended Data Figure 1. 

\bigskip

\noindent All the clusters in our sample were carefully visually inspected in the full-resolution image to search for extended radio emission. 
To emphasize the presence of diffuse radio emission, we also produced low-resolution source-subtracted images. We first applied an $uv$-cut to the data, to filter out emission associated to linear sizes larger than 500 kiloparsecat the cluster redshift and 
to create a clean component model of the compact sources. For PSZ2\,G089.39+69.36 we apply an $uv$-cut corresponding to 400 kpc.
During this step, we employed multiscale deconvolution \cite{offringa+17}, using scales of $\rm [0,4,8,16]\times pixelscale$ (with the pixel size of $1.5''$) to properly include and subtract somewhat extended radio galaxies. In addition, for the automatic deconvolution we lower the mask threshold to $1\sigma_{\rm rms}$ to subtract the faintest contaminating sources (see Supplementary Figure 1). 
Finally, we subtracted the compact sources model from the visibilities 
and tapered the $uv$-data to $\sim15''$ resolution. It is important to note that in case of an extended radio galaxy, with linear size $\gtrsim 500$~kpc, this method cannot properly subtract the radio emission from the $uv$-data. For that reason, the extended double-lobed radio galaxy just north of PLCKG147.3--16.6 has been manually excluded and blanked in the low-resolution source-subtracted image (Figure 1 and Extended Data Figure 1). 
\bigskip

\noindent We used these low-resolution source-subtracted images to define the largest linear size (LLS) of the diffuse radio emission, following the $2\sigma_{\rm rms}$ contour. Systems with $\rm LLS<500~kpc$ were not included in our statistics.
Roundish, centrally-located significantly detected sources (see the next paragraph) that follow the ICM distribution were classified as radio halos; elongated, peripheral structures were classified as radio relics. We did not identify clear examples of fossil radio plasma sources (i.e. ``radio phoenices''), these typically have small sizes (i.e. 200~kpc or smaller), irregular shapes, and are not centered on the cluster center \cite{degasperin+17,mandal+20}. Similarly, we did not find clear examples of  ``mini-halos'', as these diffuse sources usually have LLS smaller than 500 kpc and are located in dynamically relaxed systems \cite{giacintucci+17}. Uncertain classifications, %possibly 
due to contamination from extended radio galaxies in the cluster region and/or  ambiguous shapes, %$\frac{S_{\rm 144MHz}}{\Delta S}<5$,} 
were excluded from our final statistical analysis. 
If we measure the LLS following the $3\sigma_{\rm rms}$ contour, we would classify 8 radio halos instead of 9 (see Supplementary Table 3).

\bigskip

\noindent To determine the integrated flux densities for the radio halos, we first measured the total cluster radio flux densities (i.e. compact sources and diffuse emission) from the low-resolution image, following the $2\sigma_{\rm rms}$ radio contour to fully cover the extension of the diffuse radio emission (see Extended Data Figure 1). In the Supplementary Table 3, we also report the integrated flux density measurements following the $3\sigma_{\rm rms}$ radio contour. %
We then mathematically subtracted the flux densities of the compact sources within the cluster region, measured in the full-resolution $uv$-cut image, to obtain the flux density associated to the diffuse radio sources \cite{cassano+19}. 
For PSZ2\,G091.83+21.16, which also hosts a bright radio relic, we visually separated it from the radio halo region (see Supplementary Figure 2). 
The uncertainties on the flux densities are estimated adding in quadrature the $\sigma_{\rm rms}$ statistical uncertainty, the 15\% systematic error associated to the LOFAR flux scale calibration, $f_{\rm cal}$ \cite{shimwell+19}, and the uncertainty due to the source subtraction in the cluster region, $\sigma_{\rm sub}$ \cite{cassano+13}, according the following Equation:  

\begin{equation}
\Delta S = \sqrt{(f_{\rm cal} S_{\rm 144MHz})^2 + \sigma_{\rm rms}^2 N_{\rm beam,h} + \sigma_{\rm sub}^2} \, .
\end{equation}
Here, $\sigma_{\rm sub}^2=\sum_i\sigma_{\rm rms}^2 N_{{\rm beam,s}_i}$, and $N_{\rm beam, h}$ and  $N_{{\rm beam, s}_i}$ are the number of beams covering the halo and the subtracted $i$ sources, respectively. Uncertainties associated to possibly missed faint compact sources are not included, but these should be smaller than the systematic error associated to the LOFAR flux scale calibration. %\ref{fig:g091.83_vla}).} 

\bigskip

\noindent We confirmed the presence of diffuse radio emission when the measured integrated radio flux density is larger than five times the flux uncertainty, i.e. $\frac{S_{\rm 144MHz}}{\Delta S}\geq 5$ (see Column 5 in Table 1). %\ref{tab:occurrence}). 
Flux density measurements directly obtained from the low-resolution source-subtracted images are also provided in the Supplementary.

\bigskip

\noindent Finally, we calculated the $k$-corrected radio luminosity at 1.4 GHz:

\begin{equation}
P_{\rm 1.4GHz} =  4\pi D_L^2 \dfrac{S_{\rm 1.4GHz}}{(1+z)^{\alpha+1}} \, ,
\end{equation}
with $D_L$ the luminosity distance assuming standard $\Lambda$CDM cosmology (i.e. $H_{0} = 70$ km s$^{-1}$~Mpc$^{-1}$, $\Omega_{m} = 0.3$, and $\Omega_{\Lambda} = 0.7$) and $S_{\rm 1.4GHz}=S_{\rm 144MHz}\left (\frac{\rm 1.4GHz}{\rm 144MHz} \right )^\alpha$. Since the spectral indices of the radio halos are not measured, we assume a Gaussian distribution of %standard
spectral indices $\alpha$ with mean value $-1.5$ and scatter $0.3$ for computing the radio luminosity. These values were chosen to encompass the typical range of spectral indices found in radio halos \cite{vanweeren+19}, and to take into account that they could be steeper than local systems \cite{cassano+06}. Only for PLCK\,G147.3--16.6, we used the spectral index obtained combining our observations with the GMRT literature data \cite{vanweeren+14} (see Supplementary). The radio luminosities are listed in Table 1, %\ref{tab:occurrence}, 
whose uncertainties are obtained by performing Monte Carlo simulations including the flux density uncertainties and the spectral index uncertainty. 

\bigskip
\bigskip

\noindent\textbf{X-ray observations and data reduction.}
Twelve clusters among our LOFAR sample have also {\it Chandra} and/or {\it XMM-Newton} observations (see Supplementary Table 4 and Extended Data Figure 2). %(see Supplementary Table \ref{tab:xray} and Supplementary Figure \ref{fig:allxrays}).

\bigskip

\noindent {\it Chandra} data were reduced using the \texttt{chav} software package (\url{http://hea-www.harvard.edu/\textasciitilde alexey/CHAV/}) with \texttt{CIAO v4.6} and applying the \texttt{CALDB v4.7.6} calibration files \cite{vikhlinin+05}. The processing includes the application of gain maps to calibrate photon energies, filtering out counts with ASCA grade 1, 5, or 7 and bad pixels, and a correction for the position-dependent charge transfer inefficiency (CTI). Periods with count rates with a factor of 1.2 above and 0.8 below the mean count rate in the 6--12 keV band were also removed. Standard blank-sky files were used for background subtraction. The final exposure-corrected images were made in the 0.5--2.0~keV band using a pixel binning of a factor of 4 (i.e $2^{\prime\prime}$) and combining the different ObsIDs together where relevant (i.e., PSZ2\,G160.83+81.66).

\bigskip

\noindent {\it XMM-Newton} observations were reduced using version 17.0.0 of the dedicated Science Analysis System (SAS). After converting the observation data files to unfiltered event lists, we extracted light curves in bins of 100 seconds in the energy ranges 10--12~keV for the MOS and 10--14~keV for the pn detectors onboard. Good time intervals were created excluding periods when the mean count rates in these light curves were different from the corresponding mean by more than 2 standard deviations. Images were extracted in the 0.4--7.0 keV band, using only the highest quality (FLAG==0) single to quadruple MOS events (PATTERN$\leq$12) and single to double pn events (PATTERN$\leq$4). Images were weighted by the respective exposure maps for each detector and combined, accounting for a factor $\sim$2.5 difference between the typical expected count rates for MOS and pn at a given source flux (due to the slightly different effective areas and the fact that the light focused by two of the XMM-Newton telescopes is split evenly between the MOS and RGS detectors). Since we are interested mostly in the morphology of high-redshift, thus relatively compact, clusters positioned close to the aim point, the effects of vignetting and instrumental particle background are minimal and have not been corrected for this analysis.

\bigskip
\bigskip

\noindent\textbf{Turbulent energy flux at high redshift.}
According to re-acceleration models, relativistic electrons in radio halos are re-accelerated by turbulence. Assuming that turbulence in the ICM is mainly injected by cluster mergers, then the turbulence injection rate is $\rho v_i^3/L_{\rm inj}$ \cite{casano+brunetti05}, with $v_i$ the impact velocity between the two merging clusters, $\rho$ the cluster mean density and $L_{\rm inj}$ the injection scale of turbulence (which can be assumed to be of the same order of the cluster size, about 1 Mpc at both $z=0.2$ and $z=0.7$).
The relative impact velocity of two subclusters with virial masses $M_v$ and $\Delta M_v$ which collide (at a distance $R_{v}$ between the centers) starting from an initial distance $d_0$ with zero velocity is given by \cite{sarazin02}:

\begin{equation}
v_i\simeq \sqrt{2G\frac{(M_{v}+\Delta M_v)}{R_{v}}
\left (1-\frac{1}{\eta_{v}}\right)} \, .
\label{eq:vi}
\end{equation}
Here, $d_0=\eta_v R_{v}$, $\eta_v\simeq 4\big(\frac{ M_{v}+ \Delta M} {M_{v}}\big)^{1/3}$, and $R_{v}$ is the virial radius of the main cluster, i.e., the radius at which the ratio between the average density in the cluster and the mean cosmic density at the redshift of the cluster is given by \cite{kitayama+suto96}:

\begin{equation}
\Delta_{c}(z)=
18\pi^2(1+0.4093\,\omega(z)^{0.9052}) \, ,
\label{Dc}
\end{equation}
where $\omega(z)\equiv \Omega_{f}(z)^{-1}-1$ with:

\begin{equation}
\Omega_{f}(z)
=\frac{\Omega_{m,0}(1+z)^3}{\Omega_{m,0}(1+z)^3+\Omega_{\Lambda}} \, .
\label{Omz}
\end{equation}

\bigskip

\noindent The virial mass, $M_v$, and the virial radius are thus related by:

\begin{equation}
R_{v}=\left [\frac{3M_{v}}{4\pi\Delta_{c}(z)\rho_{m}(z)}
\right]^{1/3}
\label{Rv}
\end{equation}
where $\rho_{m}(z)=2.78\times10^{11}\,\Omega_{m,0}\,(1+z)^3~h^2$ M$_{\odot}$ Mpc$^{-3}$ is the mean mass density of the Universe at redshift $z$. The ratio of the gas density at high and low redshift can be assumed to scale as the ratio of the two virial densities, i.e., $\frac{\rho_{v,z=0.7}}{\rho_{v,z=0.2}}=\frac{M_{v,z=0.7}}{\left (R_{v,z=0.7} \right)^3}\times \frac {\left (R_{v,z=0.2} \right )^3}{M_{v,z=0.2}}$. In the case $M_{v,z=0.7}=M_{v,z=0.2}$, the density ratio becomes simply proportional to $\left (R_{v,z=0.2}/R_{v,z=0.7} \right )^3$. Specifically, considering both at low and high-$z$ $M_v\simeq 1.0\times 10^{15}$ M$_{\odot}$ and $\Delta M_v\simeq M_{v}/3$, the above equations give a ratio $\left (v_{i,z=0.7}/v_{i,z=0.2} \right)^3\sim 1.5$ and $\rho_{v,z=0.7}/\rho_{v,z=0.2}\sim 2.2$. The resulting ratio of turbulent energy flux derived at high and low redshift is then $\sim 3.3$.

\bigskip

\noindent The above calculation is approximate, because we only took kinematic effects and $\Lambda$CDM cosmology into account. We thus also use simulations in \cite{casano+brunetti05} to calculate the injection of turbulence during simulated cluster merging histories, and derived the turbulent energy injected in the volume swept by the infalling subcluster. We found that in this case the distributions of the turbulent energy fluxes derived for clusters with $M_v\simeq 0.9-1.4\times 10^{15}\,M_{\odot}$ in the two redshift ranges $0.2-0.3$ and $0.7-0.8$ differ by a factor of  $\sim 3-4$, in a good agreement with the simpler derivation reported above.

\bigskip
\bigskip

\noindent\textbf{Radiative lifetime and electrons acceleration in high-$z$ clusters.} 
In the paper we have assumed that $\eta_{\rm rel}$, i.e. the fraction of turbulent energy flux that is absorbed by relativistic electrons (Eq. 1), is constant with redshift. This depends on the energy budget of the population of re-accelerated seed electrons, and on the microphysics of the nonlinear interaction between turbulence and particles. The latter is assumed to be redshift-independent (e.g. \cite{brunetti+lazarian07}). Thus, a constant $\eta_{\rm rel}$ implies a similar energy budget of the seed electrons that are accumulated in high- and low-$z$ ICM.

\bigskip

\noindent More specifically, we used Eq. 1 to infer a lower limit to the ratio of the magnetic field in high- and low-$z$ radio halos, thus our conclusions may change in the case that clusters at higher redshift host a population of seed electrons with a much larger energy budget. In this section we show that this possibility is very unlikely, as the higher injection rate of electrons in the ICM expected from high-redshift sources is balanced by the stronger electron energy losses.

\bigskip

\noindent The budget of seed electrons that accumulate in the ICM can be estimated as $N_e \propto Q_e \times T_{\rm max}$ (with $Q_e$ the injection rate and $T_{\rm max}$ the maximum electron lifetime). The lifetime of relativistic electrons in the ICM, which is determined by the energy losses due to Coulomb interaction with the thermal plasma and to Inverse Compton scattering with the CMB photons and synchrotron losses, can be expressed as \cite{brunetti+jones14}:

\begin{equation}
T=4\left \{\frac{1}{3}\tilde{\gamma}\beta_{z}+\frac{\tilde{n}}{\tilde{\gamma}}\left[1.2+\frac{1}{75}\ln \left (\frac{\tilde{\gamma}}{\tilde{n}}\right )\right]\right\}^{-1} \, ,
\label{eq:Tlife}
\end{equation}
where we put $\tilde{\gamma}=\gamma/300$, $\tilde{n}= n/10^{-3}$, $\beta_z=\left ( \frac{B}{3.2} \right )^2+(1+z)^4$. Here, $\gamma$ is the Lorentz factor, $n$ the number density in cm$^{-3}$ and $B$ the magnetic field strength in $\mu$G. This lifetime increases with the energy $\gamma$ of relativistic electrons at lower energy, it reaches a maximum and then decreases as a function of $\gamma$.

\bigskip

\noindent The maximum of $T$ is obtained for $\tilde{\gamma}\approx\sqrt{\frac{3\tilde{n}}{\beta_z}}$. For $B< B_{\rm CMB}$, it is determined by the combination of IC and Coulomb losses. Replacing this in Eq. \ref{eq:Tlife} gives the typical lifetime of relativistic electrons that can be accumulated in the ICM on cluster lifetime and then re-accelerated during mergers:

\begin{equation}
T_{\rm max}\,[{\rm Gyr}]=4\left \{\frac{\beta_z}{3}\sqrt{\frac{3\tilde{n}}{\beta_z}}+\tilde{n}\sqrt{\frac{\beta_z}{3\tilde{n}}}\right\}^{-1} \, .
\label{eq:Tmax}
\end{equation}
Considering that between $z=0.2$ and $z=0.7$ the virial density (and hence the thermal gas density $n$) increases by a factor of $\sim2$, the maximum lifetime ratio for electrons radiating at these two redshifts is $\sim 2.7$, considering $B \sim 1-2 ~\mu$G.

\bigskip

\noindent Several sources can inject electrons in the ICM including AGN, galaxies (galactic winds, GW) and cosmological shocks.
Most of the energy of the relativistic plasma of GW and shocks is in the form of supra-thermal and relativistic protons with only a negligible fraction in the form of electrons (e.g. \cite{brunetti+jones14}).

\bigskip

\noindent AGN are thought to be the dominant sources of relativistic electrons in the ICM (e.g. \cite{brunetti+jones14}); this hypothesis is also supported by recent LOFAR observations, where sources of fossil radio plasma from old AGN-electrons are commonly found in galaxy clusters \cite{degasperin+17,mandal+20}. The injection rate of electrons from a number of AGN in clusters, $N_{\rm AGN}$, is $Q_e \sim N_{\rm AGN} n_{\rm rel} V / \Theta$, where $n_{\rm rel}$ is the number density of radio-emitting electrons in the jets and lobes, $V$ is the volume of the radio plasma and $\Theta$ is the AGN life-time. The number density of radio-emitting  electrons in the AGN lobes can be estimated assuming equipartition between particles and magnetic fields, i.e. $n_{\rm rel}\propto \sqrt{L_{\rm syn} / V}$, with $L_{\rm syn}$ the AGN synchrotron luminosity. This gives $Q_e \propto \sqrt{L_{\rm syn} V} N_{\rm AGN}/\Theta$. The linear size of radio galaxies is observed to decline with redshift, probably due to the higher density of the medium in which radio lobes and jets expand \cite{neeser+95,blundell+99}. Thus we attempt to account for the expected change of $V$ with redshift assuming approximate pressure equilibrium of the lobes with the surrounding medium. This would imply $P_{\rm ICM} \propto n_{\rm rel} \propto \sqrt{L_{\rm syn} / V}$ and consequently $Q_e \propto L_{\rm syn} N_{\rm AGN} / ( P_{\rm ICM} \Theta ) 
\approx \rho_L (1+z)^3/ ( P_{\rm ICM} \Theta ) $, where $\rho_L$ is the luminosity density of radio AGN per comoving-volume. The luminosity density increases by a factor $\sim 2-2.5$ from $z\sim 0.2$ to $z\sim 0.7$ (\cite{smolciv+17} their Fig. 5). The increase of $P_{\rm ICM}$ with redshift can be estimated using virial quantities, i.e. $P_{\rm ICM} \propto n_{\rm ICM} T_{\rm ICM} \propto \frac{M_v}{R_v^3}\frac{M_v}{R_v}$, implying that pressure increases by a factor $\sim 2.5$ from $z\sim 0.2$ to $z\sim 0.7$ for clusters with the same virial mass.
We find that $Q_e$ is only about 2-3 times larger in clusters at $z\sim0.7$ if we assume that the life-time of AGN does not change much with redshift. Combining this result with the maximum lifetime of the radiating electrons, that is $\sim 2.7$ times longer at lower redshift, suggests that the budget of seed electrons at low and high redshift is similar.

\bigskip

\noindent Seed electrons in the ICM can also be generated by the decay chain of inelastic proton-proton collisions (e.g. \cite{brunetti+lazarian11,pinzke+17}). The importance of this channel depends on the energy budget of CR in the ICM, which is poorly known. A relevant contribution of re-accelerated secondary particles to radio halos is still allowed by current gamma-ray limits from FERMI-LAT (e.g. \cite{brunetti+17}). For this reason, we also compare the budget of secondary electrons accumulated in galaxy clusters at low and high redshift considering cosmological shocks as the main sources of relativistic protons. In this case, numerical simulations suggest the ratio of CR to thermal ICM pressures ($X=P_{\rm CR}/P_{\rm ICM}$) in high-$z$ clusters is generally smaller than that at low redshift (e.g. \cite{vazza+14}). The ratio of the energy budget of secondary electrons at high and low redshift can be approximately estimated as:
\begin{equation} 
\frac{N_{e,z=0.7}}{N_{e,z=0.2}} \sim 
\frac{X_{z=0.7}}{X_{z=0.2}} \frac{ n_{{\rm ICM},z=0.7} P_{{\rm ICM,}z=0.7}}{n_{{\rm ICM},z=0.2} P_{{\rm ICM,}z=0.2}} \frac{T_{{\rm max},z=0.7}}{T_{{\rm max},z=0.2}} \, ,
\label{secondari}
\end{equation}
implying $N_{e,z=0.7}/N_{e,z=0.2}\sim 0.4-1.8$ if we assume typical CR pressure ratios measured in simulations, i.e. $\sim 0.2-1$ (\cite{vazza+12,vazza+14}, Figs. 14 and Fig. 12 respectively), and the ratios between physical quantities as derived above: $T_{{\rm max},z=0.7}/T_{{\rm max},z=0.2}\sim 1/2.7$, $P_{{\rm ICM},z=0.7}/P_{{\rm ICM},z=0.2} \sim 2.5$ and  $n_{{\rm ICM},z=0.7}/n_{{\rm ICM},z=0.2} \sim 2$. 
Similarly to the case of AGN, this implies that the budget of secondary particles that is available at higher redshift is very similar to that at low redshift. 

\bigskip
\bigskip

\noindent\textbf{Reynolds number estimation.}
In Figure 3b %\ref{fig:prad_m500_1.3}b 
we have used a simple model of magnetic field amplification to infer combined constraints on the initial magnetic field and on the Reynolds number.

\bigskip

\noindent We follow the method as outlined in \cite{beresnyak12}. We assume that the amplification initially operates in a kinematic regime, where the magnetic field grows exponentially with time:

\begin{equation}
B^2(t) \sim B^2_0 \exp \left (t \, \Gamma \right)  \, ,
\label{bexp}
\end{equation}
where the time-scale of the magnetic growth is $\Gamma^{-1} \sim 30 L_d/\delta v_d$. This depends on the turbulent eddy turnover time 
at the viscous dissipation scale, $L_d \sim L_{\rm inj}\, {\rm Re}^{-3/4}$,  $\delta v_d \sim v_t (L_d/L_{\rm inj})^{1/3}$. The factor $\sim 30$ is derived from simulations \cite{cho14,beresnyak+miniati16},  and accounts for the less effective stretching of the field lines due to the turbulent diffusion. When the magnetic and the kinetic energy densities become comparable at the viscous dissipation scale ($B(t)^2/8 \pi \sim 1/2 \rho \delta v_d^2$), the turbulent dynamo transits to a phase  where the magnetic field grows linearly with time. This phase is known as the non-linear regime because the magnetic field (Reynolds stress) becomes dynamically important. From Eq. \ref{bexp}, this transition occurs at a time:

\begin{equation}
t_* \sim 60 \frac{L_{\rm inj}}{v_t} {\rm Re}^{-1/2} \ln  \left ( \frac{B_*}{B_0} \right )
\label{timeexp}
\end{equation}
where $B_* \sim \sqrt{4 \pi \rho}\, v_t {\rm Re}^{-1/4}$.
According to simulations, in the non-linear regime, i.e. for time $t \geq t_*$, we assume that a constant fraction of the turbulent kinetic energy flux, $\eta_B$, is channeled into magnetic field energy ($\eta_B=0.05$ \cite{beresnyak12}). Therefore, the evolution of the magnetic field with time is:

\begin{equation}
B^2(t) \sim B^2_* + 4 \pi \frac{\rho v_t^3}{L_{\rm inj}} \eta_B (t - t_*) \, .
\label{evolution}
\end{equation}

\bigskip

\noindent In the calculations shown in Figure 3b, %\ref{fig:prad_m500_1.3}b, 
we assume a continuous injection of turbulence with $v_t = 500$ km s$^{-1}$ at the injection scale $L_{\rm inj}=1$ Mpc, and $n=\rho/m_p=3\times10^{-3}$ cm$^{-3}$ (with $m_p$ the proton mass). We note that a situation of continuous injection of turbulence with these parameters is appropriate for a dynamically active and massive cluster \cite{dolag+05,donnert+18,hitomicoll18,markevitch+vikhlinin07} and consequently calculations would overestimate the magnetic field strength if extrapolated for a time scale that is much larger than that of a typical merger.

\noindent 
An important parameter in the model is the time when the turbulent dynamo starts, as it fixes the time available for magnetic field amplification up to the epoch of observation. Here, we assume that the turbulent dynamo starts when the clusters have formed a quarter of their mass. This implies that the time available for the amplification is of about 3.7~Gyr \cite{fakhouri+10,giocoli+12}. Assuming a turbulent velocity of $v_t = 500$~km~s$^{-1}$ at an earlier stage in the cluster life-time is optimistic, as the cluster would be too small. Since the turbulent energy flux is proportional to $v_t^3$, the effect of much weaker turbulence at earlier epochs is not relevant for the magnetic field evolution. Conversely, if we assume that the dynamo started later in the cluster life-time, e.g. when it has assembled half of its mass, it results in more stringent Reynolds numbers, since the time available for the  amplification is shorter (about 2.7~Gyr). Lower turbulent velocities and lower number densities would imply higher values of the Reynolds number. On the other hand, higher turbulent velocities would imply a smaller value of Re. However, even if we consider $v_t=800$ km s$^{-1}$ \cite{dolag+05,donnert+18}, i.e. the case of a large turbulent pressure of about 30--40\% the ICM pressure, we obtain Reynolds numbers that are only three times smaller than the case with $v_t = 500$ km s$^{-1}$. We also mention that in real clusters turbulence is induced in a medium that is highly stratified due to gravity \cite{roh+19}. In this case, the transport of magnetised turbulent eddies toward the outskirts could make the turbulent dynamo slightly less efficient than in our model, implying that our constraints on the Reynolds number are conservative.

%\begin{comment}
% NEED TO BE IN THE CORRECT ORDER

\newpage
%\twocolumn

\noindent\textbf{Correspondence.}  Correspondence and requests for materials should be addressed to G.D.G. (digennaro$@$strw.leidenuniv.nl)

\bigskip

\noindent\textbf{Acknowledgements.}
The authors thank C. Giocoli and his team for the discussion on the cosmological derivations in the manuscript. 
This manuscript is based on data obtained with the International LOFAR Telescope (ILT). LOFAR (van Haarlem et al. 2013) is the Low Frequency Array designed and constructed by ASTRON. It has observing, data processing, and data storage facilities in several countries, which are owned by various parties (each with their own funding sources), and which are collectively operated by the ILT foundation under a joint scientific policy. The ILT resources have benefited from the following recent major funding sources: CNRS-INSU, Observatoire de Paris and Universit{\'e} d'Orl{\'e}ans, France; BMBF, MIWF-NRW, MPG, Germany; Science Foundation Ireland (SFI), Department of Business, Enterprise and Innovation (DBEI), Ireland; NWO, The Netherlands; The Science and Technology Facilities Council, UK; Ministry of Science and Higher Education, Poland; The Istituto Nazionale di Astrofisica (INAF), Italy.  This research made use of the Dutch national e-infrastructure with support of the SURF Cooperative (e-infra 180169) and the LOFAR e-infra group. The J{\"u}lich LOFAR Long Term Archive and the German LOFAR network are both coordinated and operated by the J{\"u}lich Supercomputing Centre (JSC), and computing resources on the supercomputer JUWELS at JSC were provided by the Gauss Centre for Supercomputing e.V. (grant CHTB00) through the John von Neumann Institute for Computing (NIC). This research made use of the University of Hertfordshire high-performance computing facility and the LOFAR-UK computing facility located at the University of Hertfordshire and supported by STFC [ST/P000096/1], and of the Italian LOFAR IT computing infrastructure supported and operated by INAF, and by the Physics Department of Turin university (under an agreement with Consorzio Interuniversitario per la Fisica Spaziale) at the C3S Supercomputing Centre, Italy.
The National Radio Astronomy Observatory is a facility of the National Science Foundation operated under cooperative agreement by Associated Universities, Inc. 
This work is based on observations obtained with XMM?Newton, an ESA science mission with instruments and contributions directly funded by ESA Member States and NASA. 
The scientific results reported in this manuscript are based in part on data obtained from the Chandra Data Archive.
The Pan-STARRS1 Surveys (PS1) and the PS1 public science archive have been made possible through contributions by the Institute for Astronomy, the University of Hawaii, the Pan-STARRS Project Office, the Max-Planck Society and its participating institutes, the Max Planck Institute for Astronomy, Heidelberg and the Max Planck Institute for Extraterrestrial Physics, Garching, The Johns Hopkins University, Durham University, the University of Edinburgh, the Queen's University Belfast, the Harvard-Smithsonian Center for Astrophysics, the Las Cumbres Observatory Global Telescope Network Incorporated, the National Central University of Taiwan, the Space Telescope Science Institute, the National Aeronautics and Space Administration under Grant No. NNX08AR22G issued through the Planetary Science Division of the NASA Science Mission Directorate, the National Science Foundation Grant No. AST-1238877, the University of Maryland, Eotvos Lorand University (ELTE), the Los Alamos National Laboratory, and the Gordon and Betty Moore Foundation.
G.D.G. and R.J.v.W. acknowledge support from the ERC Starting Grant ClusterWeb 804208. G.B., R.C., F.G., M.R. acknowledge support from INAF through the mainstream project ``Galaxy clusters science with LOFAR''. A.Bot. and R.J.v.W. acknowledge support from the VIDI research programme with project number 639.042.729, which is financed by the Netherlands Organisation for Scientific Research (NWO). H.J.A.R. acknowledge support from the ERC Advanced Investigator programme NewClusters 32127. A. Bon. acknowledges support from the ERC Stg ``DRANOEL'' no. 714245 and from the MIUR grant FARE ``SMS''.

\bigskip

\noindent\textbf{Author Contributions.} G.D.G. coordinated the research, performed the radio imaging, reduced the VLA and Chandra data and wrote the manuscript. R.J.v.W., A.Bot., and F.d.G. performed the additional calibration on the LOFAR data and wrote the data reduction software. G.B. and R.C. performed the %radio halo and 
magnetic field evolution modeling. R.J.v.W., G.B. and R.C. helped with the writing of the manuscript. M.B. and M.H. helped with the interpretation of the radio and modeling results and provided extensive feedback on the manuscript. H.J.A.R. and T.W.S. lead the LoTSS survey and coordinated the LOFAR data reduction. A.S. carried out the XMM-Newton data reduction. F.G. and M.R. helped with the interpretation of the X-ray data and sample selection. A.Bon. helped with designing the experiment and observing proposal. V.C., D.D., P.D.F., T.A.E. and S.M., helped with the interpretation of the radio and modeling results, and gave feedback on the manuscript.
All the authors of this manuscript are members of the LOFAR Surveys Key Science Project.% and contributed on the manuscript.

\bigskip

\noindent{\bf Competing interests.}
The authors declare no competing financial interests.

\bigskip

\noindent\textbf{Data availability.} The radio observations are available in the LOFAR Long Term Archive (LTA; \url{https://lta.lofar.eu/}) and in the VLA archive (\url{https://archive.nrao.edu/archive/advquery.jsp}, project code 15A\_270). The X-ray observations are available in XMM and Chandra data archives (\url{http://nxsa.esac.esa.int/nxsa-web/#search} and \url{https://cda.harvard.edu/chaser/}). The data that support the plots within this paper and other findings of this study are available from the corresponding author upon reasonable request.

\bigskip

\noindent\textbf{Codes availability.} The codes that support the plots within this paper and other findings of this study are available from the corresponding author upon reasonable request.

\clearpage
\newpage

\newpage
%\onecolumn
%\input{High-z_clusters_supplementary.tex}

\null\vspace{\stretch{1}}
\begin{center}
\null\vspace{\stretch{1}}
{\bf \Large SUPPLEMENTARY INFORMATION}
\vspace{\stretch{1}}\null

%\null\vspace{\stretch{1}}
{\Large \bf Fast magnetic field amplification in distant galaxy clusters}
\end{center}

\bigskip

{\noindent 
\textbf{Gabriella Di Gennaro$^1$, 
Reinout J. van Weeren$^1$, 
Gianfranco Brunetti$^2$, 
Rossella Cassano$^2$, 
Marcus Br\"uggen$^3$, 
Matthias Hoeft$^4$, 
Timothy W. Shimwell$^{1,5}$, 
Huub J.A. R\"ottgering$^1$, 
Annalisa Bonafede$^{2,3,6}$, 
Andrea Botteon$^{1,2}$,
Virginia Cuciti$^3$, 
Daniele Dallacasa$^{2,6}$, 
Francesco de Gasperin$^3$, 
Paola Dom\'{i}nguez-Fern\'{a}ndez$^3$, 
Torsten A. En{\ss}lin$^7$, 
Fabio Gastaldello$^8$, 
Soumyajit Mandal$^1$, 
Mariachiara Rossetti$^8$, 
Aurora Simionescu$^{1,9,10}$}}\null

{\noindent$^1$\emph{Leiden Observatory, Leiden University, PO Box 9513, 2300 RA Leiden, The Netherlands}

\noindent$^2$\emph{Istituto Nazionale di Astrofisica-Istituto di Radioastronomia, Bologna Via Gobetti 101, I40129 Bologna, Italy}

\noindent$^3$\emph{Hamburger Sternwarte, Universit\"at Hamburg, Gojenbergsweg 112, 21029 Hamburg, Germany}

\noindent$^4$\emph{Th\"uringer Landessternwarte, Sternwarte 5, 07778 Tautenburg, Germany}

\noindent$^5$\emph{ASTRON, The Netherlands Institute for Radio Astronomy, Postbus 2, 7990 AA, Dwingeloo, The Netherlands}

\noindent$^6$\emph{Dipartimento di Fisica e Astronomia, Università degli Studi di Bologna, Via Gobetti 93/2, I40129 Bologna, Italy}

\noindent$^7$\emph{MPI for Astrophysics, Karl Schwarzschildstr. 1, 85741 Garching}

\noindent$^8$\emph{IASF-Milano/INAF, via Corti 12, 20133 Milano}

\noindent$^9$\emph{SRON Netherlands Institute for Space Research, Sorbonnelaan 2, 3584 CA Utrecht, The Netherlands}

\noindent$^{10}$\emph{Kavli Institute for the Physics and Mathematics of the Universe (WPI), The University of Tokyo, Kashiwa, Chiba 277-8583, Japan}}\null
\null\vspace{\stretch{1}}

\newpage

\subsection*{Physical properties of the galaxy clusters in the sample}
Here we report the table with the properties of the galaxy clusters selected from the {\it Planck} sample. 

\begin{table*}[h!]
\begin{center}
\caption*{\bf Supplementary Table 1: List of galaxy clusters from Planck SZ Catalog with $\rm DEC\geq20^\circ$ and $z\geq0.6$}
\resizebox{0.99\textwidth}{!}{\begin{tabular}{lcccccccc} %\checkmark
\hline
\hline
PSZ name 			& $z$	& RA$_{\rm J2000}$	& DEC$_{\rm J2000}$	& $M_{\rm SZ,500} $			& LOFAR 		& {\it Chandra}	& {\it XMM}	    & Ref.\\
		 			&		& [deg]			& [deg]				& [$10^{14}~{\rm M}_\odot$] &	 	    	& 			    & 			    & \\
\hline
PSZ2G032.31+66.07	& 0.609 & 219.35404		& +24.3986			& $5.6\pm0.8$		        &		    	&		    	&		    	& \cite{planckcoll16,streblyanska+18,zohren+19} \\
PSZ2G045.87+57.70	& 0.611	& 229.58737		& +29.46957  		& $7.0\pm0.7$				& \checkmark	& 		    	& \checkmark	& \cite{planckcoll16} \\
PSZ2G066.34+26.14	& 0.630	& 270.277		& +39.86853			& $4.8\pm0.7$		    	&       		& 		    	& 		    	& \cite{planckcoll16,streblyanska+18,zohren+19} \\
PSZ2G069.39+68.05	& 0.762	& 215.40979		& +38.35494			& $-$					    &       		&		    	&			    & \cite{planckcoll16,burenin+18,streblyanska+18} \\
PSZ2G070.89+49.26	& 0.610 & 239.21558		& +44.63012			& $6.5\pm0.7$				& \checkmark 	& 		    	& \checkmark	& \cite{planckcoll16} \\ 
PSZ2G073.31+67.52	& 0.609	& 215.16250		& +39.91500	 		& $6.7\pm0.6$				&           	& 			    & \checkmark 	& \cite{planckcoll16} \\
PSZ2G084.10+58.72	& 0.731	& 222.25554		& +48.55556			& $5.4\pm0.6$				& \checkmark 	& \checkmark	& \checkmark	& \cite{planckcoll16} \\ 
PSZ2G085.95+25.23	& 0.782	& 277.6164 		& +56.8823 			& $5.4\pm0.6$   			&  		    	& 	    		& 		    	& \cite{planckcoll16,amodeo+18,zohren+19}\\
PSZ2G086.28+74.76	& 0.699	& 204.47446		& +38.90189			& $5.3^{+0.7}_{-0.8}$    	& \checkmark	& 	    		& 		    	& \cite{planckcoll16,streblyanska+18,zohren+19} \\
PSZ2G086.93+53.18	& 0.675	& 228.50446		& +52.81074			& $5.4\pm0.5$				& \checkmark 	& \checkmark	& \checkmark	& \cite{planckcoll16} \\ 
PSZ2G087.39+50.92	& 0.748	& 231.63821		& +54.15210			& $5.2^{+0.5}_{-0.6}$		& \checkmark 	& 	    		& \checkmark    & \cite{planckcoll16} \\ 
PSZ2G088.98+55.07	& 0.702	& 224.74384		& +52.81730			& $4.9\pm0.6$				& \checkmark 	& \checkmark	& \checkmark	& \cite{planckcoll16} \\
PSZ2G089.39+69.36	& 0.680	& 208.43748		& +43.48476 		& $5.7\pm0.7$				& \checkmark 	& 	    		& 			    & \cite{planckcoll16} \\
PSZ2G091.83+26.11	& 0.822	& 277.78430		& +62.24770			& $7.4\pm0.4$				& \checkmark 	& \checkmark	& \checkmark	& \cite{planckcoll16,amodeo+18} \\ 
PSZ2G092.69+59.92	& 0.848 & 216.65041		& +51.26417			& $4.5\pm0.5$		    	& \checkmark	& \checkmark	& \checkmark    & \cite{planckcoll16,burenin+18,zohren+19} \\
PSZ2G097.52+51.70	& 0.700	& 223.8588		& +58.85514			& $5.2\pm0.5$				& \checkmark 	& 		    	& \checkmark	& \cite{planckcoll16} \\
PSZ2G099.86+58.45	& 0.616	& 213.6909		& +54.78029			& $6.8\pm0.5$ 				& \checkmark 	& 		    	& \checkmark	& \cite{planckcoll16,cassano+19,sereno+18} \\
PSZ2G100.22+33.81	& 0.620	& 258.43971		& +69.36335			& $-$				    	&       		&	    		&			    & \cite{planckcoll16,streblyanska+18}\\
PSZ2G104.74+40.42	& 0.690	& 236.6082		& +69.95975			& $4.3^{+0.5}_{-0.6}$		& \checkmark 	& 		    	& 		    	& \cite{planckcoll16,barrena+18} \\
PSZ2G106.15+25.75	& 0.630	& 284.25043		& +74.93208			& $4.6\pm0.7$				&		    	&		    	&		    	& \cite{planckcoll16,vanderburg+16} \\
PSZ2G108.27+48.66	& 0.674	& 52.79720		& +65.65932			& $4.9\pm0.5$				&           	& 		    	& 			    & \cite{planckcoll16,barrena+18} \\
PSZ2G126.28+65.62	& 0.820	& 190.5975		& +51.43944			& $5.0\pm0.7$		    	& \checkmark	& 	    		& 		    	& \cite{planckcoll16,burenin+18,zohren+19} \\
PSZ2G126.57+51.61	& 0.815	& 187.4492		& +65.35361			& $5.8\pm0.6$		    	&           	& 	    		& 		    	& \cite{planckcoll16,burenin+18,zohren+19} \\
PSZ2G127.01+26.21	& 0.630	& 89.6057       & +86.2544			& $7.3^{+0.8}_{-0.9}$		& \checkmark 	& \checkmark	& 		    	& \cite{planckcoll16} \\
PSZ2G139.00+50.92	& 0.600	& 170.070701	& +63.24996			& $4.6^{+0.7}_{-0.8}$    	& \checkmark	& 		    	& 		    	& \cite{planckcoll16,streblyanska+18,zohren+19} \\
PSZ2G141.77+14.19	& 0.830	& 70.27167  	& +68.22275			& $7.7\pm0.9$				& \checkmark 	& \checkmark 	& 		    	& \cite{planckcoll16,vanderburg+16} \\ 
PSZ2G141.98+69.31	& 0.714	& 183.16929		& +46.35641			& $4.1^{+0.9}_{-0.9}$    	& \checkmark 	& 		    	& 		    	& \cite{planckcoll16,streblyanska+18,zohren+19} \\
PLCKG147.3--16.6	& 0.645	& 44.105898	    &+40.290140			& $6.3\pm0.4$				& \checkmark	& 		    	& \checkmark	& \cite{planckcoll16,amodeo+18,vanweeren+14} \\
PSZ2G147.88+53.24	& 0.600	& 164.37923 	& +57.99591			& $6.5\pm0.6$				& \checkmark	& 		    	& 		    	& \cite{planckcoll16} \\ 
PSZ2G160.83+81.66	& 0.888	& 186.74267		& +33.54682 		& $5.7^{+0.6}_{-0.7}$		& \checkmark	& \checkmark	& \checkmark	& \cite{planckcoll16} \\ 
\hline
\end{tabular}}
\end{center}

\end{table*}

\subsection*{LOFAR images}

\begin{figure}
\centering
{\includegraphics[width=0.3\textwidth]{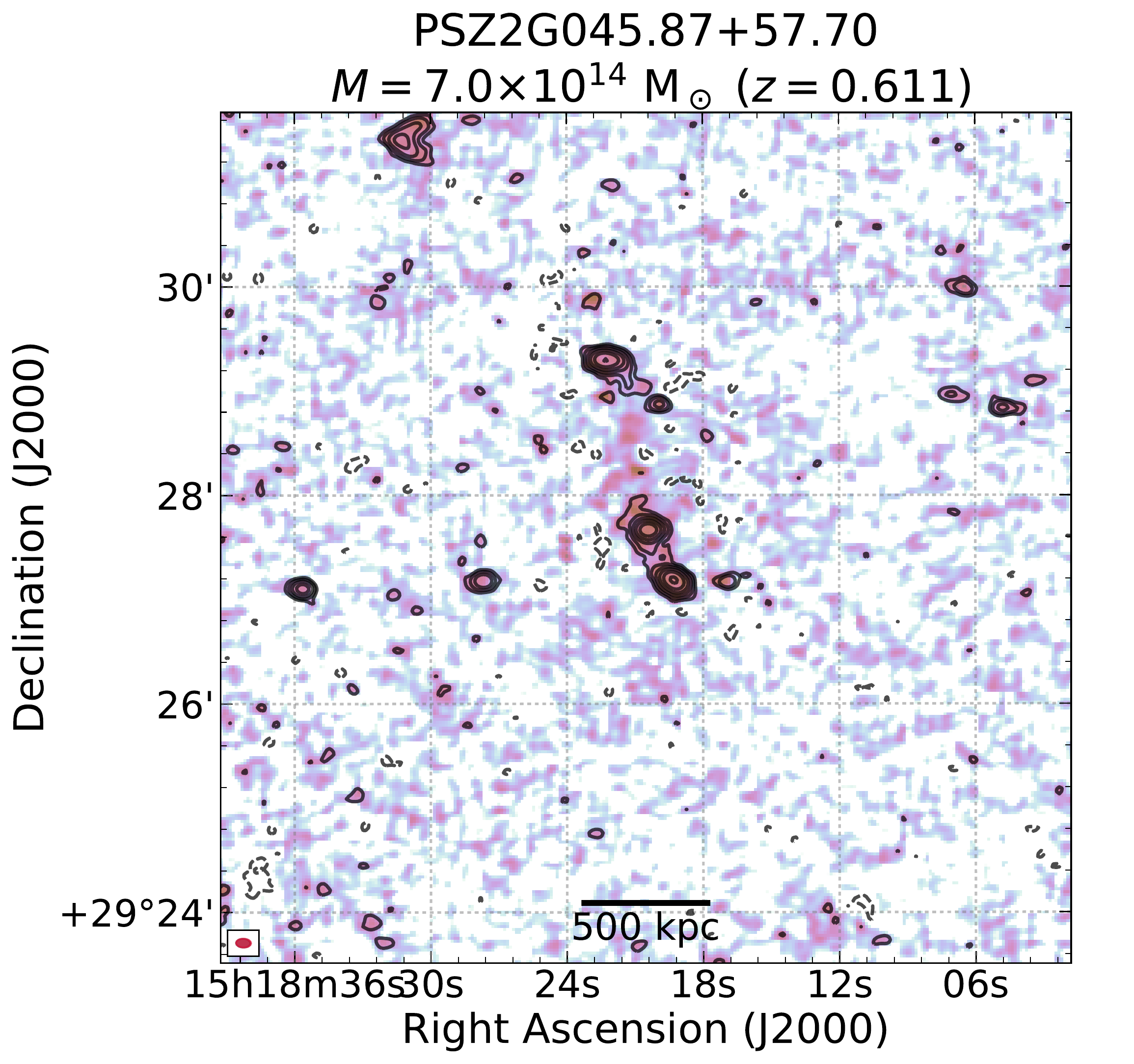}}
{\includegraphics[width=0.3\textwidth]{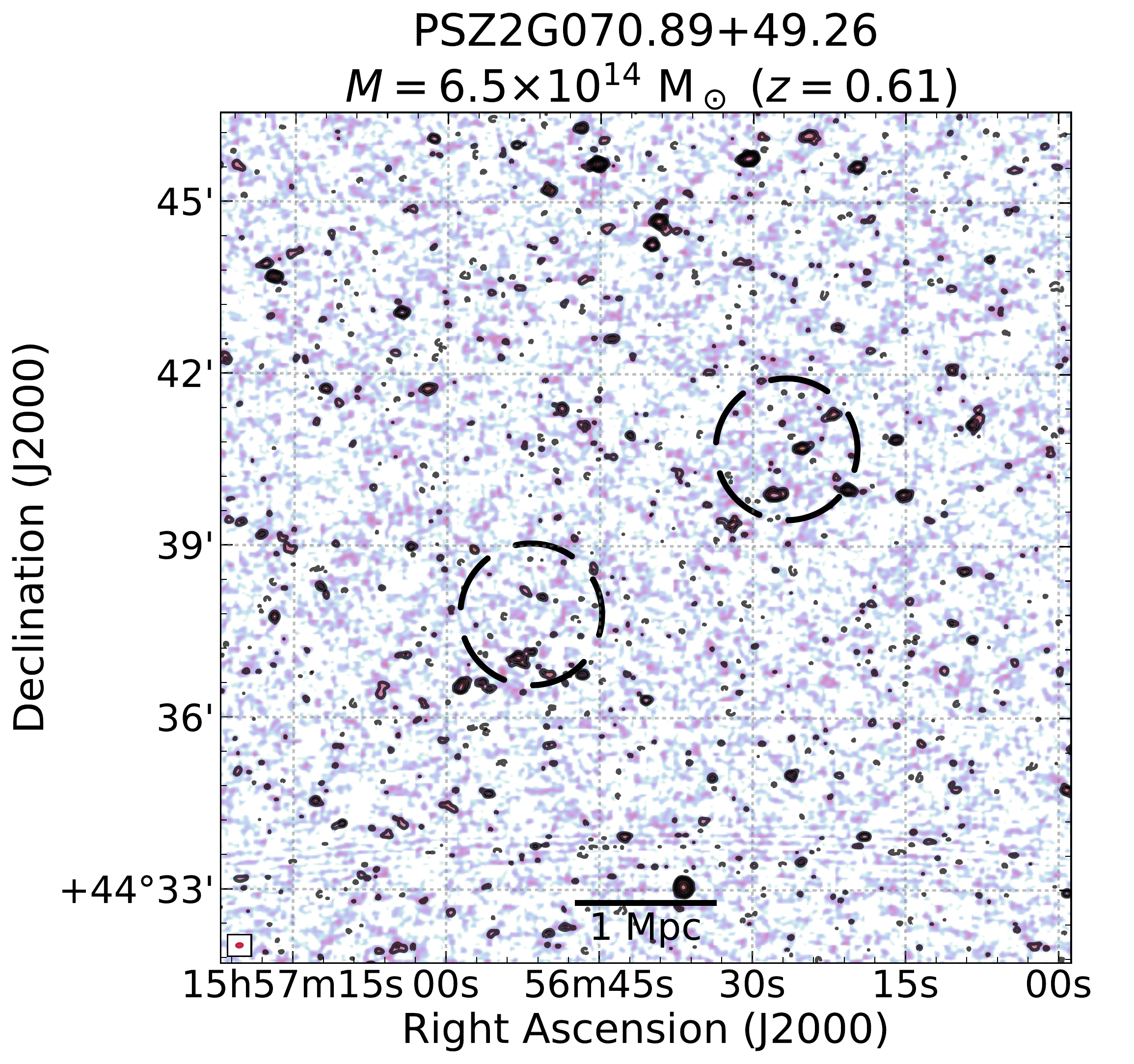}}
{\includegraphics[width=0.3\textwidth]{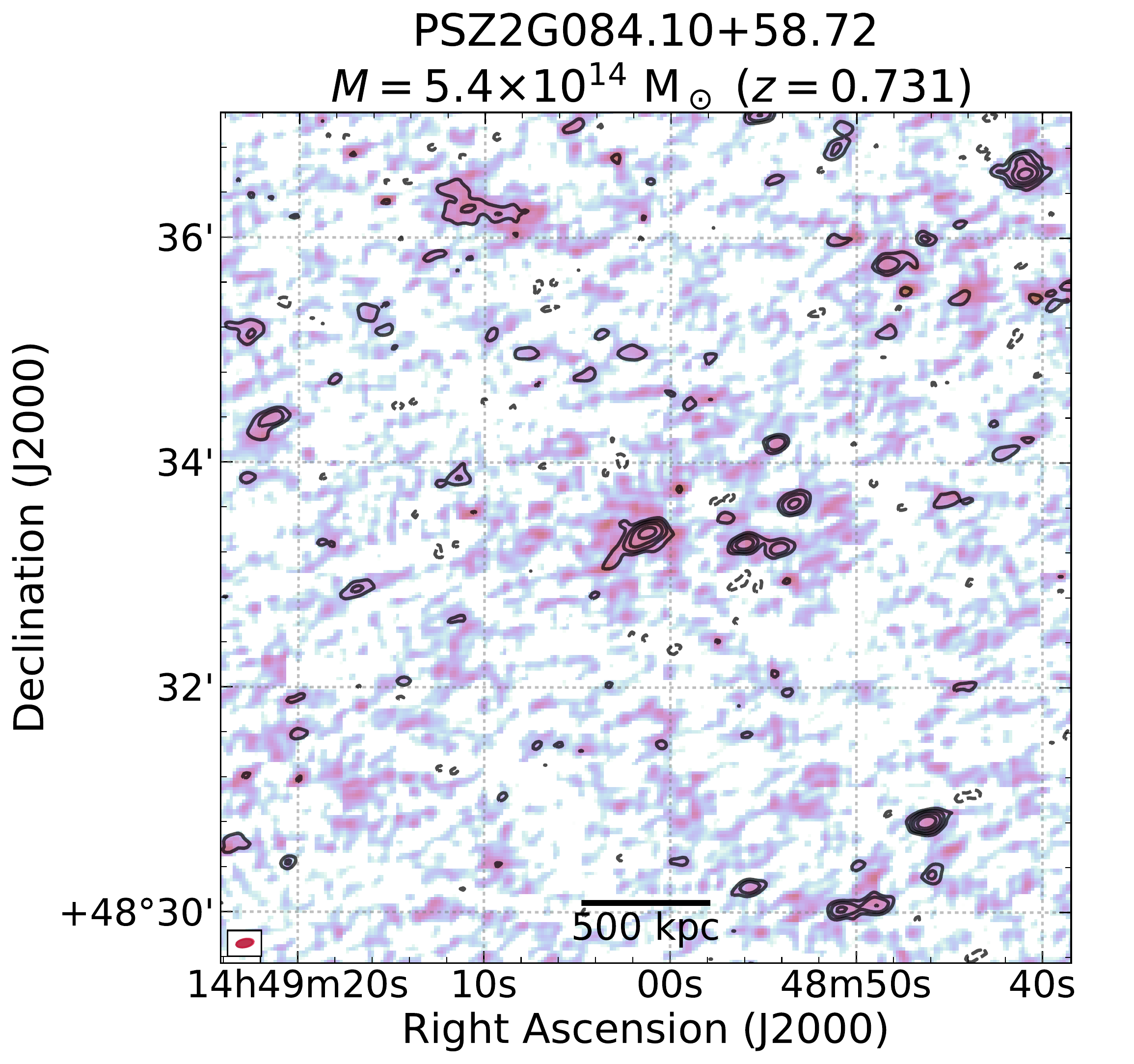}}
{\includegraphics[width=0.3\textwidth]{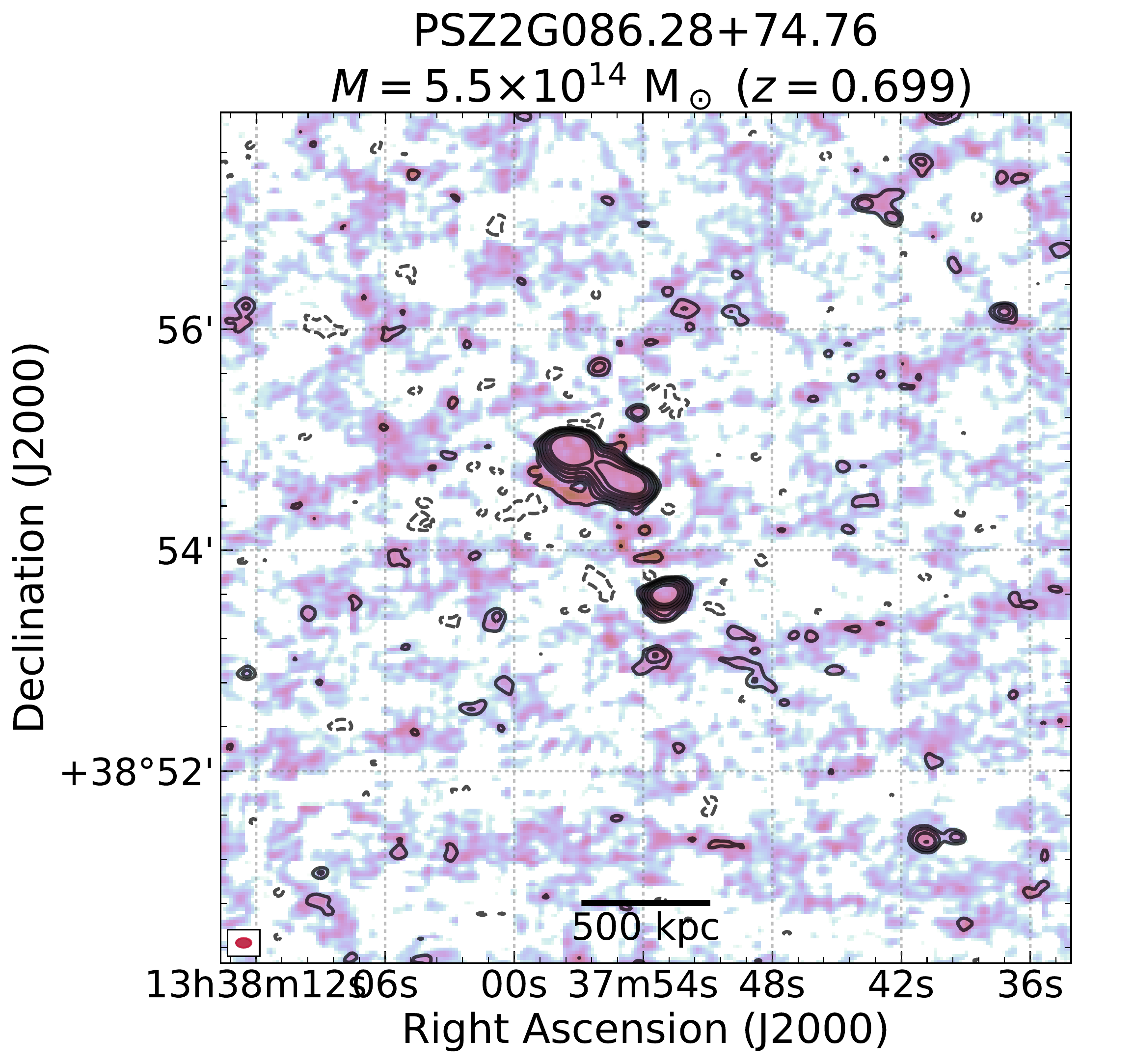}}
{\includegraphics[width=0.3\textwidth]{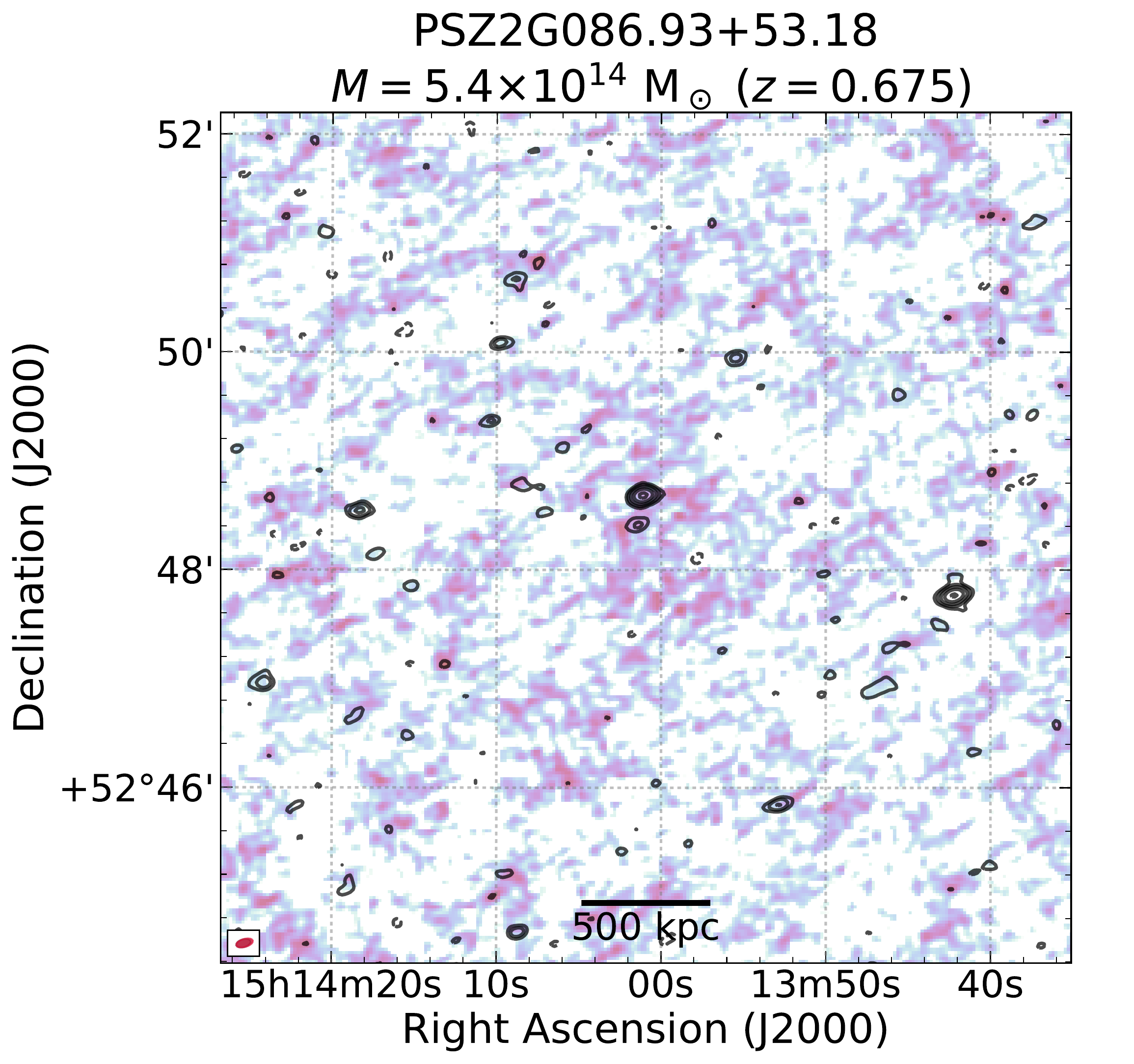}}
{\includegraphics[width=0.3\textwidth]{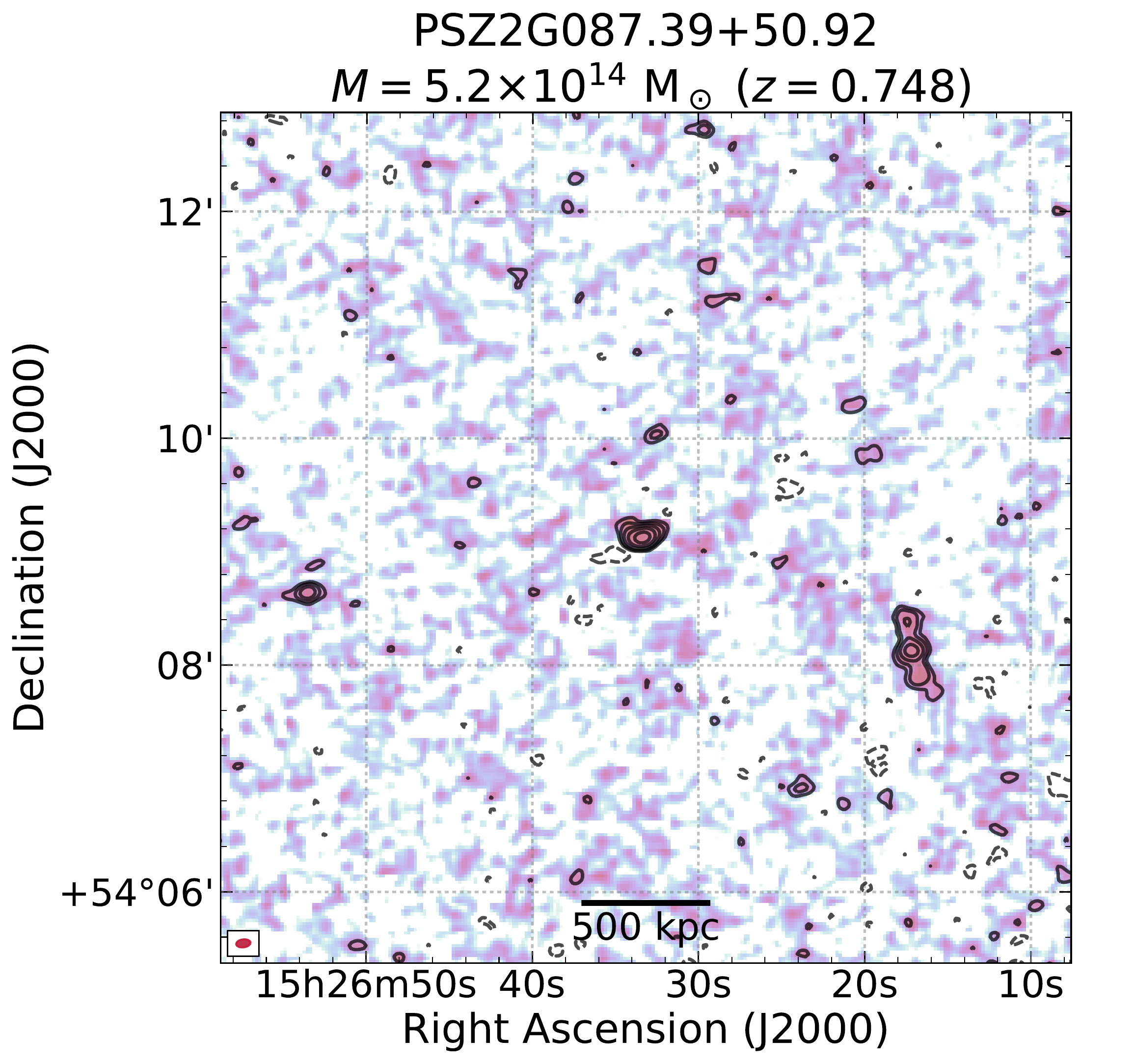}}
{\includegraphics[width=0.3\textwidth]{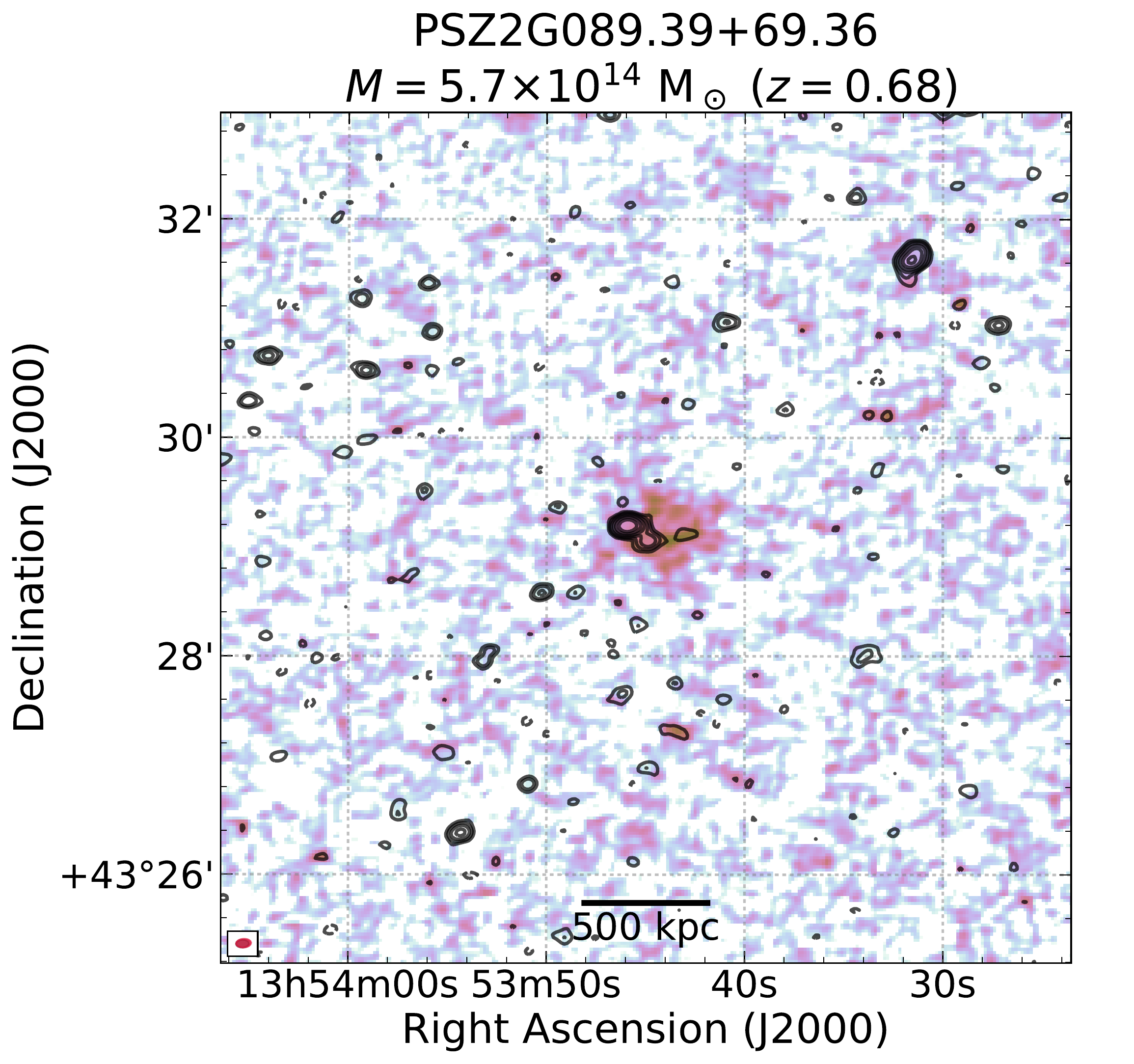}}
{\includegraphics[width=0.3\textwidth]{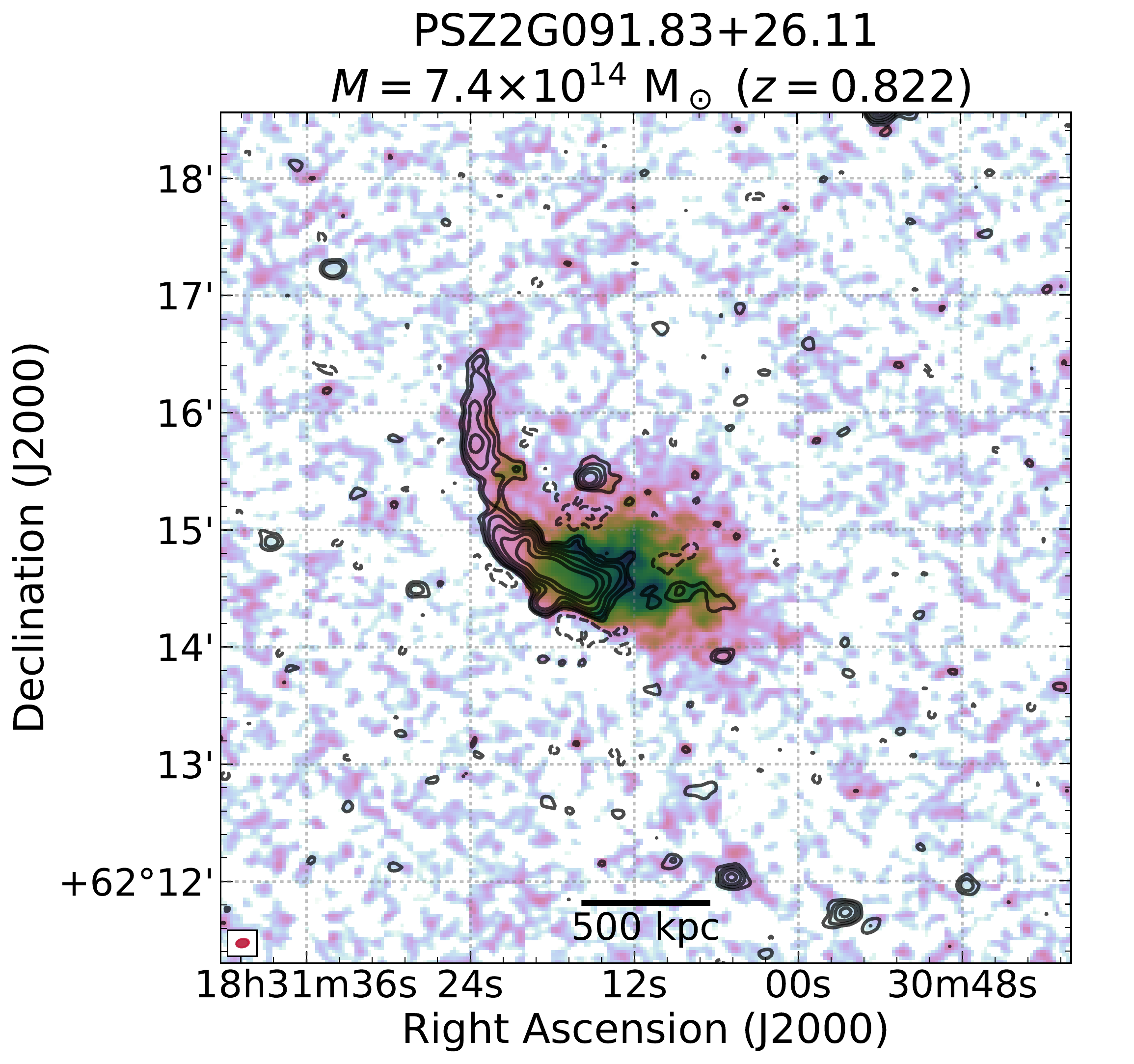}}
{\includegraphics[width=0.3\textwidth]{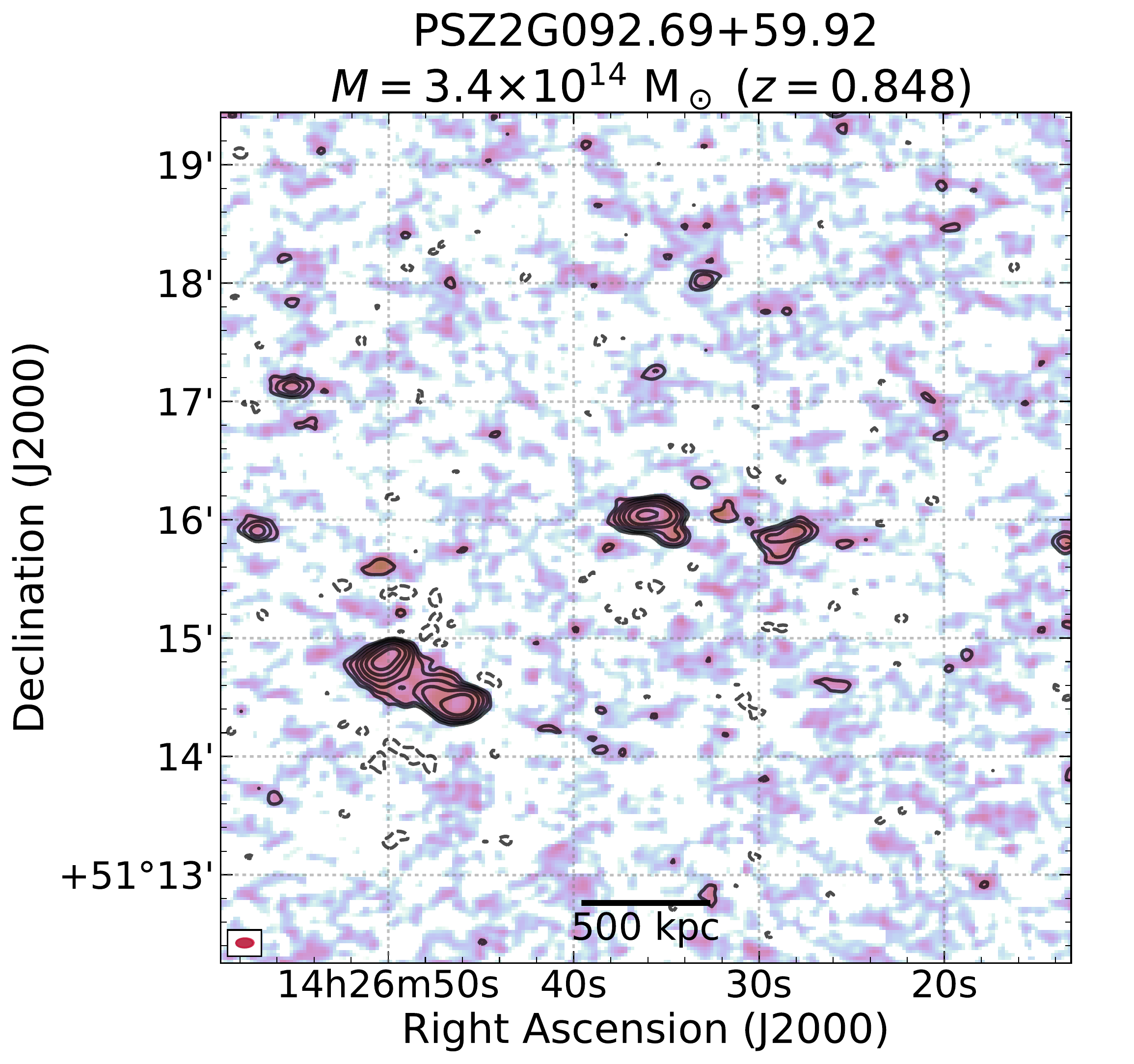}}
{\includegraphics[width=0.3\textwidth]{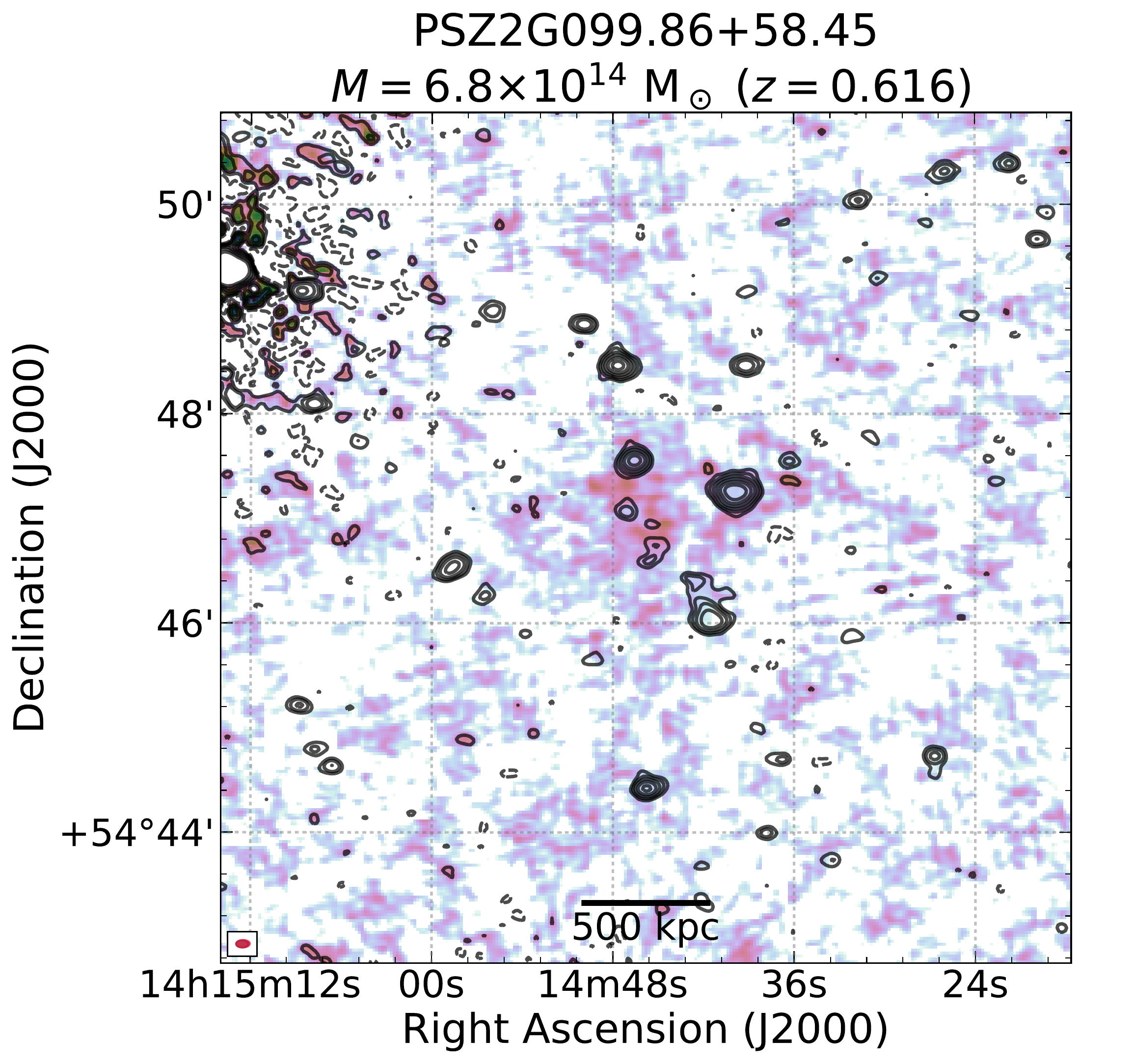}}
{\includegraphics[width=0.3\textwidth]{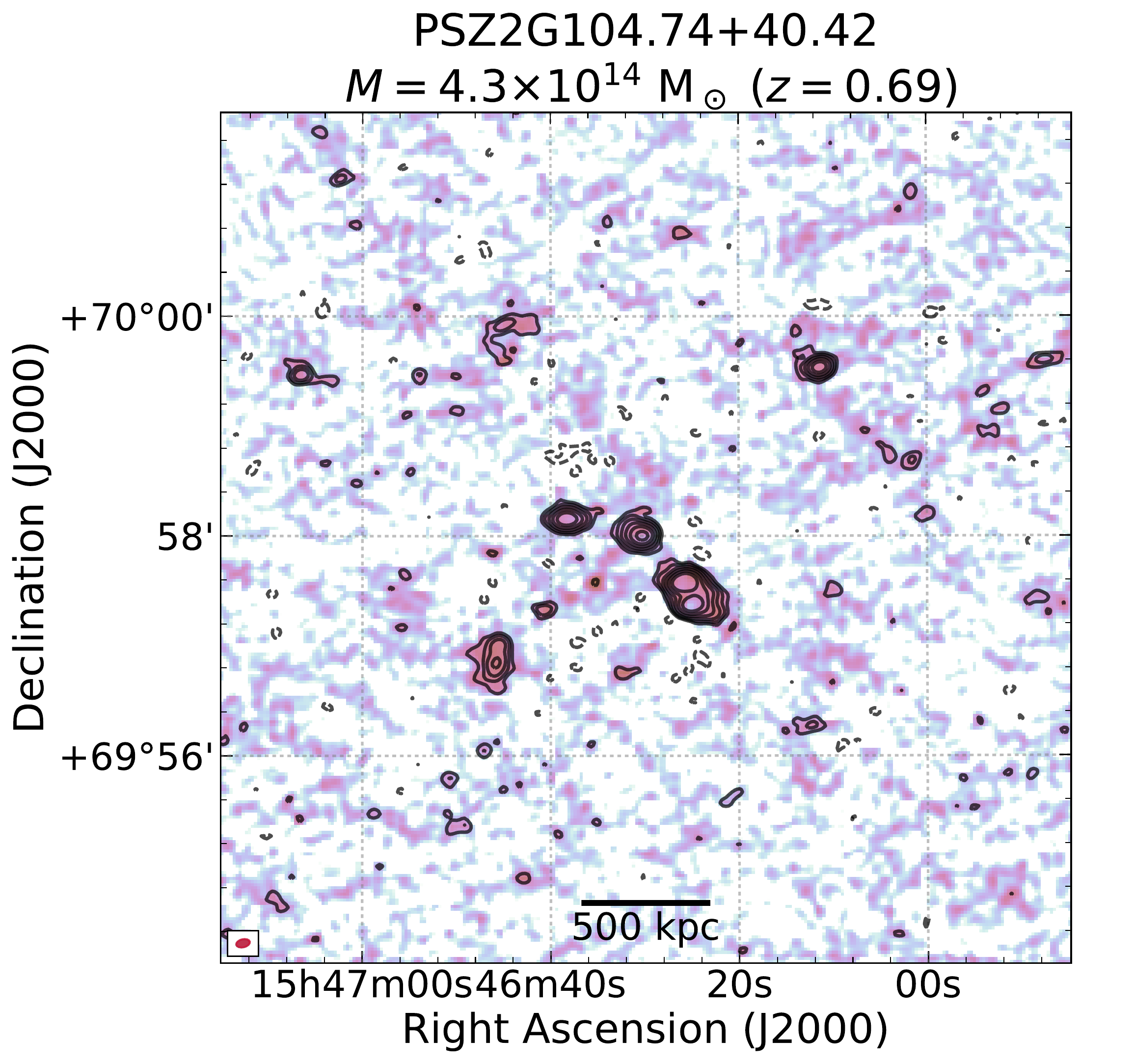}}
{\includegraphics[width=0.3\textwidth]{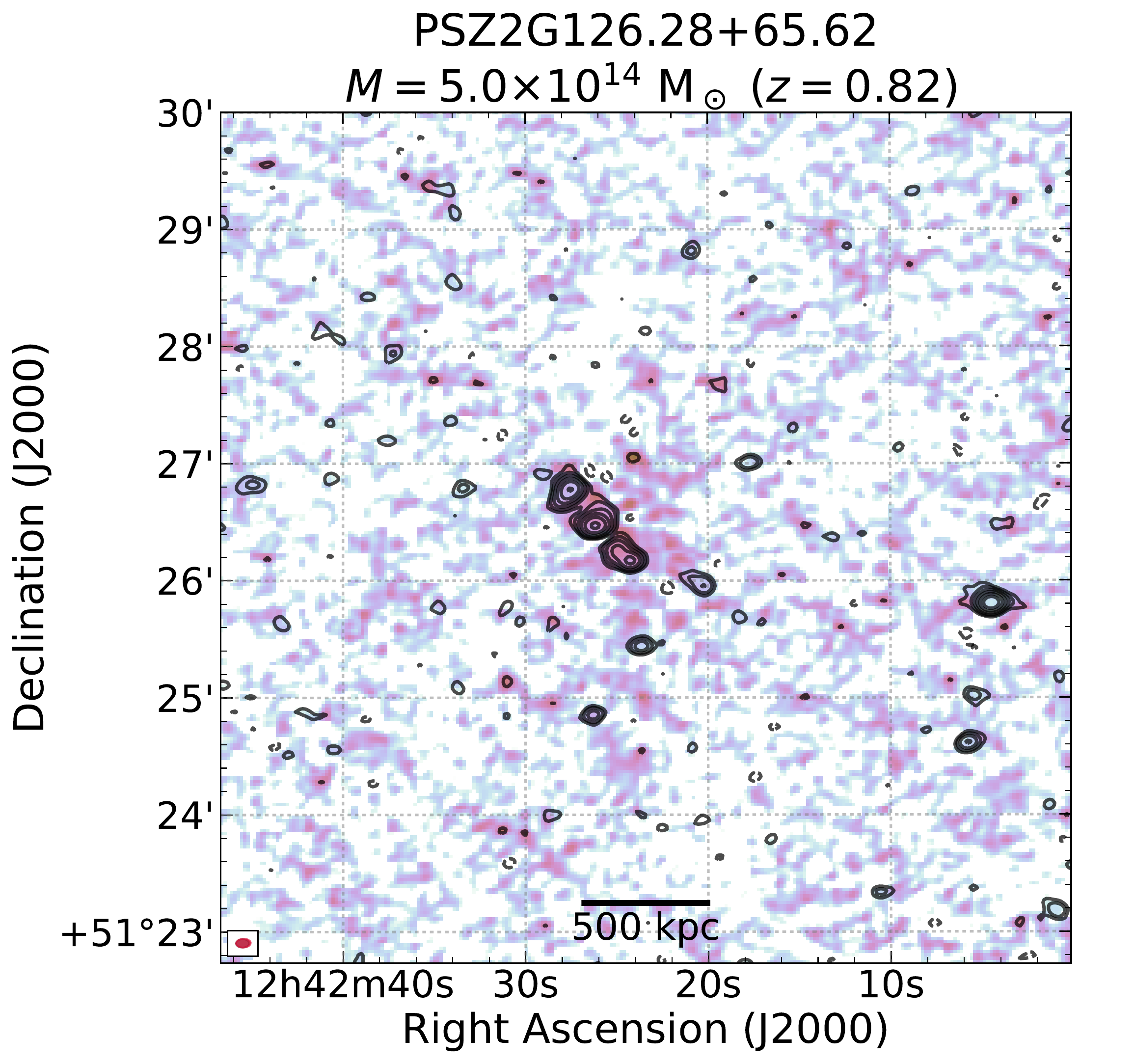}}
\caption*{{\bf Supplementary Figure 1: Observed diffuse radio emission in our high-$z$ galaxy cluster sample.} In colorscale we show the full-resolution LOFAR images of the clusters in our sample after the subtraction of the compact sources. Radio contours show the subtracted compact sources, at the same resolution. The contour levels are drawn at $2\sigma_{\rm rms}\times[-1,1, 2, 4, \dots]$, with $\sigma_{\rm rms}$ the noise level of the full-resolution $uv$-cut image (with the short-dashed line displaying the negative contour). In the header of each image, the galaxy cluster name, mass and redshift are reported. The dashed black circle in each map shows the $R=0.5R_{\rm SZ,500}$ region, obtained from $M_{\rm SZ,500}$.}\label{fig:radio_highressub_sample}
\end{figure}

\begin{figure}[t!]
\centering
\ContinuedFloat
{\includegraphics[width=0.3\textwidth]{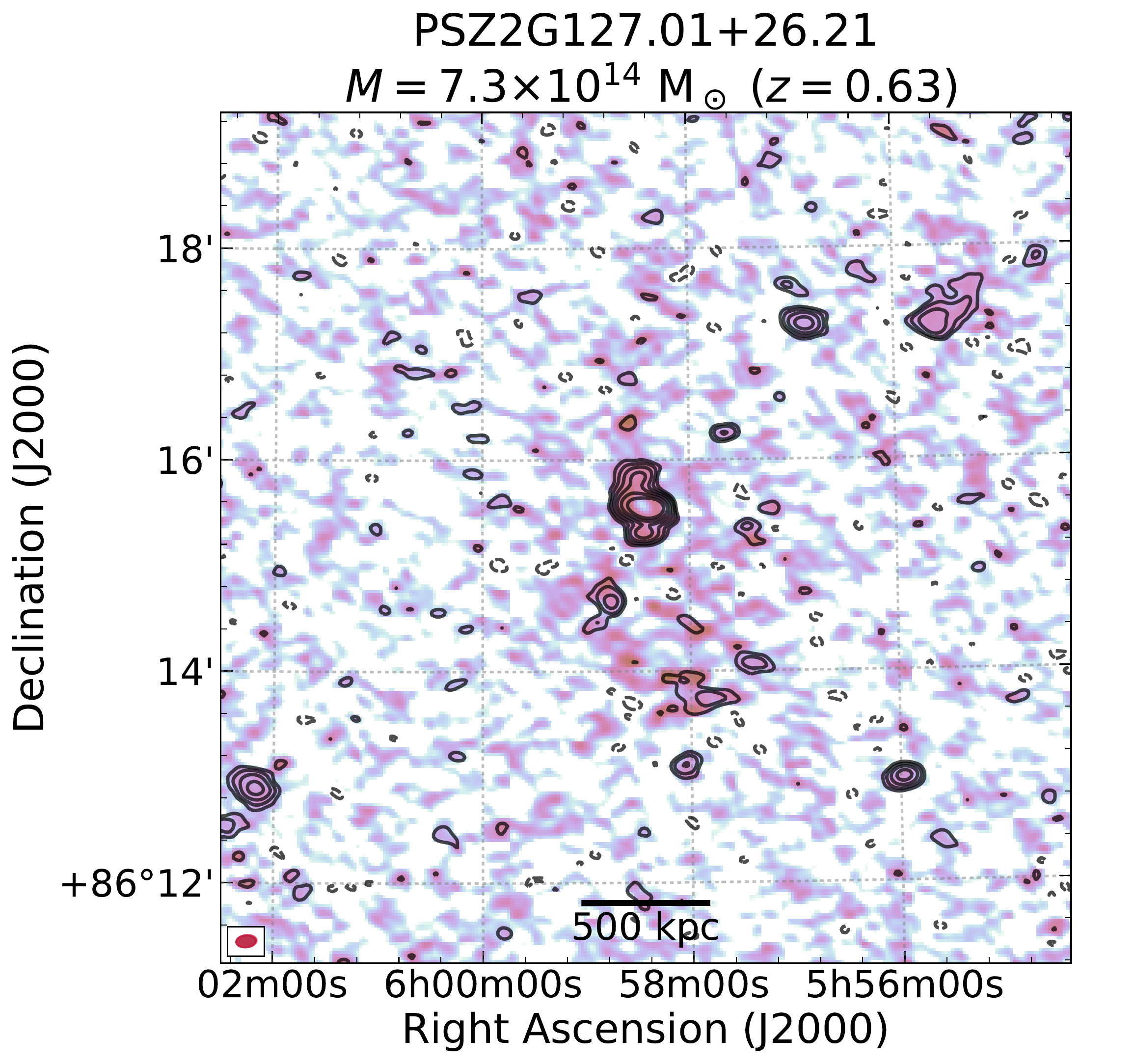}}
{\includegraphics[width=0.3\textwidth]{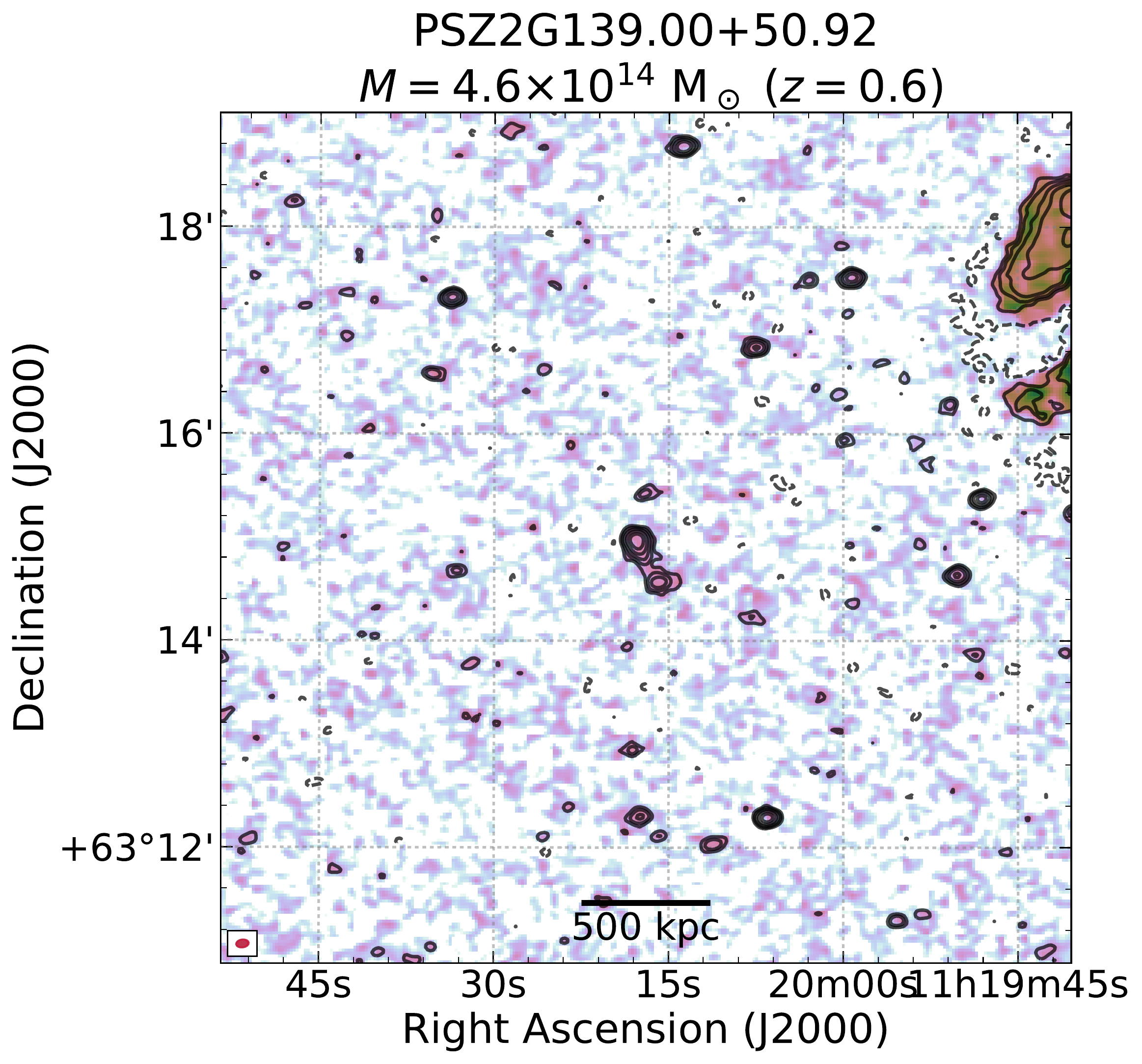}}
{\includegraphics[width=0.3\textwidth]{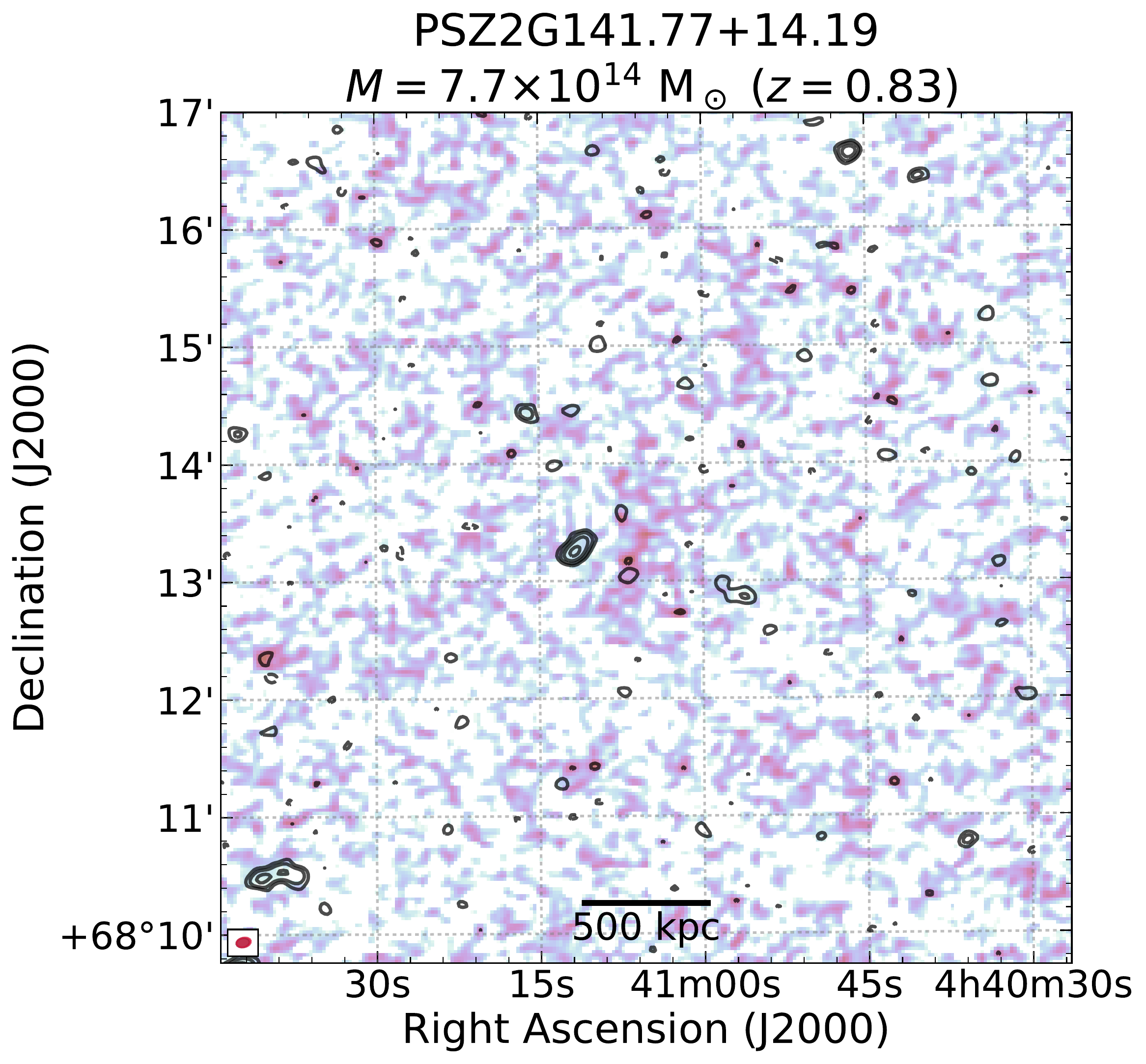}}
{\includegraphics[width=0.3\textwidth]{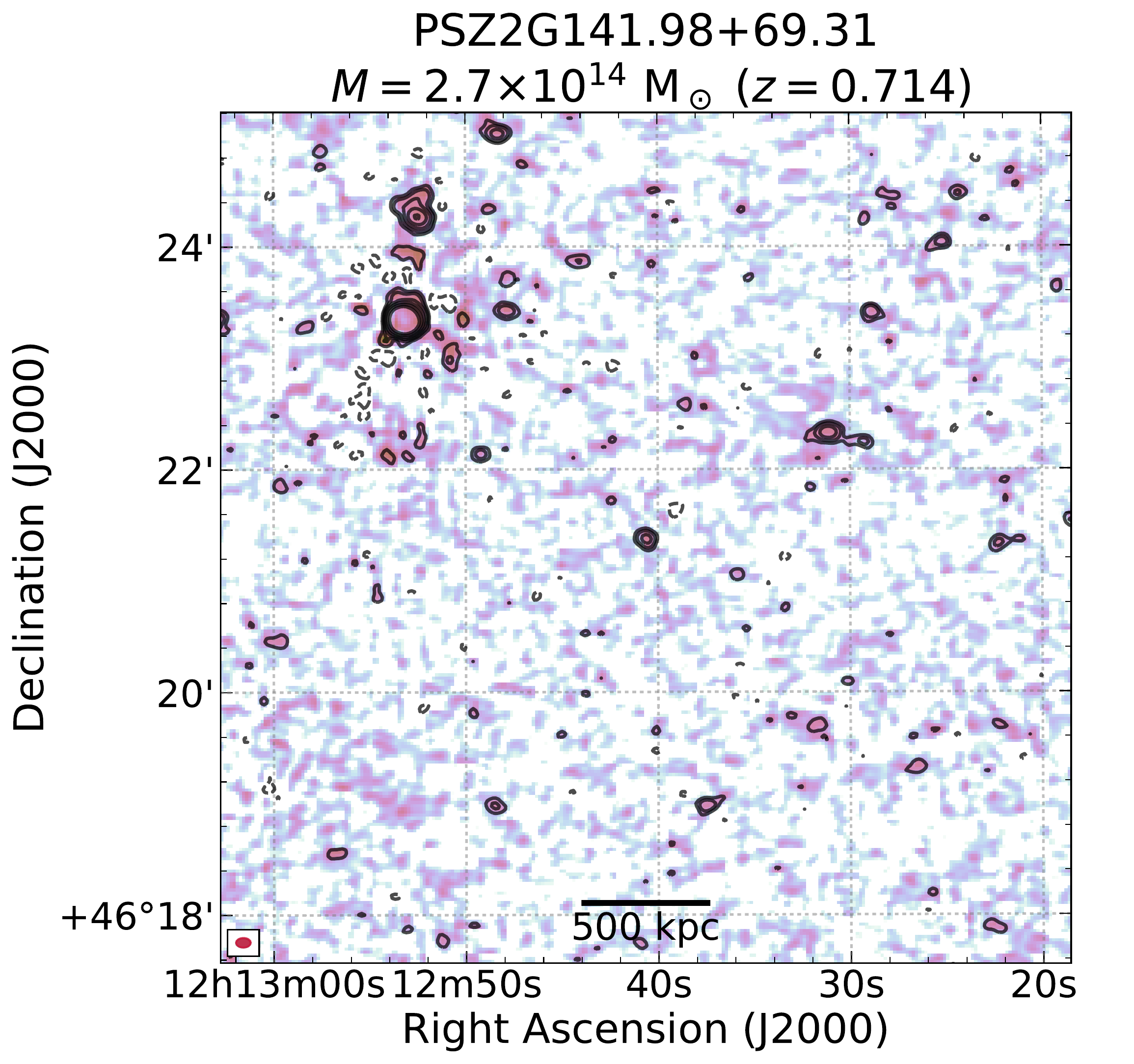}}
{\includegraphics[width=0.3\textwidth]{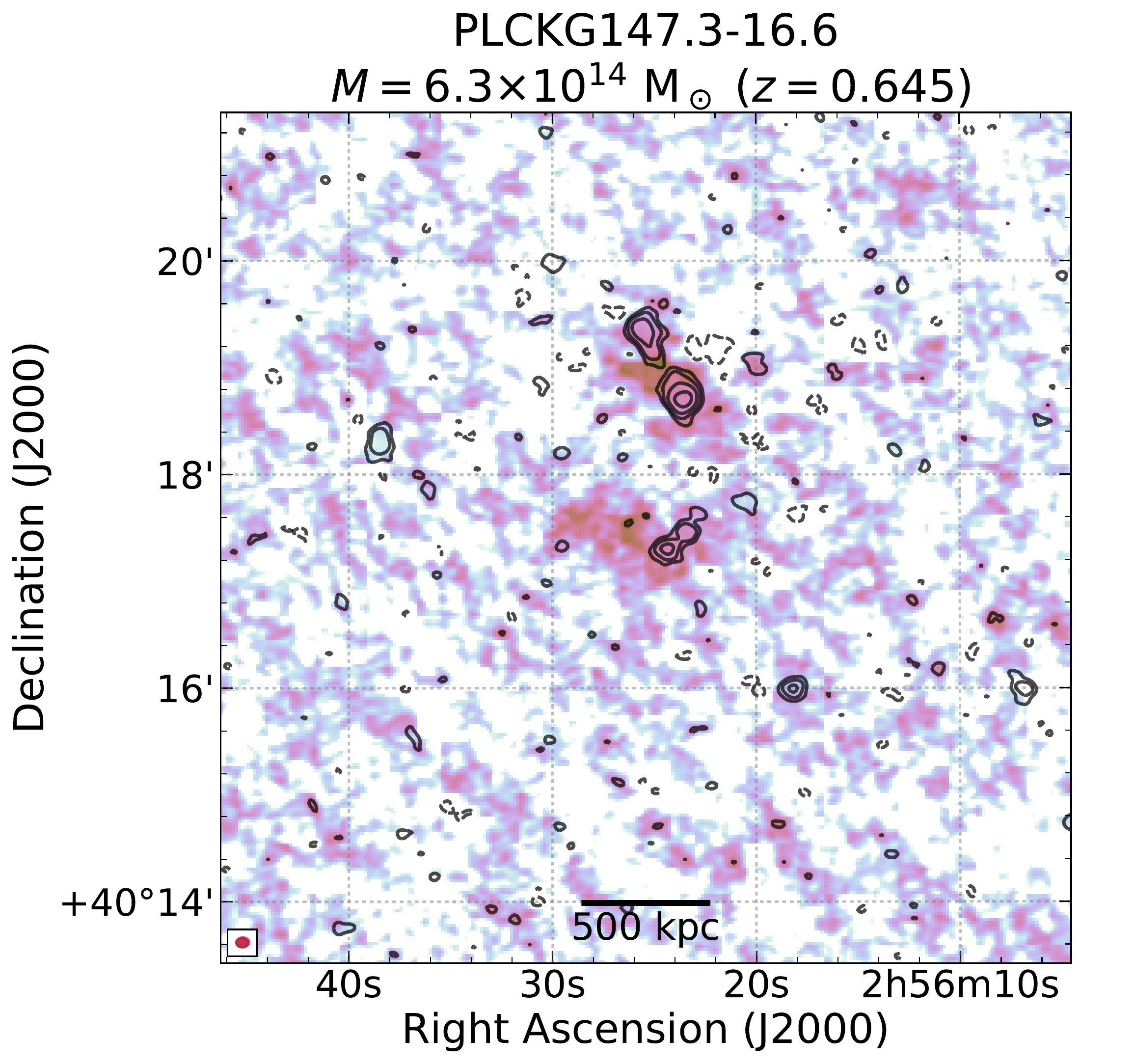}}
{\includegraphics[width=0.3\textwidth]{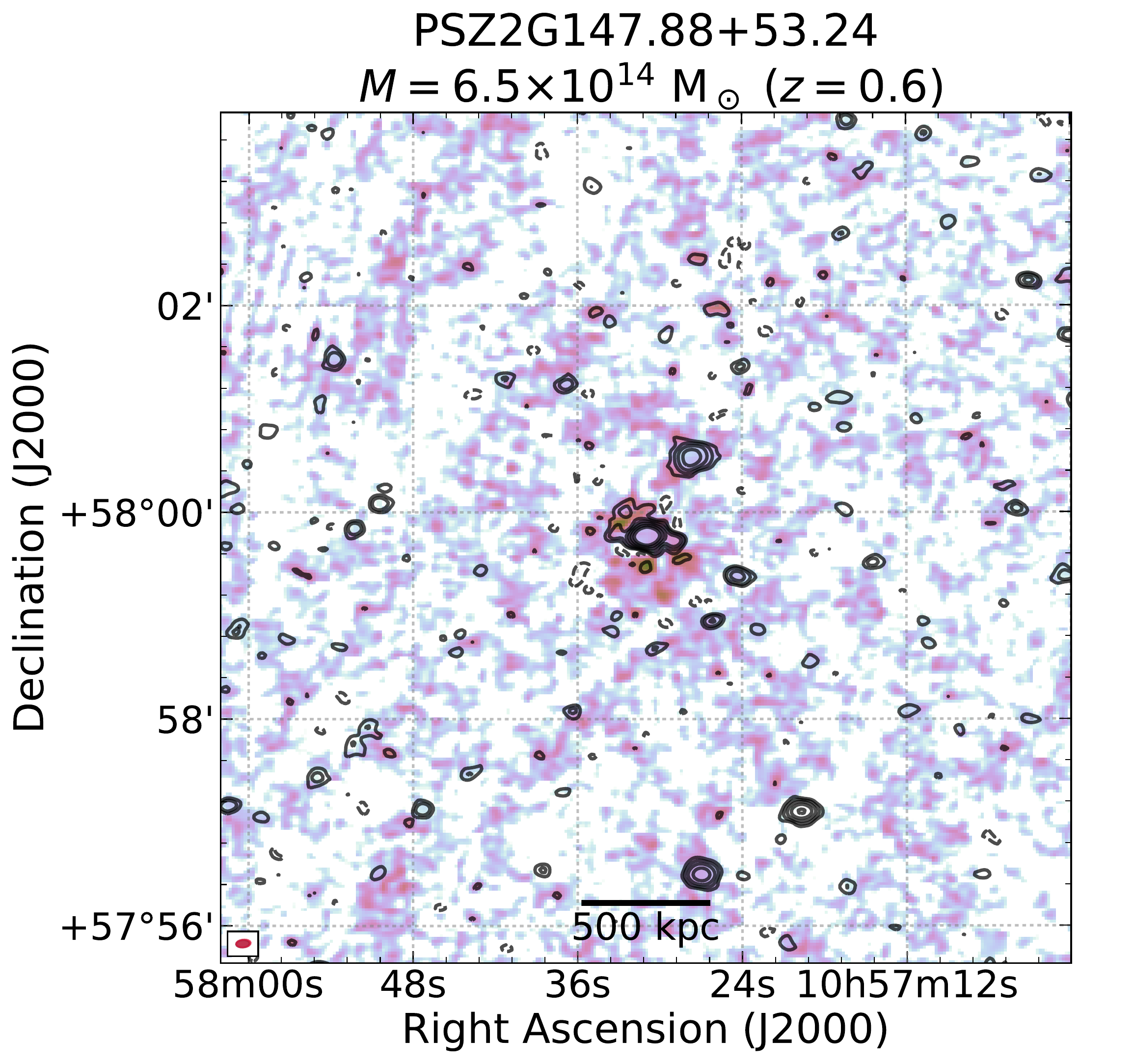}}
{\includegraphics[width=0.3\textwidth]{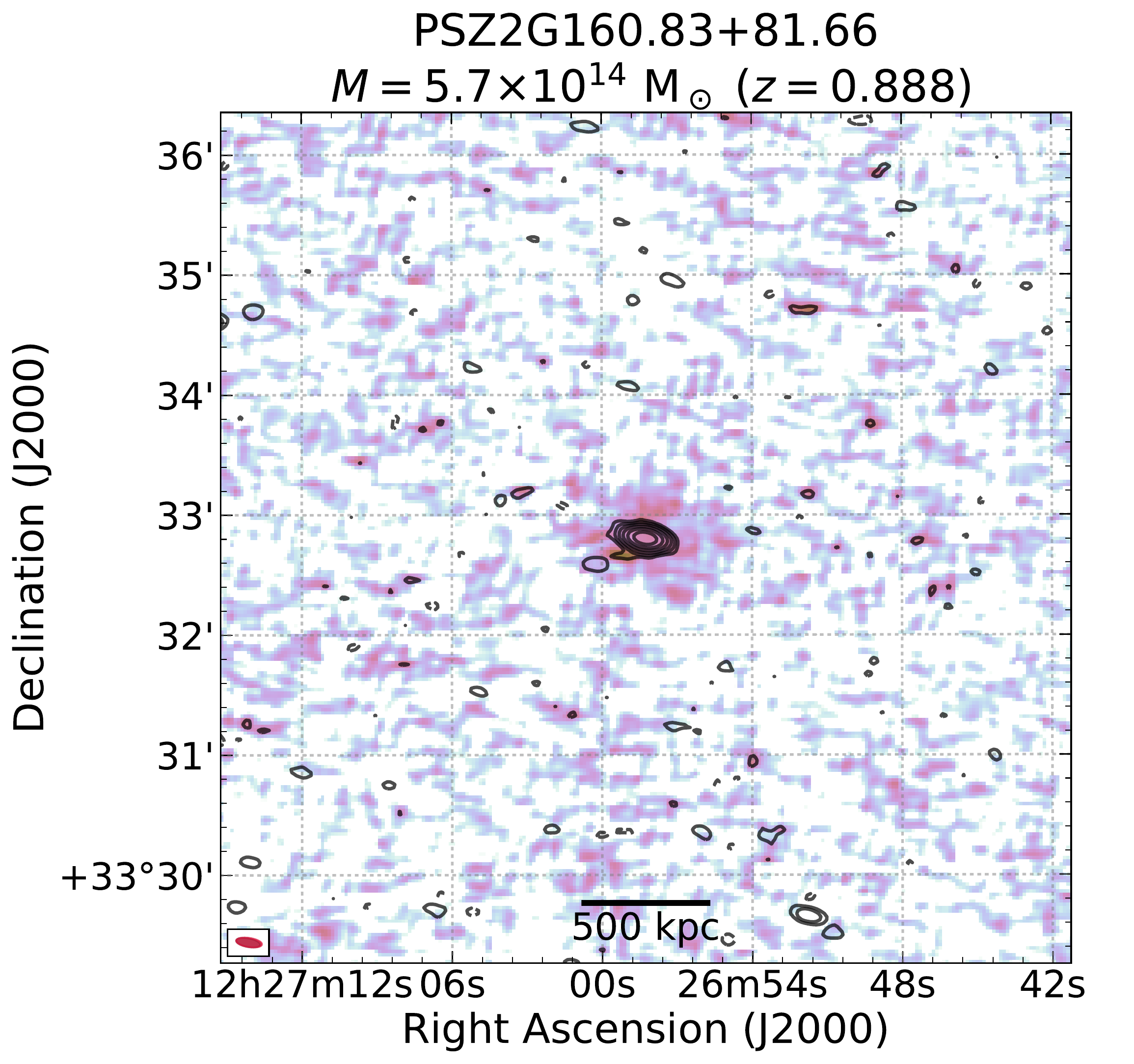}}
\caption*{\bf Supplementary Figure 1 continued.}
\end{figure}

In this section we present the LOFAR full resolution source-subtracted images of the clusters in our sample (Supplementary Figure 1).
The radio contours of the subtracted radio galaxies overlaid, starting from $3\sigma_{\rm rms}$, with $\sigma_{\rm rms}$ the noise of the full-resolution $uv$-cut map, spaced by a factor of 2. The dashed contours display the $-3\sigma_{\rm rms}$ contour level. 

Below we provide a brief description of the galaxy clusters (ObsIDs list is reported in Supplementary Table 2).

\begin{itemize}
\item PSZ2\,G045.87+57.70 ($z=0.611$) shows extended diffuse radio emission in the north-south direction, which is unresolved in the east-west direction. For this reason, and since it is not located in the cluster center, we classify this emission as ``uncertain''. The X-ray emission also looks elongated in the north-south direction.

\item PSZ2\,G070.89+49.26 ($z=0.610$) is actually a double cluster \cite{planckcoll15}. In both systems, there is no evidence of diffuse radio emission (see Extended Data Figure 1). 

\item PSZ2\,G084.10+58.72 ($z=0.731$) does not show clear presence of extended diffuse emission in the full resolution image. The X-ray emission suggests a dynamically disturbed system.

\item An extended, probably foreground, 
radio galaxy contaminates the radio emission from PSZ2\,G086.28+74.76 ($z=0.699$). We cannot classify this cluster hosting diffuse radio emission. No X-ray observations are available.

\item PSZ2\,G086.93+53.18 ($z=0.675$) is the faintest candidate halos in our sample. We find hints of diffuse radio emission on the $2\sigma_{\rm rms}$ level, located southward of the BCG. The X-ray emission is spherically symmetric within the $R=0.5R_{500}$ region.

\item PSZ2\,G087.39+50.92 ($z=0.748$) does not show extended radio emission. The X-ray image suggests this is a dynamically relaxed system.

\item PSZ2\,G089.39+69.36 ($z=0.680$) hosts a bright radio halo with a size of about 1 Mpc. No X-ray observations are available for this cluster.

\item PSZ2\,G091.83+26.11 ($z=0.822$) hosts the brightest radio halo in our sample ($S_{\rm 144MHz}=84.3\pm12.7$ mJy). Moreover, it also hosts an elongated 1.2~Mpc arc-like radio source. No clear optical counterparts is found for this source. The brightest region of this arc is also visible in a short Very Large Array (VLA) observation, which also allowed us to define the relic region for the flux density measurements (see the blue polygon in Supplementary Figure 3). 
Future spectral index and polarization studies would help to verify the different contributions of the halos and relic. Combining our LOFAR flux density  measurements with that of the VLA yields an integrated spectral index of $\alpha\sim-1.6$ (also see the following Section). The steep spectral index, the lack of clear optical counterparts and the elongated shape, strongly suggest we are observing a radio relic. 
Shallow X-ray observations also show disturbed thermal emission, elongated in the north-south direction, suggesting that the cluster is in a merging state (second panel in Figure 2). 
The radio relic is located parallel to the putative merger axis, and not perpendicularly to it as it is usually observed in low-redshift systems. This suggests a merger that is more complex than a head-on binary merger.

\item PSZ2\,G092.69+59.92 ($z=0.848$) does not show extended radio emission. The dynamical state of this cluster is unclear from the X-ray image.

\item The radio halo in PSZ2\,G099.86+58.45 ($z=0.616$) was recently studied \cite{cassano+19}, combining LOFAR, VLA and {\it XMM-Newton} observations. The cluster is thought to undergo a merger event. Additionally, it has been claimed \cite{sereno+18} that this cluster is located in a particularly high-density environment, which might have favored the formation of the halo.

\item PSZ2\,G104.74+40.42 ($z=0.690$) shows hints of presence of extended radio emission. After subtracting the flux density of the radio galaxies in the cluster volume, the resulting diffuse radio emission result to be $S_{\rm 144MHz}<5\Delta S$, hence it is classified as ``uncertain'' detection. No X-ray observations are available.

\item Although the radio emission is dominated by three radio galaxies, PSZ2\,G126.28+65.62 ($z=0.820$) shows evidence of diffuse radio emission, spanning a region of about 800~kpc. No X-ray observations are available.

\item All the radio emission detected in PSZ2\,G127.01+26.21 ($z=0.630$) is associated with the radio galaxies. The X-ray emission is peaked on the BCG, appearing elongated in the north-south direction.

\item A small double-lobed radio galaxy is observed in the center of  PSZ2\,G139.00+50.92, without hints of diffuse emission. No X-ray observations are available.

\item The radio halo in PSZ2\,G141.77+14.19 extends over 600~kpc. The X-ray emission appears very disturbed and elongated in the northwest-southeast direction. 

\item PSZ2\,G141.98+69.31 ($z=0.714$) does not show extended radio emission. No X-ray observations are available.

\item Diffuse radio emission in PLCKG147.3--16.6 ($z=0.645$) extends in the east-west direction over 800~kpc.  Optical and X-ray data suggest the cluster is undergoing a merger event. The radio halo was earlier discovered using GMRT observations \cite{vanweeren+14}. Using the GMRT and LOFAR halo flux densities, we calculate an integrated spectral index $\alpha=-0.77\pm0.15$. This value is flatter than what we assume for the other radio halos in our sample, but it is consistent, within the uncertainties, with other literature radio halos \cite{feretti+12}. Further studies on the spectral index of this source are however needed.

\item The radio emission in PSZ2\, G147.88+53.24 ($z=0.600$) has a largest size of about 600 kpc. The emission encompasses the BCG in the south-west direction. No X-ray observations are available for this cluster.

\item PSZ2\,G160.83+81.66, at $z=0.888$, represents the highest redshift object of our sample. Diffuse radio emission is detected extending over 700~kpc centered around the central BCG. The radio emission follows the thermal gas emission (see last panel in Figure 2). 
This is the most distant radio halo discovered so far. Deep X-ray observations \cite{maughan+07} suggest this cluster has a relaxed morphology (see also last panel in Figure 2). 
However, no signs of central cooling have been found from an X-ray temperature analysis, suggesting a possible earlier merger event \cite{maughan+07}. In agreement with this, a weak lensing analysis revealed a bimodal mass distribution  \cite{jee+taylor09}.
\end{itemize}

\begin{table*}[h!]
\begin{center}
\caption*{\bf Supplementary Table 2: ObsIDs list of the LOFAR observations.}
%\resizebox{\textwidth}{!}{
\begin{tabular}{ll} %\checkmark
\hline
\hline
PSZ name 			& ObsID(s) \\
		 			& 	\\
\hline
PSZ2G045.87+57.70	& P230+27, P228+30, P231+30 \\
PSZ2G070.89+49.26   & P236+45, P240+45 \\
PSZ2G084.10+58.72	& P221+47,P223+50,P225+47 \\ 
PSZ2G086.28+74.76	& P203+37, P204+40 \\
PSZ2G086.93+53.18	& P227+55, P227+53, P231+53\\ 
PSZ2G089.39+69.36	& P207+45, P209+42 \\
PSZ2G091.83+26.11	& P275+60, P281+63, P275+63, P280+60\\ 
PSZ2G092.69+59.92	& P214+52, P215+50, P219+52, P219+50 \\
PSZ2G099.86+58.45	& P209+55, P214+55 \\
PSZ2G104.74+40.42	& P240+70, P232+70 \\ 
PSZ2G126.28+65.62	& P33Hetdex08, P29Hetdex19, P30Hetdex06\\
PSZ2G127.01+26.21	& P098+84, P113+87, P066+87 \\
PSZ2G139.00+50.92	& P168+62, P168+65 \\ 
PSZ2G141.77+14.19	& P075+69, P072+67, P068+69\\ 
PSZ2G141.98+69.31	& P20Hetdex17, P23Hetdex20, P19Hetdex17 \\
PLCKG147.3--16.6	& P044+39, P045+41, P042+41\\
PSZ2G147.88+53.24	& P166+60, P165+57\\ 
PSZ2G160.83+81.66	& P185+35, P188+35, P188+32, P185+32\\
\hline
\end{tabular}\label{tab:lofarIDs}%}
\end{center}
\end{table*}

\subsection*{Additional flux density measurements}
In Table 1 
we reported the flux density measurements and the largest linear size of the candidate radio halos obtained following the $2\sigma_{\rm rms}$ radio contour in the low-resolution source-subtracted image. Here, we report the same analysis but following the $3\sigma_{\rm rms}$ radio contour (see Supplementary Table 3). . 
Moreover, we also provide flux density measurements from the low-resolution source-subtracted images, following both the 2 and $3\sigma_{\rm rms}$ radio contours. The uncertainty on the flux densities are calculated as Equation~2 but, in this case, the term $\sigma_{\rm sub}^2$ refers to the uncertainty due to the source subtraction in the $uv$ plane, i.e. a fraction (about 3\%) of the total flux of the subtracted radio galaxies, see \cite{cassano+13,hoang+17}.

\begin{table*}[h!]
\begin{center}
\caption*{{\bf Supplementary Table 3: Integrated flux density and radio luminosity of the galaxy clusters observed with LOFAR.} Column 1: {\it Planck} cluster name. Column 2: Cluster redshift. Column 3: Largest linear size (LLS), following the  2 and $3\sigma_{\rm rms}$ radio contours. Columns 4 and 5: The integrated flux density and radio luminosity measurements, following the $2\sigma_{\rm rms}$ and $3\sigma_{\rm rms}$ contours (from Figure 4, 
and, for both, the values using the flux-subtraction (see Method) and the $uv$-subtraction (see above) strategies.}
\resizebox{\textwidth}{!}{%0.48\textwidth
\begin{tabular}{lccclcccccccc}
\hline
\hline
\noalign{\smallskip}
{\it Planck} (PSZ) name & $z$ & \multicolumn{2}{c}{LLS} & Classification	&  \multicolumn{4}{c}{$S_{\rm 144MHz}$} & \multicolumn{4}{c}{$P_{\rm 1.4GHz}$} \\
		 			&   &\multicolumn{2}{c}{[Mpc]}& & \multicolumn{4}{c}{[mJy]} 			& \multicolumn{4}{c}{[$10^{24}$ W Hz$^{-1}$]} \\
\cmidrule(lr){3-4}\cmidrule(lr){6-9}\cmidrule(lr){10-13}	 			
		 			& & $2\sigma_{\rm rms}$ & $3\sigma_{\rm rms}$ & & \multicolumn{2}{c}{$2\sigma_{\rm rms}$} & \multicolumn{2}{c}{$3\sigma_{\rm rms}$} & \multicolumn{2}{c}{$2\sigma_{\rm rms}$} & \multicolumn{2}{c}{$3\sigma_{\rm rms}$} \\
\cmidrule(lr){6-7}\cmidrule(lr){8-9}\cmidrule(lr){10-11}\cmidrule(lr){12-13}
		 			& & & & & flux sub & $uv$-sub & flux sub & $uv$-sub & flux sub & $uv$-sub & flux sub & $uv$-sub\\
%\cline{2-4}
\hline
\noalign{\smallskip}
PSZ2G045.87+57.70	& 0.611 & -- & -- & Uncertain	 & $-$ & $-$ & $-$ & $-$ & $-$ & $-$ & $-$ & $-$ \\
PSZ2G070.89+49.26   & 0.610 & -- & -- & --       	& $-$ & $-$ & $-$ & $-$ & $-$ & $-$ & $-$ & $-$ \\
PSZ2G084.10+58.72	& 0.731 &  -- &-- & Uncertain        & $-$ & $-$ & $-$ & $-$ & $-$ & $-$ & $-$ & $-$ \\ 
PSZ2G086.28+74.76	& 0.699 &  -- &-- & Uncertain        & $-$ & $-$ & $-$ & $-$ & $-$ & $-$ & $-$ & $-$  \\
PSZ2G086.93+53.18	& 0.675 & 0.5 & 0.3 & Halo/Uncertain  & $7.2\pm1.5$ & $7.9\pm1.6$ & $4.3\pm0.8$ & $5.2\pm1.0$ & $0.7\pm0.4$ & $0.8\pm0.5$ & $0.5\pm0.2$ & $0.5\pm0.3$ \\ 
PSZ2G087.39+50.92	& 0.748 & -- & -- & --            	& $-$ & $-$ & $-$ & $-$ & $-$ & $-$ & $-$ & $-$ \\ 
PSZ2G089.39+69.36	& 0.680 & 1.0 & 0.9 & Halo          	& $12.5\pm1.9$ & $11.9\pm1.7$ & $10.3\pm1.5$ & $10.1\pm1.5$ & $1.3\pm0.7$ & $1.1\pm0.6$ & $1.0\pm0.6$ & $0.9\pm0.5$ \\
\noalign{\smallskip}
\multirow{2}{*}
{PSZ2G091.83+26.11}	&
\multirow{2}{*}{0.822} & 1.2 & 1.2 & Halo              & $84.3\pm12.7$     & $73.0\pm9.0$ & $83.9\pm11.4$ & $72.8\pm9.0$ & $13.8\pm8.4$ & $12.4\pm5.5$ & $12.6\pm5.6$ & $11.8\pm4.0$\\ 
					& & 1.2 & & Relic	            & $259.4\pm38.9$   & -- & -- & -- & $-$ & -- & -- & -- \\
\noalign{\smallskip}
PSZ2G092.69+59.92	& 0.848 & & -- & -- & --	            & $-$ & $-$ & $-$ & $-$ & $-$ & $-$ & $-$ \\
PSZ2G099.86+58.45	& 0.616 & 1.2 & 1.0 & Halo              & $27.8\pm4.3$ & $21.5\pm3.6$ & $21.0\pm3.2$ & $17.6\pm3.0$ & $2.2\pm1.3$ & $1.7\pm0.9$ & $1.6\pm1.0$ & $1.4\pm0.9$ \\
PSZ2G104.74+40.42	& 0.690 & -- & -- & Uncertain         & $-$ & $-$ & $-$ & $-$ & $-$ & $-$ & $-$ & $-$ \\ 
PSZ2G126.28+65.62	& 0.820 & 0.8 & 0.8 & Halo       	    & $8.8\pm1.7$ & $9.2\pm1.8$ & $5.5\pm1.1$ & $6.3\pm1.2$ & $1.4\pm0.6$ & $1.7\pm0.9$ & $0.9\pm0.4$ & $1.4\pm0.9$\\
PSZ2G127.01+26.21	& 0.630 & -- & -- & Uncertain         & $-$ & $-$ \\
PSZ2G139.00+50.92	& 0.600 & -- & -- & --        	   & $-$ & $-$ & $-$ & $-$ & $-$ & $-$ & $-$ & $-$  \\ 
PSZ2G141.77+14.19	& 0.830 & 0.6 & 0.5 & Halo      	    & $8.8\pm1.4$ & $9.0\pm1.6$ & $4.7\pm0.8$ & $6.4\pm1.2$ & $1.4\pm0.7$ & $1.6\pm0.7$ & $0.7\pm0.4$ & $1.1\pm0.6$\\ 
PSZ2G141.98+69.31	& 0.714 & -- & -- & --        	   & $-$ & $-$ & $-$ & $-$ & $-$ & $-$ & $-$ & $-$  \\
PLCKG147.3--16.6	& 0.645 & 0.8 & 0.8 & Halo      	    & $22.5\pm3.7$ & $23.2\pm3.9$ & $19.0\pm3.0$ & $19.4\pm3.8$ & $6.4\pm3.4$ & $6.7\pm2.7$ & $4.9\pm1.7$ & $5.9\pm2.0$\\
PSZ2G147.88+53.24	& 0.600 & 0.6 & 0.6 & Halo              & $14.4\pm2.3$	& $11.8\pm2.3$ & $11.1\pm1.7$ & $10.0\pm2.1$ & $0.9\pm0.5$ & $0.7\pm0.5$ & $0.8\pm0.4$ & $0.8\pm0.6$\\ 
PSZ2G160.83+81.66	& 0.888 & 0.7 & 0.5 & Halo      	    & $13.0\pm2.1$ & $10.6\pm2.0$ & $10.2\pm1.6$ & $7.9\pm1.6$ &  $2.7\pm1.5$ & $2.2\pm1.1$ & $1.9\pm1.0$ & $1.7\pm0.6$\\
\noalign{\smallskip}
\hline
\end{tabular}\label{tab:occurrence_new}}
\end{center}
\end{table*}

\subsection*{Additional radio maps and integrated spectral index for PSZ2 G091.83+26.11}
PSZ2\,G091.83+26.11 was  observed with the Very Large Array (VLA), in the 1--2 GHz frequency band, in the B-array configuration on the 22nd of March 2015 (project code: 15A-270). The total time on source is about 40 minutes. Standard VLA data reduction \cite{digennaro+18} has been performed on the dataset  with \texttt{CASA v5.3} \cite{mcmullin+07}. Applied corrections included the  antenna delays,  bandpass, cross-hand  delays, and  polarisation leakage and angles using the primary calibrators 3C286 and 3C147. These calibration solutions were applied to the target, and self-calibration was  performed to refine the amplitudes and phases. The final image was produced using w-projection \cite{cornwell+05,cornwell+08}, Briggs weighting with \texttt{robust=0} and \texttt{nterms=3} \cite{rau+cornwell11}. The correction for the primary beam attenuation has also been applied. The final image resolution is $3.8''\times3.0''$, with a noise of $\sigma_{\rm rms}=50.7~\mu$Jy beam$^{-1}$. Only the brightest part of the relic is visible in this observation (Supplementary Figure 3).  
We measured a radio flux density of $S_{\rm 1.5GHz}\sim5.5$ mJy. Combined  LOFAR-VLA images yields an integrated spectral index of $\alpha\sim-1.6$. The lack of short baselines and short on-source integration time, means that the radio halo cannot be detected. 

\begin{figure}[h!]
\centering
{\includegraphics[width=0.6\textwidth]{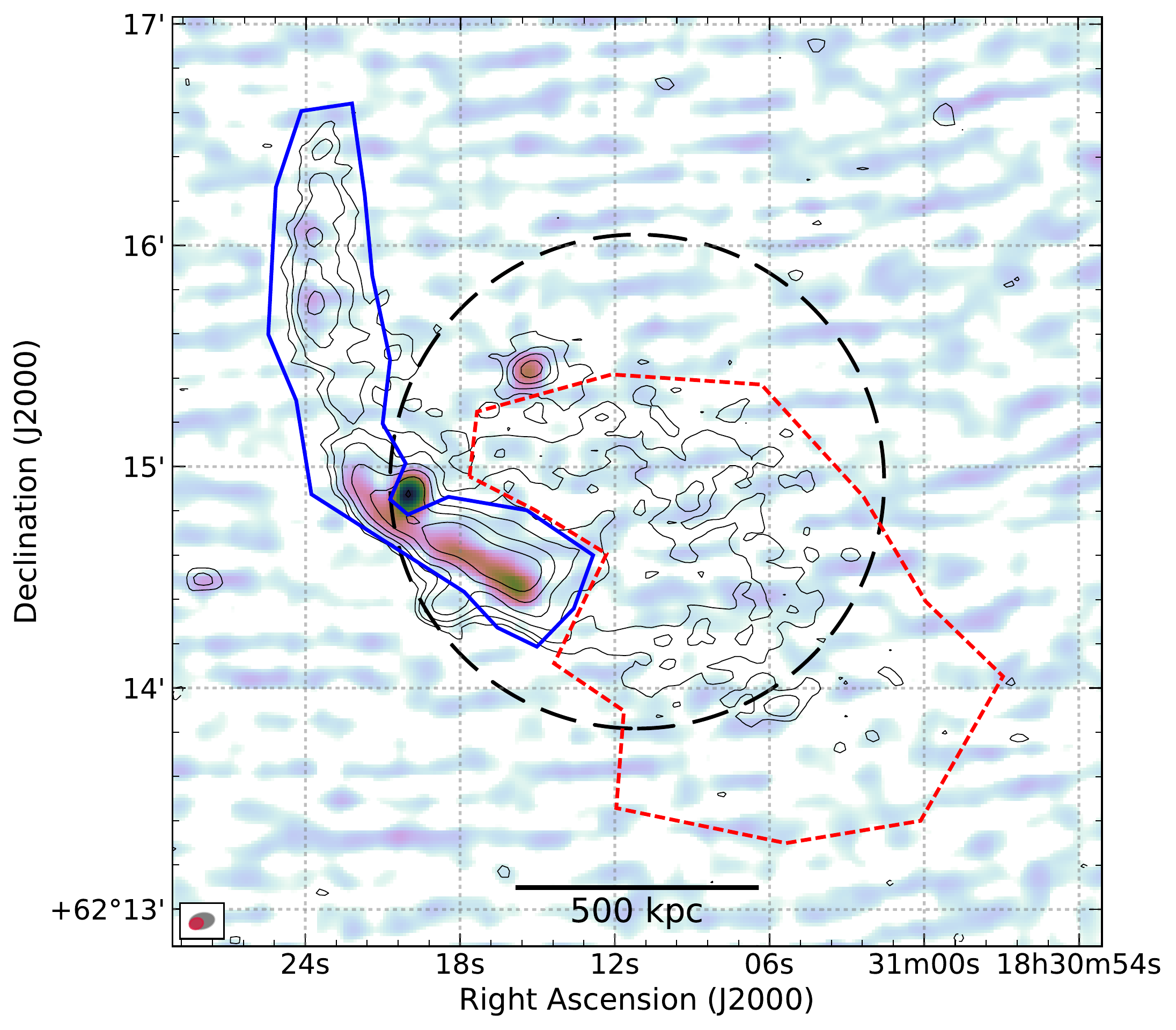}}
\caption*{{\bf Supplementary Figure 2: VLA-LOFAR comparison of PSZ2G091.83+26.11}. In colorscale we show the 1--2 GHz B-array VLA radio image of the cluster. The full-resolution LOFAR radio contours are displayed at $3\sigma_{\rm rms}\times[-1,1,2,4, \dots]$ level, with $\sigma_{\rm rms}$ the map noise and negative contours displayed with the short-dashed line, as comparison. The VLA and LOFAR beams are displayed in the bottom left corner, in pink and black respectively. The blue and red boxes show the regions where the flux densities were extracted, for the relic and the halo respectively.}
\label{fig:g091.83_vla}
\end{figure}

\subsection*{X-ray observations and images}
Here we list the summary (Supplementary Table 4) 
of the clusters with X-ray observations. For those clusters which had both {\it Chandra} and {\it XMM-Newton} observations, we only show the best case, i.e., a compromise between high resolution and exposure time.

\begin{table*}[h!]
\begin{center}
\caption*{\bf Supplementary Table 4: ObsIDs list of the X-ray observations.}
%\resizebox{\textwidth}{!}{
\begin{tabular}{lccc} %\checkmark
\hline
\hline
PSZ name 			& Satellite			&ObsID(s)		& Total exposure time \\
		 			& 				    & 			& [ks]	 \\
\hline
PSZ2G045.87+57.70	& {\it XMM-Newton}	& 0693661101	& 32 \\
PSZ2G070.89+49.26	& {\it XMM-Newton}	& 0693661301 	& 50 \\ 
PSZ2G084.10+58.72	& {\it XMM-Newton}	& 783880901    	& 86 \\
PSZ2G086.93+53.18	& {\it XMM-Newton}	& 0783880701   	& 81 \\
PSZ2G087.39+50.92	& {\it XMM-Newton}	& 0783881201	& 22 \\
PSZ2G091.83+26.11	& {\it Chandra}		& 18285	    	& 23 \\
\noalign{\smallskip}
\multirow{2}{*}{PSZ2G092.69+59.92}	& \multirow{2}{*}{\it XMM-Newton}	& 0783880401 & \multirow{2}{*}{107} \\
                                    &                                   & 0783881901 & \\
\noalign{\smallskip}
\noalign{\smallskip}
\multirow{3}{*}{PSZ2G099.86+58.45}	& \multirow{3}{*}{\it XMM-Newton}	& 0693660601 & \multirow{3}{*}{63} \\
							    	&					        		& 0693662701 & \\
							    	&					        		& 0723780301 & \\
\noalign{\smallskip}
PSZ2G127.01+26.21	& {\it Chandra}		& 18286		& 16 	\\
PSZ2G141.77+14.19	& {\it Chandra}		& 18289		& 21 \\
PLCKG147.3--16.6		& {\it XMM-Newton}	& 0679181301	& 10 \\
\noalign{\smallskip}
\multirow{2}{*}{PSZ2G160.83+81.66}	& \multirow{2}{*}{\it Chandra}		& 3180	& \multirow{2}{*}{67} \\
					    			&		        					& 5014  & \\
\hline
\end{tabular}\label{tab:xray}%}
\end{center}
\end{table*}


\begin{thebibliography}{99}
\bibitem{carilli+taylor02} Carilli, C.~L. \& Taylor, G.~B. Cluster Magnetic Fields. {\it Annnual Review Astron. Astrophys.} {\bf 40}, 319 (2002).

\bibitem{bonafede+10} Bonafede, A., Feretti, L., Murgia, M., Govoni, F., Giovannini, G. et al. The Coma cluster magnetic field from Faraday rotation measures. {\it Astron. Astrophys} {\bf 513}, 30 (2010).

\bibitem{vanweeren+19} van Weeren, R. J., de Gasperin, F., Akamatsu, H., Br\"uggen, M., Feretti, L.  et al. Diffuse Radio Emission from Galaxy Clusters. {\it Space Sci. Rev.} {\bf 215}, 16 (2019).

\bibitem{cassano+10a} Cassano, R., Ettori, S., Giacintucci, S., Brunetti, G., Markevitch, M. et al. On the Connection Between Giant Radio Halos and Cluster Mergers. {\it Astrophys. J.} {\bf 721}, 82--85 (2010).

\bibitem{brunetti+jones14} Brunetti, G. \& Jones, T. W. Cosmic Rays in Galaxy Clusters and Their Nonthermal Emission. {\it Int. J. Mod. Phys. D.} {\bf 23} 1430007--98 (2014).

%\bibitem{markevitch+vikhlinin07} Markevitch, M. \& Vikhlinin, A. Shocks and cold fronts in galaxy clusters. {\it Physics Reports}, {\bf 443}, 1-53 (2007).

\bibitem{dolag+05} Dolag, K., Grasso, D., Springel, V. \& Tkachev, I. Constrained simulations of the magnetic field in the local Universe and the propagation of ultrahigh energy cosmic rays. {\it Journal of Cosmology and Astropart. Phys.} {\bf 2005}, 9 (2005).

\bibitem{subramanian+06} Subramanian, K., Shukurov, A. \& Haugen, N. E. L. Evolving turbulence and magnetic fields in galaxy clusters. {\it Mon. Not. R. Astron. Soc} {\bf 366}, 1437--1454 (2006).

\bibitem{ryu+08} Ryu, D., Kang, H., Cho, J. \& Das, S. Turbulence and Magnetic Fields in the Large-Scale Structure of the Universe. {\it Science} {\bf 320}, (5878):909 (2008).

\bibitem{miniati+beresnyak15} Miniati, F. \& Beresnyak, A. Self-similar energetics in large clusters of galaxies. {\it Nature} {\bf 523}, (7558):59--62 (2015).

\bibitem{vazza+18} Vazza, F., Brunetti, G., Br\"uggen, M. \& Bonafede, A. Resolved magnetic dynamo action in the simulated intracluster medium. {\it Mon. Not. R. Astron. Soc} {\bf 474}, 1672--1687 (2018).

\bibitem{dominguez-fernandez+19} Dom\'{i}nguez-Fern\'{a}ndez, P., Vazza, F., Br\"uggen, M. \& Brunetti, G. Dynamical evolution of magnetic fields in the intracluster medium. {\it Mon. Not. R. Astron. Soc} {\bf 486}, 623--638 (2019).

\bibitem{donnert+18} Donnert, J., Vazza, F., Br\"uggen, M. \& ZuHone, J. Magnetic Field Amplification in Galaxy Clusters and Its Simulation. {\it Space Sci. Rev.} {\bf 214}, 122 (2018).


\bibitem{vanhaarlem+13} van Haarlem, M.~P., Wise, M.~W., Gunst, A.~W., Heald, G., McKean, J.~P. et al. LOFAR: The LOw-Frequency ARray. {\it Astron. Astrophys} {\bf 556}, 2 (2013).

\bibitem{shimwell+19} Shimwell, T. W. , Tasse, C., Hardcastle, M. J., Mechev, A. P., Williams, W. L. et al. The LOFAR Two-metre Sky Survey. II. First data release. {\it Astron. Astrophys} {\bf 622}, 1 (2019).

\bibitem{planckcoll16} Planck Collaboration: Ade, P. A. R., Aghanim, N., Arnaud, M., Ashdown, M., Aumont, J. et al. Planck 2015 results. XXVII. The second Planck catalogue of Sunyaev-Zeldovich sources. {\it Astron. Astrophys} {\bf 594}, 27 (2016).


\bibitem{cassano+13} Cassano, R., Ettori, S., Brunetti, G., Giacintucci, S., Pratt, G. W. et al. Revisiting Scaling Relations for Giant Radio Halos in Galaxy Clusters. {\it Astrophys. J.} {\bf 777}, 141 (2013).


\bibitem{cassano+19} Cassano, R., Botteon, A., Di Gennaro, G., Brunetti, G., Sereno, M. et al. LOFAR Discovery of a Radio Halo in the High-redshift Galaxy Cluster PSZ2 G099.86+58.45. {\it Astrophys. J.} {\bf 881}, 18 (2019).

\bibitem{brunetti+vazza20} Brunetti, G. \& Vazza, F.\ econd-order Fermi Reacceleration Mechanisms and Large-Scale Synchrotron Radio Emission in Intracluster Bridges. {\it Phys. Rev. Lett.} {\bf 124}, 051101 (2020).

\bibitem{vanweeren+14} van Weeren, R. J., Intema, H. T., Lal, D. V., Bonafede, A., Jones, C. et al. The Discovery of a Radio Halo in PLCK G147.3-16.6 at $z = 0.65$. {\it Astrophys. J.}, {\bf 781} 32 (2014).

\bibitem{lindner+14} Lindner, R. R., Baker, A. J., Hughes, J. P., Battaglia, N., Gupta, N. et al. The Radio Relics and Halo of El Gordo, a Massive $z = 0.870$ Cluster Merger. {\it Astrophys. J.} {\bf 786}, 49 (2014).

%\bibitem{fakhouri+10} Fakhouri, O., Ma, C.-P., \& Boylan-Kolchin, M.\ he merger rates and mass assemblyhistories of dark matter haloes in the two Millennium simulations. {\it Mon. Not. R. Astron. Soc}, {\bf 406}, 2267, (2010).

%\bibitem{giocoli+12} Giocoli, C., Tormen, G., and Sheth, R. K.. Formation times, mass growth histories and concentrations of dark matter haloes. {\it Mon. Not. R. Astron. Soc}, {\bf 422}, 185--198, (2012).

\bibitem{cho14} Cho, J. Origin of Magnetic Field in the Intracluster Medium: Primordial or Astrophysical? {\it Astrophys. J.} {\bf 797}, 133 (2014).

\bibitem{beresnyak+miniati16} Beresnyak, A. \& Miniati, F. Turbulent Amplification and Structure of the Intracluster Magnetic Field. {\it Astrophys. J.} {\bf 817}, 127 (2016).


\bibitem{hitomicoll18} Hitomi Collaboration: Aharonian, F., Akamatsu, H., Akimoto, F., Allen, S.~W., Angelini, L. et al. Atmospheric gas dynamics in thePerseus cluster observed with Hitomi. {\it Publ. Astron. Soc. Japan} {\bf 70}, 9 (2018).


\bibitem{markevitch+vikhlinin07} Markevitch, M. \& Vikhlinin, A. Shocks and cold fronts in galaxy clusters. {\it Physics Reports} {\bf 443}, 1-53 (2007).


\bibitem{brunetti+lazarian07} Brunetti, G. \& Lazarian. A. Compressible turbulence in galaxy clusters: physics and stochastic particle re-acceleration. {\it Mon. Not. R. Astron. Soc} {\bf 378} 245--275 (2007).

\bibitem{schekochihin+cowley06} Schekochihin, A. A.  \& Cowley, S. C. Turbulence, magnetic fields, and plasma physics in clusters of galaxies. {\it Phys. Plasmas} {\bf 13}, 056501--056501 (2006).

\bibitem{zhuravleva+19} Zhuravleva, I., Churazov, E., Schekochihin, A. A., Allen, S. W., Vikhlinin, A. et al. Suppresse deffective viscosity in the bulk intergalactic plasma. {\it Nat. Astron.} {\bf 3}, 832--837 (2019).


\bibitem{xu+11} Xu, H., Li, H., Collins, D. C., Li, S. \& Norman, M. L. Evolution and Distribution of Magnetic Fields from Active Galactic Nuclei in Galaxy Clusters. II. The Effects of Cluster Size and Dynamical State. {\it Astrophys. J.} {\bf 739}, 77 (2011).





\bibitem{eckert+11} Eckert, D., Molendi, S. \& Paltani, S. The cool-core bias in X-ray galaxy cluster samples. I. Methodand application to HIFLUGCS. {\it Astron. Astrophys} {\bf 526}, 79 (2011).

\bibitem{rossetti+17} Rossetti, M., Gastaldello, F., Eckert, D., Della Torre, M., Pantiri, G. et al. The cool-core state of Planck SZ-selected clusters versus X-ray-selected samples: evidence for cool-core bias. {\it Mon. Not. R. Astron. Soc} {\bf 468}, 1917 (2017).

\bibitem{andrade-santos+2017} Andrade-Santos, F., Jones, C., Forman, W.~R., Lovisari, L., Vikhlinin, A. et al. The Fraction of Cool-core Clusters in X-Ray versus SZSamples Using Chandra Observations. 2017, {\it Astrophys. J.} {\bf 843}, 76 (2017).

\bibitem{amodeo+18} Amodeo, S., Mei, S., Stanford, S. A., Lawrence, C. R., Bartlett, J. G. et al. Spectroscopic Confirmation and Velocity Dispersions for 20 Planck Galaxy Clusters at $0.16 < z < 0.78$. {\it Astrophys. J.} {\bf 853}, 36 (2018).

\bibitem{barrena+18} Barrena, R., Streblyanska, A., Ferragamo, A., Rubi{\~n}o-Mart\'{i}n, J. A., Aguado-Barahona, A. et al. Optical validation and characterization of Planck PSZ1 sources at the Canary Islands observatories. I. First year of ITP13 observations. {\it Astron. Astrophys} {\bf 616}, 42 (2018).

\bibitem{burenin+18} Burenin, R. A., Bikmaev, I. F., Khamitov, I. M., Zaznobin, I. A., Khorunzhev, G. A. et al. Optical Identifications of High-Redshift Galaxy Clusters from the Planck Sunyaev-Zeldovich Survey. {\it Astronomy Letters} {\bf 44}, 297--308 (2018).

\bibitem{sereno+18} Sereno, M., Giocoli, C., Izzo, L., Marulli, F., Veropalumbo, A. et al. Gravitational lensing detection of an extremely dense environment around a galaxy cluster. {\it Nat. Astron.} {\bf 2}, 744--750 (2018).

\bibitem{streblyanska+18} Streblyanska, A., Barrena, R., Rubi{\~n}o-Mart\'{i}n, J. A., van der Burg, R. F. J., Aghanim,  N. et al. Characterization of a sub-sample of the Planck SZ source cluster catalogues using optical SDSS DR12 data. {\it Astron. Astrophys} {\bf 617}, 71 (2018).

\bibitem{vanderburg+16} van der Burg, R. F. J., Aussel, H., Pratt, G. W., Arnaud, M., Melin, J. B. et al. Prospects for high-z cluster detections with Planck, based on a follow-up of 28 candidates using MegaCam at CFHT. {\it Astron. Astrophys} {\bf 587}, 23, (2016).

\bibitem{zohren+19} Zohren, H., Schrabback, T., van der Burg, R. F. J., Arnaud, M., Melin, J. et al. Optical follow-up study of 32 high-redshift galaxy cluster candidates from Planck with the William Herschel Telescope. {\it Mon. Not. R. Astron. Soc} {\bf 488}, 2523--2542 (2019).

%
%\bibitem{vanweeren+14} van Weeren, R. J., Intema, H. T., Lal, D. V., Bonafede, A., Jones, C., et al. The Discovery of a Radio Halo in PLCK G147.3-16.6 at $z = 0.65$. {\it Astrophys. J.}, {\bf 781}, 32, (2014).
%

\bibitem{panstarss16} Chambers, K.~C., Magnier, E.~A., Metcalfe, N., Flewelling, H.~A., Huber, M.~E. et al. The Pan-STARRS1 Surveys. arXiv e-prints, arXiv:1612.05560 (2016).

\bibitem{vanweeren+16} van Weeren, R. J., Williams, W. L., Hardcastle, M. J., Shimwell, T. W., Rafferty, D. A. et al. LOFAR Facet Calibration. {\it Astrophys. J. Suppl.} {\bf 223}, 2 (2016).

\bibitem{williams+16} Williams, W. L., van Weeren, R. J., R\"ottgering, H. J. A., Best, P., Dijkema, T. J. et al. LOFAR 150-MHz observations of the Bo\"otes field: catalogue and source counts. {\it Mon. Not. R. Astron. Soc} {\bf 460}, 2385--2412 (2016).

\bibitem{degasperin+19} de Gasperin, F., Dijkema, T. J., Drabent, A., Mevius, M., Rafferty, D. et al. Systematic effects in LOFAR data: A unified calibration strategy. {\it Astron. Astrophys} {\bf 622}, 5 (2019).

\bibitem{tasse14} Tasse, C. Nonlinear Kalman filters for calibration in radio interferometry. {\it Astron. Astrophys}, {\bf 566}, 127  (2014).

\bibitem{smirnov+tasse15} Smirnov, O. M. \& Tasse, C.  Radio interferometric gain calibration as a complex optimization problem. {\it Mon. Not. R. Astron. Soc} {\bf 449}, 2668--2684 (2015).

\bibitem{tasse+18} Tasse, C., Hugo, B., Mirmont, M., Smirnov, O., Atemkeng M. et al. Faceting for direction-dependent spectral deconvolution. {\it Astron. Astrophys} {\bf 611}, 87 (2018).

\bibitem{offringa+14} Offringa, A. R., McKinley, B., Hurley-Walker, N., Briggs, F. H., Wayth, R. B. et al. WSCLEAN: an implementation of a fast, generic wide-field imager for radio astronomy. {\it Mon. Not. R. Astron. Soc} {\bf 444}, 606--619 (2014).

\bibitem{offringa+17} Offringa, A. R. \& Smirnov,O. An optimized algorithm for multiscale wide-band deconvolution of radio astronomical images. {\it Mon. Not. R. Astron. Soc} {\bf 471}, 301--316 (2017).

\bibitem{degasperin+17} de Gasperin, F., Intema, H.~T., Shimwell, T.~W., Gianfranco, B., Br\"uggen, M. et al. Gentle re-energization of electrons in merging galaxy clusters. {\it Science Advances} {\bf 3}, 1701634 (2017).

\bibitem{mandal+20} Mandal, S., Intema, H.~T., van Weeren, R.~J., Shimwell, T.~W., Botteon, A. et al. Revived fossil plasma sources in galaxy clusters. {\it Astron. Astrophys} {\bf 634}, 4 (2020)

\bibitem{giacintucci+17} Giacintucci, S., Markevitch, M., Cassano, R., Venturi, T., Clarke, T. et al. Occurrence of Radio Minihalos in a Mass-limited Sample of Galaxy Clusters. {\it Astrophys. J.} {\bf 841}, 71 (2017)

\bibitem{cassano+06} Cassano, R., Brunetti, G. \& Setti, G. Constraining B in galaxy clusters from statistics of giantradio halos, {\it Astronomische Nachrichten} {\bf 327}, 557 (2006).

\bibitem{vikhlinin+05} Vikhlinin, A., Markevitch, M., Murray, S. S., Jones, C., Forman, W. et al. Chandra Temperature Profiles for a Sample of Nearby Relaxed Galaxy Clusters. {\it Astrophys. J.} {\bf 628}, 655--672 (2005).


\bibitem{casano+brunetti05} Cassano, R. \& Brunetti, G. Cluster mergers and non-thermal phenomena: a statistical magneto-turbulent model. {\it Mon. Not. R. Astron. Soc} {\bf 357}, 1313--1329 (2005).

\bibitem{sarazin02} Sarazin, C. L. The Physics of Cluster Mergers. Merging Processes in Galaxy Clusters. {\it Astrophysics and Space Science Library}, {\bf 272} 1--38 (2002).

\bibitem{kitayama+suto96} Kitayama, T. \& Suto, Y.  Semianalytic Predictions for Statistical Properties of X-Ray Clusters of Galaxies in Cold Dark Matter Universes. {\it Astrophys. J.} {\bf 469}, 480 (1996).


\bibitem{neeser+95} Neeser, M.~J., Eales, S.~A., Law-Green, J.~D., Leahy, J.~P., Rawlings, S. The Linear-Size Evolution of Classical Double Radio Sources. {\it Astrophys. J.} {\bf 451}, 76 (1995).


\bibitem{blundell+99} Blundell, K.~M., Rawlings, S. \& Willott, C.~J. The Nature and Evolution of Classical Double Radio Sources from Complete Samples. {\it Astron. J.} {\bf 117}, 677 (1999).


\bibitem{smolciv+17} Smol{\v{c}}i{\'c}, V., Novak, M., Delvecchio, I., Ceraj, L., Bondi, M. et al. The VLA-COSMOS 3 GHz Large Project: Cosmic evolution of radio AGN and implications for radio-mode feedback since $z=5$. {\it Astron. Astrophys} {\bf 602}, 6 (2017).

\bibitem{brunetti+lazarian11} Brunetti, G. \& Lazarian, A. Acceleration of primary and secondary particles in galaxy clusters by compressible MHD turbulence: from radio haloes to gamma-rays. {\it Mon. Not. R. Astron. Soc} {\bf 410}, 127 (2011).

\bibitem{pinzke+17} Pinzke, A., Peng Oh, S. \& Pfrommer, C. Turbulence and particle acceleration in giant radio haloes: the origin of seed electrons. {\it Mon. Not. R. Astron. Soc} {\bf 465}, 4800--4816 (2017).

\bibitem{brunetti+17} Brunetti, G., Zimmer, S. \& Zandanel, F. Relativistic protons in the Coma galaxy cluster: first gamma-ray constraints ever on turbulent. {\it Mon. Not. R. Astron. Soc} {\bf 472}, 1506 (2017).

\bibitem{vazza+14} Vazza, F., Gheller, C. \& Br{\"u}ggen, M. Simulations of cosmic rays in large-scale structures: numerical and physical effects. {\it Mon. Not. R. Astron. Soc} {\bf 439}, 2662 (2014).

\bibitem{vazza+12} Vazza, F., Br{\"u}ggen, M., Gheller, C. \& Brunetti, G. Modelling injection and feedback of cosmic rays in grid-based cosmological simulations: effects on cluster outskirts. {\it Mon. Not. R. Astron. Soc} {\bf 421}, 3375 (2012).


\bibitem{beresnyak12} Beresnyak, A. Universal Nonlinear Small-Scale Dynamo. {\it Phys. Rev. Lett.} {\bf 108}, 035002 (2012).

%\bibitem{overzier16} Overzier, R.~A. The realm of the galaxy protoclusters. A review  {\it Astron. Astrophys. Reviews}, {\bf 24}, 14, (2016).


\bibitem{fakhouri+10} Fakhouri, O., Ma, C.-P., \& Boylan-Kolchin, M. The merger rates and mass assembly histories of dark matter haloes in the two Millennium simulations. {\it Mon. Not. R. Astron. Soc} {\bf 406}, 2267 (2010).

\bibitem{giocoli+12} Giocoli, C., Tormen, G., \& Sheth, R. K. Formation times, mass growth histories and concentrations of dark matter haloes. {\it Mon. Not. R. Astron. Soc} {\bf 422}, 185--198 (2012).


\bibitem{roh+19} Roh, S., Ryu, D., Kang, H., Ha, S. \& Jang, H. Turbulence Dynamo in the Stratified Medium of Galaxy Clusters. {\it Astrophy. J.} {\bf 883}, 138 (2019).

%\newpage
%\onecolumn
%\input{supplementary.tex}

%SUPPLEMENTARY
%\bibitem{planckcoll15} Planck Collaboration: Ade, P. A. R., Aghanim, N., Arnaud, M., Ashdown, M., Aumont, J., et al. Planck intermediate results. XXVI. Optical identification and redshifts of Planck clusters with the RTT150 telescope. {\it Astron. Astrophys}, {\bf 582}, 29 (2015).

%\bibitem{feretti+12} Feretti, L., Giovannini, G., Govoni, F., \& Murgia, M., Clusters of galaxies: observational properties of the diffuse radio emission. {\it Astron. Astrophy. Reviews}, {\bf 20}, 54 (2012).


%\bibitem{maughan+07} Maughan, B.~J., Jones, C., Jones, L.~R., \& Van Speybroeck, L. Deep XMM-Newton and Chandra Observations of ClJ1226.9+3332: A Detailed X-Ray Mass Analysis of a $z=0.89$ Galaxy Cluster. {\it Astrophys. J.}, {\bf 659}, 1125 (2007).

%\bibitem{jee+taylor09} Jee, M.~J., \& Tyson, J.~A. Dark Matter in the Galaxy Cluster CL J1226+3332 at $z=0.89$. {\it Astrophys. J.}, {\bf 691}, 1337 (2009).

%\bibitem{hoang+17} Hoang, D.~N., Shimwell, T.~W., Stroe, A., Akamatsu, H., Brunetti, G., et al. Deep LOFAR observations of the merging galaxy cluster CIZA J2242.8+5301. {\it Mon. Not. R. Astron. Soc}, {\bf 471}, 1107 (2017).


%\bibitem{digennaro+18} Di Gennaro, G., van Weeren, R.~J., Hoeft, M., Kang, H., Ryu, D., et al. Deep Very Large Array Observations of the Merging Cluster CIZAJ2242.8+5301: Continuum and Spectral Imaging. {\it Astrophys. J.}, {\bf 865}, 24 (2018).

%\bibitem{mcmullin+07} McMullin, J.~P., Waters, B., Schiebel, D., Young, W., \& Golap, K. CASA Architecture and Applications. {\it Astronomical Data Analysis Software and Systems XVI}, {\bf 376}, 127 (2007).

%\bibitem{cornwell+05} Cornwell, T.~J., Golap, K., \& Bhatnagar, S., W-Projection: A New Algorithm for Wide Field Imaging with Radio Synthesis Arrays. {\it Astronomical Data Analysis Software and Systems XVI}, {\bf 347}, 86 (2007).

%\bibitem{cornwell+08} Cornwell, T.~J., Golap, K., \& Bhatnagar, S., The Noncoplanar Baselines Effect in Radio Interferometry: The W-Projection Algorithm. {\it IEEE Journal of Selected Topics in Signal Processing}, {\bf 2}, 647-657 (2008).

%\bibitem{rau+cornwell11} Rau, U., \& Cornwell, T.~J.  multi-scale multi-frequency deconvolution algorithm for synthesisimaging in radio interferometry. {\it Astron. Astrophys}, {\bf 532}, 71 (2011).


\end{thebibliography}

\begin{thebibliography}{99}
\bibitem{planckcoll16} Planck Collaboration: Ade, P. A. R., Aghanim, N., Arnaud, M., Ashdown, M., Aumont, J. et al. Planck 2015 results. XXVII. The second Planck catalogue of Sunyaev-Zeldovich sources. {\it Astron. Astrophys}, {\bf 594}, 27 (2016).

\bibitem{streblyanska+18} Streblyanska, A., Barrena, R., Rubi{\~n}o-Mart\'{i}n, J. A., van der Burg, R. F. J., Aghanim,  N. et al. Characterization of a sub-sample of the Planck SZ source cluster catalogues using optical SDSS DR12 data. {\it Astron. Astrophys}, {\bf 617}, 71 (2018).

\bibitem{zohren+19} Zohren, H., Schrabback, T., van der Burg, R. F. J., Arnaud, M., Melin, J. et al. Optical follow-up study of 32 high-redshift galaxy cluster candidates from Planck with the William Herschel Telescope. {\it Mon. Not. R. Astron. Soc}, {\bf 488}, 2523--2542 (2019).

\bibitem{burenin+18} Burenin, R. A., Bikmaev, I. F., Khamitov, I. M., Zaznobin, I. A., Khorunzhev, G. A. et al. Optical Identifications of High-Redshift Galaxy Clusters from the Planck Sunyaev-Zeldovich Survey. {\it Astronomy Letters}, {\bf 44}, 297--308 (2018).

\bibitem{amodeo+18} Amodeo, S., Mei, S., Stanford, S. A., Lawrence, C. R., Bartlett, J. G. et al. Spectroscopic Confirmation and Velocity Dispersions for 20 Planck Galaxy Clusters at $0.16 < z < 0.78$. {\it Astrophys. J.}, {\bf 853}, 36 (2018).

\bibitem{cassano+19} Cassano, R., Botteon, A., Di Gennaro, G., Brunetti, G., Sereno, M. et al. LOFAR Discovery of a Radio Halo in the High-redshift Galaxy Cluster PSZ2 G099.86+58.45. {\it Astrophys. J.}, {\bf 881}, 18 (2019).

\bibitem{sereno+18} Sereno, M., Giocoli, C., Izzo, L., Marulli, F., Veropalumbo, A. et al. Gravitational lensing detection of an extremely dense environment around a galaxy cluster. {\it Nat. Astr.}, {\bf 2}, 744--750 (2018).

\bibitem{barrena+18} Barrena, R., Streblyanska, A., Ferragamo, A., Rubi{\~n}o-Mart\'{i}n, J. A., Aguado-Barahona, A. et al. Optical validation and characterization of Planck PSZ1 sources at the Canary Islands observatories. I. First year of ITP13 observations. {\it Astron. Astrophys}, {\bf 616}, 42 (2018).

\bibitem{vanderburg+16} van der Burg, R. F. J., Aussel, H., Pratt, G. W., Arnaud, M., Melin, J. B. et al. Prospects for high-z cluster detections with Planck, based on a follow-up of 28 candidates using MegaCam at CFHT. {\it Astron. Astrophys}, {\bf 587}, 23, (2016).

\bibitem{vanweeren+14} van Weeren, R. J., Intema, H. T., Lal, D. V., Bonafede, A., Jones, C., et al. The Discovery of a Radio Halo in PLCK G147.3-16.6 at $z = 0.65$. {\it Astrophys. J.}, {\bf 781}, 32, (2014).

\bibitem{planckcoll15} Planck Collaboration: Ade, P. A. R., Aghanim, N., Arnaud, M., Ashdown, M., Aumont, J., et al. Planck intermediate results. XXVI. Optical identification and redshifts of Planck clusters with the RTT150 telescope. {\it Astron. Astrophys}, {\bf 582}, 29 (2015).

\bibitem{feretti+12} Feretti, L., Giovannini, G., Govoni, F., \& Murgia, M., Clusters of galaxies: observational properties of the diffuse radio emission. {\it Astron. Astrophy. Reviews}, {\bf 20}, 54 (2012).


\bibitem{maughan+07} Maughan, B.~J., Jones, C., Jones, L.~R., \& Van Speybroeck, L. Deep XMM-Newton and Chandra Observations of ClJ1226.9+3332: A Detailed X-Ray Mass Analysis of a $z=0.89$ Galaxy Cluster. {\it Astrophys. J.}, {\bf 659}, 1125 (2007).

\bibitem{jee+taylor09} Jee, M.~J., \& Tyson, J.~A. Dark Matter in the Galaxy Cluster CL J1226+3332 at $z=0.89$. {\it Astrophys. J.}, {\bf 691}, 1337 (2009).

\bibitem{cassano+13} Cassano, R., Ettori, S., Brunetti, G., Giacintucci, S., Pratt, G. W. et al. Revisiting Scaling Relations for Giant Radio Halos in Galaxy Clusters. {\it Astrophys. J.}, {\bf 777}, 141 (2013).

\bibitem{hoang+17} Hoang, D.~N., Shimwell, T.~W., Stroe, A., Akamatsu, H., Brunetti, G., et al. Deep LOFAR observations of the merging galaxy cluster CIZA J2242.8+5301. {\it Mon. Not. R. Astron. Soc}, {\bf 471}, 1107 (2017).


\bibitem{digennaro+18} Di Gennaro, G., van Weeren, R.~J., Hoeft, M., Kang, H., Ryu, D., et al. Deep Very Large Array Observations of the Merging Cluster CIZAJ2242.8+5301: Continuum and Spectral Imaging. {\it Astrophys. J.}, {\bf 865}, 24 (2018).

\bibitem{mcmullin+07} McMullin, J.~P., Waters, B., Schiebel, D., Young, W., \& Golap, K. CASA Architecture and Applications. {\it Astronomical Data Analysis Software and Systems XVI}, {\bf 376}, 127 (2007).

\bibitem{cornwell+05} Cornwell, T.~J., Golap, K., \& Bhatnagar, S., W-Projection: A New Algorithm for Wide Field Imaging with Radio Synthesis Arrays. {\it Astronomical Data Analysis Software and Systems XVI}, {\bf 347}, 86 (2007).

\bibitem{cornwell+08} Cornwell, T.~J., Golap, K., \& Bhatnagar, S., The Noncoplanar Baselines Effect in Radio Interferometry: The W-Projection Algorithm. {\it IEEE Journal of Selected Topics in Signal Processing}, {\bf 2}, 647-657 (2008).

\bibitem{rau+cornwell11} Rau, U., \& Cornwell, T.~J.  multi-scale multi-frequency deconvolution algorithm for synthesisimaging in radio interferometry. {\it Astron. Astrophys}, {\bf 532}, 71 (2011).
\end{thebibliography}
\end{document}